\documentclass[11pt,a4paper,twoside,openright]{article}
\usepackage{authblk}

\usepackage[utf8]{inputenc}
\usepackage[russian, francais, english]{babel}
\usepackage[T1]{fontenc}
\usepackage[left=3cm,right=3cm,top=3cm,bottom=3cm]{geometry}
\usepackage{amsmath}
\allowdisplaybreaks 
\usepackage{amsfonts}
\usepackage[numbers]{natbib}
\usepackage[globalcitecopy, labelstoglobalaux]{bibunits}
\usepackage{bibentry}
\nobibliography* 
\usepackage{titletoc} 
\usepackage{appendix}%
\usepackage[tabbotcap, TABBOTCAP]{subfigure}
\usepackage[hidelinks]{hyperref}
\usepackage{import}
\usepackage{amssymb}
\usepackage{graphicx}
  \usepackage[export]{adjustbox}%
\usepackage{setspace}
\usepackage{placeins}
\usepackage{changepage}
\usepackage{caption}
\usepackage{numprint}
\npstyleenglish
\npthousandsep{}
\usepackage{multicol}
\usepackage{multirow, bigdelim}
\usepackage{longtable}
\usepackage{indentfirst}
\usepackage{url}
\usepackage[outdir=fig/]{epstopdf}
\usepackage{color}
\usepackage[usenames,dvipsnames,svgnames,table]{xcolor}
\usepackage{arydshln}
\usepackage{afterpage}
\usepackage{float}
\usepackage{listings}
\usepackage{xpatch}

\usepackage{titlesec}

\titleformat{\paragraph}
{\normalfont\normalsize\bfseries}{\theparagraph}{1em}{}
\titlespacing*{\paragraph}
{0pt}{3.25ex plus 1ex minus .2ex}{1.5ex plus .2ex}

\title{\Large \textbf{Effect of the Reynolds number on turbulence kinetic energy exchanges in flows with highly variable fluid properties}}
\def\runtitle{Effect of the Reynolds number in flows with highly variable fluid properties}

\def\runauthor{D. Dupuy, A. Toutant and F. Bataille}
\author[1]{Dorian Dupuy}
\author[1]{Adrien Toutant\thanks{Corresponding author : adrien.toutant@univ-perp.fr}}
\author[1]{Fran\c coise Bataille}
\affil[1]{PROMES CNRS, Universit\'{e} de Perpignan Via Domitia, Rambla de la thermodynamique, Tecnosud, 66100 Perpignan, France}

\date{\vspace{-4ex}\itshape\small (Published version: Physics of Fluids 31, 015104 (2019); https://doi.org/10.1063/1.5080769)\vspace{-4ex}}

\usepackage{fancyhdr}
\usepackage{scalerel,stackengine}
\stackMath
\DeclareRobustCommand\reallywidehat[1]{%
\savestack{\tmpbox}{\stretchto{%
  \scaleto{%
    \scalerel*[\widthof{\ensuremath{#1}}]{\kern-.6pt\bigwedge\kern-.6pt}%
    {\rule[-\textheight/2]{1ex}{\textheight}}
  }{\textheight}%
}{0.5ex}}%
\stackon[1pt]{\displaystyle #1}{\tmpbox}%
}
\stackMath
\DeclareRobustCommand\reallywidetilde[1]{%
\savestack{\tmpbox}{\stretchto{%
  \scaleto{%
    \scalerel*[\widthof{\ensuremath{#1}}]{\kern-1.0pt\sim\kern-1.0pt}%
    {\rule[-\textheight/2]{1ex}{\textheight}}
  }{\textheight}%
}{0.5ex}}%
\stackon[1pt]{\displaystyle #1}{\tmpbox}%
}
\newcommand{\overbar}[1]{\mkern 1.5mu\overline{\mkern-1.5mu#1}}

\newcommand{\vcc}[1]{\underline{#1}}

\let\OLDthebibliography\thebibliography
\renewcommand\thebibliography[1]{
  \OLDthebibliography{#1}
  \setlength{\parskip}{0pt}
  \setlength{\itemsep}{0pt plus 0.3ex}
}

\begin{document}

\hyphenation{aniso-thermal}

\newcommand{\der}[2]{\frac{\partial #1}{\partial #2}}
\newcommand{\dd}[2]{\frac{\partial^2 #1}{\partial #2^2}}
\newcommand{\dtot}[2]{\frac{D #1}{D #2}}
\newcommand{\ov}[1]{\overline{#1}}
\renewcommand{\f}[1]{\overbar{#1}}
\newcommand{\fa}[1]{\widetilde{#1}}
\newcommand{\FA}[1]{\reallywidetilde{#1}}
\newcommand{\ff}[1]{\widehat{#1}}
\newcommand{\FF}[1]{\reallywidehat{#1}}
\newcommand{\m}[1]{\overbar{#1}}
\newcommand{\ma}[1]{\widetilde{#1}}
\newcommand{\MA}[1]{\reallywidetilde{#1}}
\newcommand{\vv}[1]{\boldsymbol{#1}}
\newcommand{\vt}[1]{\boldsymbol{#1}}
\newcommand{\w}[1]{\widehat{#1}}
\newcommand{\W}[1]{\reallywidehat{#1}}

\newcommand{\refv}[1]{#1^{b}}
\newcommand{\adim}[1]{#1^{\circ}}

\newcommand{\upi}[0]{\pi}

\newcommand\Real{\mbox{Re}}
\newcommand\Imag{\mbox{Im}}
\newcommand\Rey{\mbox{\textit{Re}}}
\newcommand\Pran{\mbox{\textit{Pr}}}
\newcommand\Pen{\mbox{\textit{Pe}}}
\newcommand\Ai{\mbox{Ai}}
\newcommand\Bi{\mbox{Bi}}

\newcommand{\corr}[1]{{\color{red} #1}}
\newcommand{\corre}[1]{{\color{blue} #1}}

\newcommand{\e}[0]{\mathrm{e}}
\newcommand{\I}[0]{\mathrm{i}}
\newcommand{\Dee}[0]{D}
\newcommand{\ft}[1]{\widehat{#1}}
\newcommand{\FT}[1]{\reallywidehat{#1}}

\newcounter{subfigcounter}
\setcounter{subfigcounter}{1}
\DeclareRobustCommand{\lecID}{\refstepcounter{subfigcounter}}
\DeclareRobustCommand{\subfigtopleft}[1]{\lecID\def\stackalignment{l}\topinset{\text{\footnotesize(\alph{subfigcounter})}}{#1}{0.1in}{-.05in}}

\maketitle

{
\let\latexthesection\thesection %
    \renewcommand{\thesection}{\arabic{section}} 
    \renewcommand{\thefigure}{\arabic{figure}} 
    \setcounter{section}{0}
\begin{abstract}
Spatial and spectral energy exchanges associated with the turbulence
kinetic energy per unit mass, or half-trace of the velocity covariance tensor,
are studied in an anisothermal low Mach number turbulent channel flow.
The
temperatures of the two channel walls are 293 K and 586 K.
This generates a strong temperature gradient in the wall-normal direction.
The effect of the temperature gradient on the energy exchanges is
investigated using two direct numerical simulations of the channel,
at the mean friction Reynolds numbers 180 and 395.
The temperature gradient creates an asymmetry between the energy exchanges at
the hot and cold sides, due to the variations of the local fluid properties
and low Reynolds number effects.
The low Reynolds number effects are smaller at higher Reynolds number,
reducing the asymmetry between the hot and cold sides.
We also decomposed the energy exchanges in order to study separately the
mean-property terms, as found in the constant-property isothermal case, and the
thermal terms, specific to flows with variable fluid properties.
The significant thermal terms have a similar effect on the flow.
Besides, low Reynolds number effects have a negligible impact on thermal
terms and only affect mean-property terms.
\end{abstract}

\section{Introduction}

Energy exchanges are fundamental processes of turbulence which drives the
evolution of the different parts of total energy. They represent the
transformation of energy between different forms and its transfer between
scales and locations.
Particularly valuable are the energy exchanges associated with turbulence
kinetic energy, strongly connected to the intensity of turbulence.
The budget of turbulence kinetic energy describes the production, spatial
transfer, interscale transport and dissipation of turbulence kinetic energy.
In homogeneous isotropic incompressible turbulence, production occurs at large
scales and dissipation at small scales, close to Kolmogorov length scale.
In the intermediate range of scales, called inertial subrange, an energy
cascade transports the energy towards small scales.
These processes are Reynolds-number dependent. Dissipation occurs at smaller
scales at higher Reynolds number, separating distinctly the scales of
production and dissipation \citep{gamard2000reynolds}.
This canonical description of the energy exchanges can be altered profoundly
in more complex flow configuration or if additional physical phenomena are
involved, for instance
in compressible turbulent reacting flows \citep{towery2016spectral}
or in magnetohydrodynamic turbulence \citep{yang2017energy, yang2017energytransfer}.
In this paper, we focus on the turbulence kinetic energy exchanges in
strongly anisothermal wall-bounded flows at low Mach number.
These flows are found in many industrial applications, including
concentrated solar power plants \citep{daguenet2012experimental, DaguenetFrick2013164, serra2011, serra2012, serra2012turbulent, toutant2013},
heat exchangers and cooling systems \citep{zonta2008direct, campolo2009computing, lee2010nanofluid, zonta2012modulation}
or fluids at supercritical pressure within a small temperature range \citep{yoo2013turbulent, peeters2016turbulence, nemati2015mean}.
They are characterised by a strong influence on the energy exchanges
of the walls and of temperature.

Wall-bounded turbulent flows are intrinsically anisotropic and inhomogeneous in
the wall-normal direction.
This provides a very different physics from
homogeneous isotropic turbulence because energy is both transferred
spatially and transported in between scales.
Near the wall, turbulence is driven by an autonomous cycle, relatively
independent of the outer flow \citep{jimenez1999autonomous, flores2010hierarchy},
and two-way interactions between the structures of the inner and outer layers
\citep{jimenez1999physics, del2004scaling, jimenez2007we, hutchins2007large, mathis2009large}.
The study of the energy exchanges between the different parts of total energy
can give a representation of the exchanges of energy in the wall-normal
direction and between scales.
Knowledge of the three-dimensional flow fields is required to compute the
turbulence kinetic energy exchanges.
Direct numerical simulations is an useful tool to obtain such data,
although it is computationally expensive because the 
length and time scales of turbulence are small near the wall \citep{smits2011high, jimenez2013near}.
Using direct numerical simulations, the energy exchanges associated with
turbulence kinetic energy have been investigated in channel flows for various
friction Reynolds number:
$180$ \citep{kim1987a, moser1999a, abe2001direct, hoyas2008reynolds, vreman2014comparison, lee2015direct},
$395$ \citep{moser1999a, abe2001direct},
$550$--$640$ \citep{moser1999a, del2003spectra, abe2001direct, hoyas2008reynolds},
$950$--$1000$ \citep{hoyas2008reynolds, lee2015direct},
$2000$ \citep{hoyas2006scaling, hoyas2008reynolds} and
$5200$ \citep{lee2015direct}.
The spectral turbulence kinetic energy budget has been studied for the friction Reynolds numbers
$170$--$210$ \citep{domaradzki1994energy, bolotnov2010, mizuno2016spectra, schiavo2017turbulent, andrade2018analyzing},
$500$--$615$ \citep{mizuno2016spectra, lee2015spectral, schiavo2017turbulent},
$900$--$1000$ \citep{mizuno2016spectra, lee2015spectral, andrade2018analyzing} and
$5200$ \citep{lee2015spectral}.
\citet{andrade2018analyzing} showed that the spectral range of energy
exchanges is broader at higher Reynolds number due to an increase in the
low wavenumber scales.
Studies of energy exchanges have also been carried out in more complex
wall-bounded geometries \citep{blackman2017turbulent, schiavo2017turbulent, mollicone2018turbulence}. 

In strongly anisothermal wall-bounded flows, the energy exchanges are also
under the influence of the variations of temperature.
Contrary to high-speed compressible flows, acoustic waves have a negligible
impact in strongly anisothermal low Mach number flows \citep{paolucci1982a, meister1999asymptotic, nicoud2000conservative}.
Velocity and temperature are nevertheless strongly coupled because the
temperature gradient leads to significant variations of the fluid properties
(density, viscosity and thermal conductivity).
The temperature dependence of the fluid properties alters the behaviour of
flow and of the energy exchanges \citep{bradshaw1977compressible, lele1994compressibility, huang1995compressible, lee2013effect}.
As noted by \citet{serra2012turbulent}, temperature acts as a strong external
agency analogous to a strong magnetic field \citep{zhou2004colloquium, zhou2010renormalization, zhou2011nonlocality}
or rotation \citep{zhou1995phenomenological, sharma2018statistical, sharma2018energy}.
In particular, the variations of the fluid properties create an asymmetry
between the two sides of strongly anisothermal turbulent channel flows \citep{serra2011, toutant2013}.
The study of the effect of the asymmetry on the energy exchanges requires the
choice of a decomposition of kinetic energy, as several approaches have been
suggested in variable-property flows \citep{cousteix1989turbulence, aupoix2000introduction, chassaing2013variable, gatski2013compressibility, chassaing2013variable, dupuy2018equations}.
\citet{aulery2015energy, aulery2016} investigated the energy exchanges
associated with turbulence kinetic energy in the spectral domain using a
weighting by the square-root of density to decompose total energy \citep{kida92, cook2002}.
\citet{dupuy2018turbulence} analysed the energy exchanges in the spatial
and spectral domains using a ternary decomposition of kinetic energy \citep{dupuy2018equations}.
For a mean friction Reynolds number $Re_{\tau} = 180$, the effect of the
temperature gradient on the energy exchanges was identified to be in large part
the combined effect of the mean local variations of the fluid properties and a
Reynolds number effect.

This paper investigates 
the effect of the Reynolds number on the energy exchanges associated with the
half-trace of the velocity covariance tensor using
direct numerical simulations of a strongly anisothermal channel flow at
the mean friction Reynolds numbers $Re_{\tau} = 180$ and $Re_{\tau} = 395$.
The temperatures of the two channel walls are 293~K and 586~K for the
two simulations.
The analysis is carried out in the spatial and spectral domains, using
the same approach as \citet{dupuy2018turbulence}, and
focus on the effect of the mean friction Reynolds number on
the energy exchanges and on the influence of the temperature gradient.
Following \citet{dupuy2018equations, dupuy2018turbulence},
the energy exchanges are decomposed into mean-property terms, formally
identical to the terms present in the constant-property isothermal configuration,
and thermal terms, specific to flows with variable fluid properties.
The effect of the mean friction Reynolds number on mean-property and
thermal terms is compared.

The numerical method and the study configuration is given in section~\ref{r001}.
The equations of the energy exchanges are presented in section~\ref{r002}.
The results are analysed in section~\ref{r003}.

\section{Study configuration}\label{r001}

\subsection{Channel flow configuration}

Let us consider a flow in an anisothermal biperiodic channel flow (figure \ref{r004}).
The channel is periodic in the streamwise ($x$) and spanwise ($z$) directions.
There is a strong temperature gradient in the wall-normal direction ($y$):
the temperature at the cold wall ($y=0$) is $T_1 = 293$~K;
the temperature at the hot wall ($y=2h$) is $T_2 = 586$~K.
There is a streamwise volume force in the channel to balance dissipative forces.
We study the channel from direct numerical simulations of
the low Mach number equations at the mean friction Reynolds numbers $Re_{\tau} = 180$
and $Re_{\tau} = 395$.
Here, the mean friction Reynolds number is the average of the friction Reynolds
numbers $Re_{\tau, \omega} = U_{\tau} h/\nu_{\omega}$ at the hot and cold
sides, with $\nu_{\omega}$ the kinetic viscosity at the wall and
$U_{\tau}= [\nu_{\omega} (\partial_y \m{U}_x)_{\!\omega}]^{0.5}$ the
friction velocity. 
The direct numerical simulation at $Re_{\tau} = 180$ has been presented in \citet{dupuy2018turbulence}.

\begin{figure}
\centerline{\includegraphics[scale=0.6]{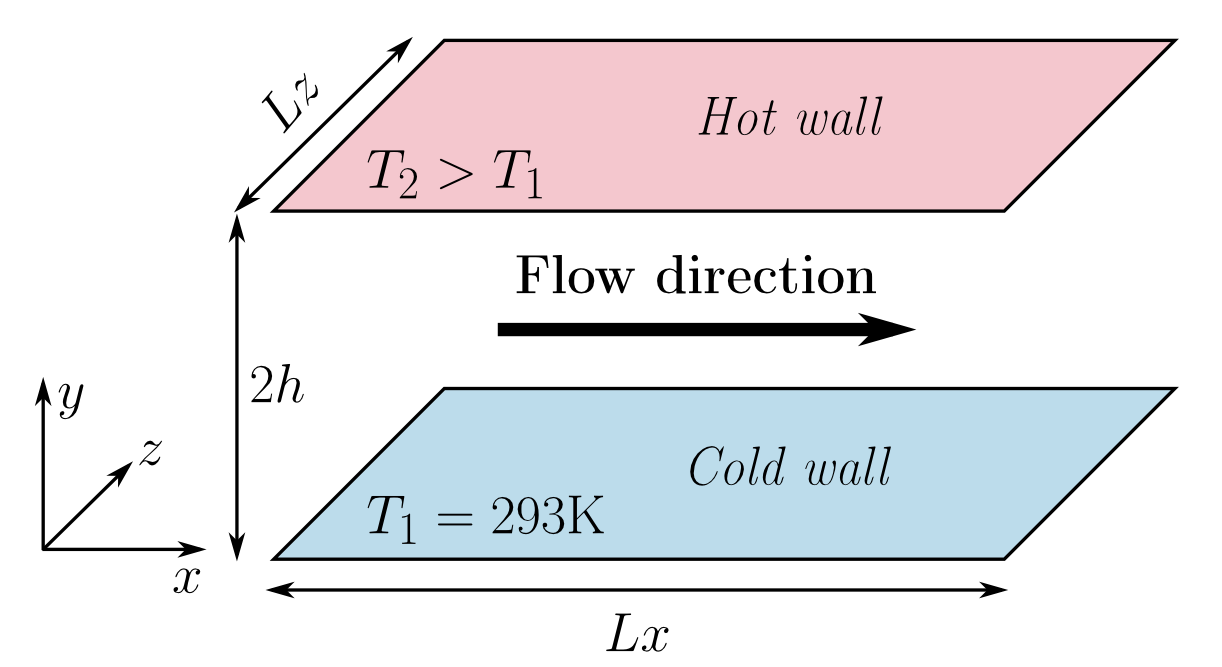}}
\caption{
Biperiodic anisothermal channel flow.
\label{r004}}
\end{figure}

\subsection{Governing equations}

The low Mach number equations are
an approximation of the Navier--Stokes equations
relevant for flows that are turbulent and strongly anisothermal but
slow compared to the speed of sound (Mach number below $10^{-2}$).
Purely compressible effects of high-speed flows, such as acoustic waves,
are neglected using the method of \citet{paolucci1982a}.
We assume in addition an ideal gas, Newton's law of viscosity and Fourier's law.
The flow governing equations are given by:
\begin{itemize}
\item Mass conservation
\begin{equation}
\der{\rho}{t} + \der{\rho U_{j}}{x_j} = 0,
\end{equation}
\item Momentum conservation
\begin{equation}
\der{\rho U_i}{t} = - \der{\rho U_j U_i}{x_j} - \der{P}{x_i} + \der{\varSigma_{ij}(\vv{U}, T)}{x_j},
\end{equation}
\item Energy conservation
\begin{equation}
\der{U_j}{x_j} = - \frac{1}{\gamma P_{0}}\left[ (\gamma - 1)\der{Q_j(T)}{x_j} + \der{P_{0}}{t} \right],
\end{equation}
\item Ideal gas law
\begin{equation}
T = \frac{P_{0}}{\rho r},
\end{equation}
\end{itemize}
where
$\rho$ is the density,
$T$ the temperature,
$t$ the temps and
$U_k$ the $k$-th component of velocity.
Einstein summation convention is used and
$\delta_{ij}$ is the Kronecker delta.
The pressure is decomposed in two terms.
The thermodynamical pressure $P_0$, constant in space, is the mean pressure in the domain.
The mechanical pressure $P$, space-dependent, is related to momentum variations.
The shear-stress tensor is given by
\begin{equation}
\varSigma_{ij}(\vv{U}, T) = \mu(T) \left[ \left( \der{U_i}{x_j} + \der{U_j}{x_i} \right) - \frac{2}{3} \der{U_k}{x_k} \delta_{ij} \right].
\end{equation}
The dynamic viscosity $\mu(T)$ is specified by Sutherland's law \citep{sutherland1893lii},
\begin{equation}
\mu(T) = \mu_0 \left(\frac{T}{T_0}\right)^{\frac{3}{2}} \frac{T_0 + S_1}{T + S_1},
\end{equation}
using $\mu_0 = \numprint{1.716} \cdot 10^{-5}$~Pa~s, $S_1 = \numprint{110.4}$~K and $T_0 = \numprint{273.15}$~K.
The heat flux is given by
\begin{equation}
Q_j(T) = - \lambda(T) \der{T}{x_j}.
\end{equation}
The thermal conductivity $\lambda(T)$ is specified assuming a constant Prandtl number,
$\lambda(T) = \mu(T) C_p/Pr$, using $Pr = 0.76$ and $C_p = \numprint{1005}$~J~kg$^{-1}$~K$^{-1}$.
The ideal gas specific constant is $r = \numprint{287}$~J~kg$^{-1}$~K$^{-1}$.
The mean value of the thermodynamic pressure is $\m{P}_0 = \numprint{1.5} \cdot 10^{5}$~Pa.

\subsection{Numerical settings}

The simulation domain is discretised using a mesh uniform in the
streamwise and spanwise directions and uneven in the wall-normal
direction.
In the wall-normal direction, the mesh follows a hyperbolic tangent law of
the form
\begin{equation}
y_k = L_y \left( 1 + \frac{1}{a} \tanh\left[ \left(\frac{k-1}{N_y-1} - 1\right)\tanh^{-1}(a)\right] \right),
\end{equation}
with $a$ the mesh dilatation parameter and
$N_y$ the number of nodes in the half-height of the channel.
The domain size and grid spacing of the simulations are given in table
\ref{r005}.
There is a similar level of refinement at $Re_{\tau} = 180$ and $Re_{\tau} = 395$.

We use a finite difference method on a staggered grid \citep{morinishi1998fully, nicoud2000conservative}
with a third-order Runge--Kutta time scheme \citep{williamson1980low},
a fourth-order centred momentum convection scheme.
The simulations are performed using the TrioCFD software \citep{calvin2002object}.
To compute the turbulence statistics, a data collection duration of
42 characteristic time ($h/U_{\tau}$) is used at $Re_{\tau} = 180$ and
40 characteristic time at $Re_{\tau} = 395$.

The numerical method has been validated by a mesh convergence study and the
comparison of the results in the constant-property isothermal case to the
reference data of \citet{moser1999a, vreman2014comparison, lee2015direct, bolotnov2010}
at $Re_{\tau} = 180$ and \citet{moser1999a} at $Re_{\tau} = 395$.
The validation of the simulation at $Re_{\tau} = 180$
may be found in \citet{dupuy2018turbulence}.
The validation of turbulence statistics at $Re_{\tau} = 395$
is presented in figure \ref{r006} in the isothermal case.
Our results agree with the reference data of \citet{moser1999a},
validating the numerical method at the constant-property limit.

\begin{table*}
\centerline{\begin{tabular}{llll}%
                                     &\multicolumn{1}{c}{ Number of grid points                }&\multicolumn{1}{c}{ Dimension of the domain                          }&\multicolumn{1}{c}{ Cell sizes in wall units}     \\
                                     &\multicolumn{1}{c}{ $N_x\times N_y\times N_z$            }&\multicolumn{1}{c}{ $L_x\times L_y\times L_z\phantom{(/3)}$        }&\multicolumn{1}{c}{ $\phantom{0}\Delta_x^+\hfill;\hfill\Delta_y^+(0)$\hfill--\hfill$\Delta_y^+(h)\hfill;\hfill\Delta_z^+$}\\[.5em]
\hline\\[-.5em]
DNS180-2   &\multicolumn{1}{c}{ $384\times266\times384$ }&\multicolumn{1}{c}{ $4\pi h\times2 h\times2\pi h\phantom{(/3)}$ }   &\multicolumn{1}{c}{ $\phantom{0}\numprint{8.5}\hfill;\hfill\phantom{0}\numprint{0.13}$\hfill--\hfill$\numprint{4.2}\hfill;\hfill\numprint{4.2}$ } \\
DNS395-2   &\multicolumn{1}{c}{ $768\times512\times512$ }&\multicolumn{1}{c}{ $4\pi h\times2 h\times(4/3)\pi h$           }   &\multicolumn{1}{c}{ $\phantom{0}\numprint{9.1}\hfill;\hfill\phantom{0}\numprint{0.25}$\hfill--\hfill$\numprint{4.1}\hfill;\hfill\numprint{4.6}$ } \\
\end{tabular}}
\caption{
Computational domain and cell sizes of the simulations.
\label{r005}}
\end{table*}

\begin{figure*}
\centerline{
\includegraphics[width=0.49\textwidth, trim={0 0 0 0}, clip]{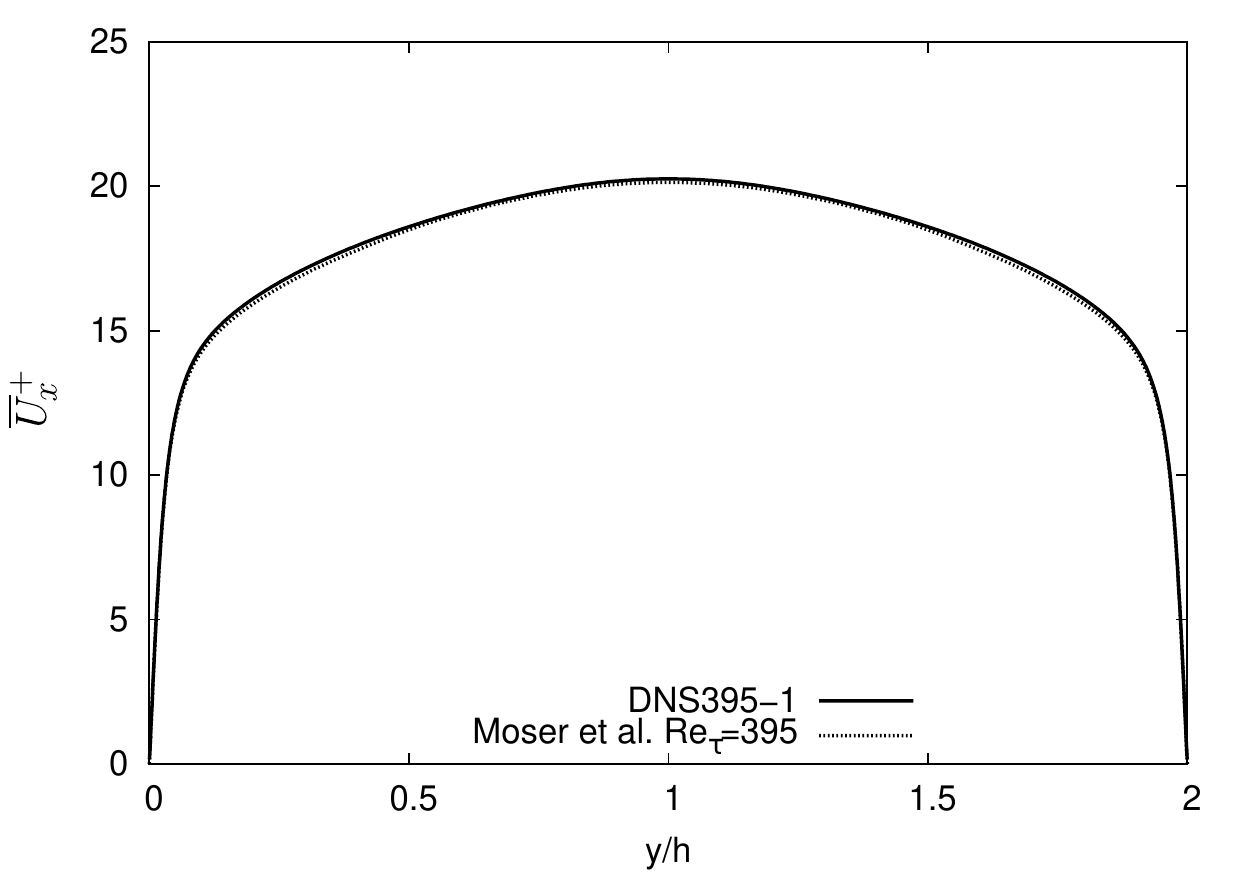}
\includegraphics[width=0.49\textwidth, trim={0 0 0 0}, clip]{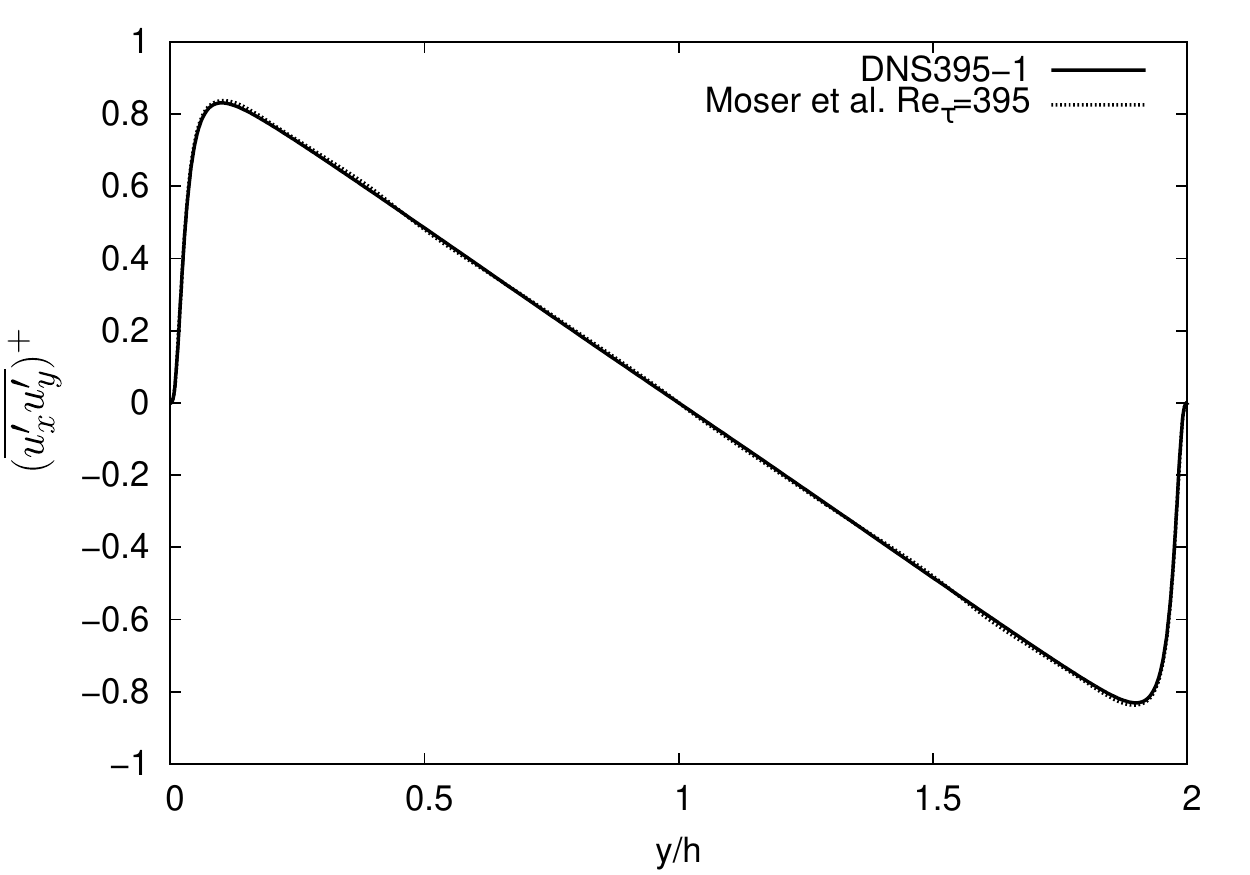}
}
\caption{
Validation of the profiles of the mean streamwise velocity $\m{U}_x$ (left) and the covariance of streamwise and wall-normal velocity $\m{u_x' u_y'}$ (right) in the isothermal channel at $Re_{\tau} = 395$.
The values are scaled by the friction velocity.
The reference data is from \citet{moser1999a}.
\label{r006}}
\end{figure*}

\section{Equations of the energy exchanges}\label{r002}

We study the terms of the evolution equation of the half-trace of the velocity covariance tensor
in the spatial and spectral domain.
Denoting ($\m{\phantom{ia}}$) the statistical average operator and ($'$) the
fluctuating part operator, the half-trace of the velocity covariance tensor is
defined as $e = \tfrac{1}{2} u_i' u_i'$ in the spatial domain.
Its evolution can formally be expressed as
\begin{align}
\der{\m{e}}{t} &=
	  \m{\varphi}
	+ \m{\mathcal{P}}
	+ \m{\zeta}
	+ \m{\varepsilon},
\end{align}
where 
$\m{\varphi} = \m{\varphi}^{c} + \m{\varphi}^{p} + \m{\varphi}^{\nu}$
is the sum of the conservative energy transfers related to convection, pressure and viscous stresses,
$\m{\mathcal{P}}$ the production, that is interaction with the other parts of kinetic energy,
$\m{\zeta} = \m{\zeta}^{c} + \m{\zeta}^{p} + \m{\zeta}^{\nu}$
the interaction with variable-density kinetic energy and
$\m{\varepsilon}$ the interaction with internal energy.
This set of equations is represented schematically on figure
\ref{r007}.
It is possible to extend
the study to the spectral domain to give the effect of the energy exchanges on
the size of turbulent structures.
Let us define the hat operator ($\w{\phantom{a}}$), which gives for any
physical quantity $g(x,y,z)$ the coefficients of the Fourier series expansion
of $g$ \citep{lesieur}:
\begin{equation}
\w{g}(\vv{k},y,t) =
	\frac{1}{L_x L_z} \int_0^{L_x}\!\!\!\! \int_0^{L_z} g(\vv{x},y,t) \e^{-\I \vv{k}\cdot\vv{x}} d\vv{x},
\end{equation}
where $\vv{x} = (x,z)$ is the position vector in the $xOz$ plane and $\vv{k} =
(k_x,k_z)$ the position vector in the $k_x O k_z$ plane.
The spectral counterpart of $e$ is
$\check{e} = \frac{1}{2} \w{u_i'}^*\w{u_i'}$.
Its evolution equation can formally be expressed as
\begin{equation}
\der{\m{\check{e}}}{t} =
	  \m{\check{\varphi}}
	+ \m{\check{\mathcal{P}}}
	+ \m{\check{\zeta}}
	+ \m{\check{\varepsilon}} 
	+ \m{\check{\Xi}}.
\end{equation}
Each term of the spatial decomposition is extended to the spectral domain.
In addition, a purely spectral term interscale transport term $\m{\check{\Xi}}$ redistributes the energy between scales
without spatial contribution.
The spectral energy exchanges are given by:
\begin{align}
\m{\check{\varphi}^{c}} ={}& \Real( - (1/2) (\partial_y \m{\w{u_i'}^*\w{u_i' U_y}}) ), \\
\m{\check{\varphi}^{p}} ={}& \Real(- \partial_y \ov{\w{u_y'}^*\W{P/\rho}} ), \\
\m{\check{\varphi}^{\nu}} ={}& \Real( \partial_y \ov{ \w{u_i'}^*\W{\varSigma_{iy}/\rho} } ), \\
\m{\check{\mathcal{P}}} ={}& \Real( - \ov{ \w{u_i'}^*\w{u_y'} } (\partial_y \m{U}_i) ), \\
\m{\check{\Xi}} ={}& \Real( (1/2) (\ov{ \w{\partial_j u_i'}^*\w{u_i' u_j'}}) - (1/2) (\ov{ \w{u_i'}^*\w{ u_j' \partial_j u_i'} }) ), \\
\m{\check{\zeta}^{c}} ={}& \Real( (1/2) (\ov{ \w{u_i'}^*\w{ u_i' \partial_j U_j} }) ), \\
\m{\check{\zeta}^{p}} ={}& \Real( - \ov{\w{u_i'}^*\W{(P/\rho^2) (\partial_i \rho)}} ), \\
\m{\check{\zeta}^{\nu}} ={}& \Real( \ov{ \w{u_i'}^*\W{(\varSigma_{ij}/\rho^2) (\partial_j \rho)} } ), \\
\m{\check{\varepsilon}} ={}& \Real( - \ov{ \w{\partial_j u_i'}^*\W{\varUpsilon_{ij}/\rho} } ).
\end{align}
These expressions take into account the symmetries of the geometry
($\partial_x \m{\,\cdot\,}=\partial_z \m{\,\cdot\,}=0$ and $\m{U}_z=0$).
The spatial terms can be obtained from the integration of the spectral terms on
the full wavenumber space.
In particular, for any $a$ and $b$, a spectral term of the form
$\Real(\m{\w{a}^*\w{b}})$
is associated with a spatial term $\m{a b}$.

In both the spatial and spectral domains, the energy exchanges are decomposed to
separate the terms specific variable-property flows (thermal terms, subscript $\varGamma$) and
terms that are non-zero in the constant-property isothermal case (mean-property terms, subscript $I$).
The mean-property terms may explicitly depend on the
local value of the mean fluid properties outside any derivative.
The decomposition is only relevant for the production $\m{\check{\mathcal{P}}}$, the conservative energy transfer $\m{\check{\varphi}}$ and the interaction with internal energy $\m{\check{\varepsilon}}$,
which are composed of at least one non-zero mean-property term.
Namely, we let
$\m{\check{\varphi}_{I}} = \m{\check{\varphi}_{I}} + \m{\check{\varphi}_{\varGamma}}$,
$\m{\check{\mathcal{P}}} = \m{\check{\mathcal{P}}_{I}} + \m{\check{\mathcal{P}}_{\varGamma}}$ and
$\m{\check{\varepsilon}} = \m{\check{\varepsilon}_{I}} + \m{\check{\varepsilon}_{\varGamma}}$.
The spectral mean-property terms are given by:
\begin{align}
\m{\check{\varphi}^{c}_{I}} ={}& \Real( - (1/2) (\partial_y \m{\w{u_i'}^*\w{u_i' u_y'}}) ), \\
\m{\check{\varphi}^{p}_{I}} ={}& \Real(- (1/\ov{\rho}) (\partial_y \ov{\w{u_y'}^*\w{P}}) ), \\
\m{\check{\varphi}^{\nu}_{I}} ={}& \Real( \m{\nu} (\partial_y \partial_y \m{\check{e}}) + \m{\nu} (\partial_y \partial_y \m{\w{u_y'}^*\w{u_y'}}) ), \\
\m{\check{\mathcal{P}}_{I}} ={}& \Real( - \ov{ \w{u_x'}^*\w{u_y'} } (\partial_y \m{U}_x) ), \\
\m{\check{\varepsilon}_{I}} ={}& \Real( - 2 \m{\nu} (\m{\w{s_{ij}'}^*\w{s_{ij}'}}) ).
\end{align}
The interscale transport $\m{\check{\Xi}}$ is not decomposed.
The thermal energy exchange $\m{\check{\zeta}}$ is zero in the constant-property
isothermal case. 

\begin{figure*}
\centerline{\includegraphics[scale=0.5]{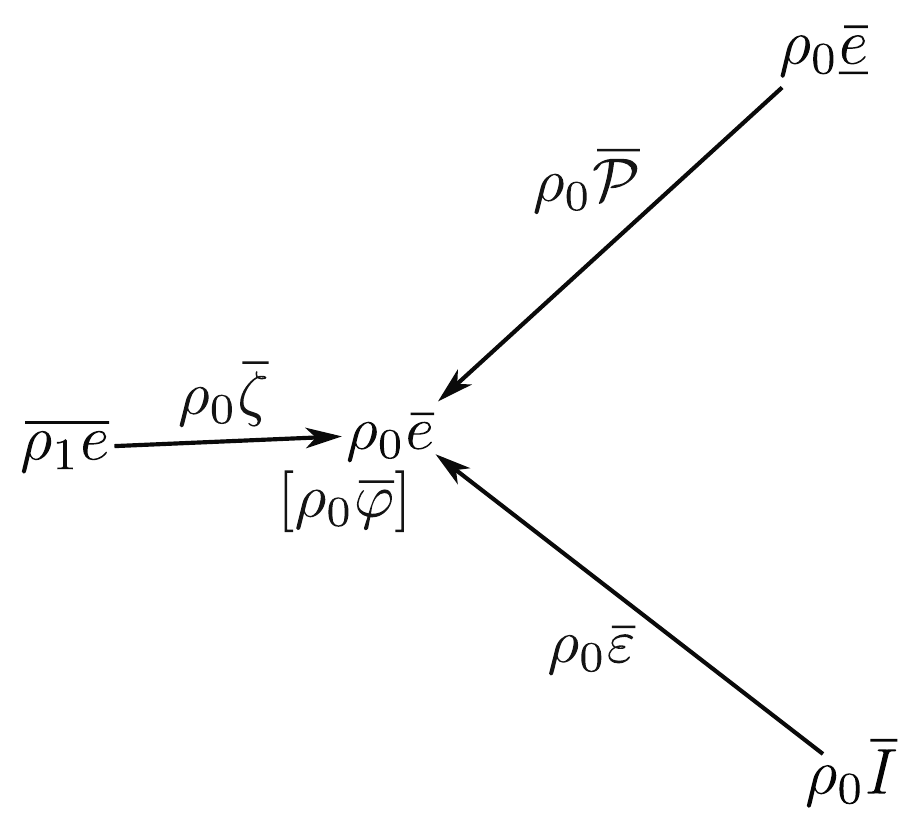}}
\caption{
Schematic representation of the energy exchanges associated with the
half-trace of the velocity covariance tensor, up to the constant
scalar factor $\rho_0$, with
$e = \tfrac{1}{2} u_i' u_i'$ the turbulence kinetic energy per unit mass,
$\vcc{e} = u_i' \m{U}_i$ the mixed kinetic energy per unit mass and
$I$ the internal energy per unit mass \citep{dupuy2018equations, dupuy2018turbulence}.
An arrow represents an interaction between two quantities.
This representation follows from the Reynolds decomposition of velocity
and the decomposition of density in a constant part $\rho_0$
and a variable part $\rho_1$.
\label{r007}
}
\end{figure*}

\subsection{Wavenumber bin}

The spectral terms are computed using a discrete two-dimensional Fourier
transform in the streamwise and spanwise directions,
\begin{equation}
\begin{aligned}
&\w{g}(k_{x,m},y,k_{z,n})  = \\
&\frac{1}{N_xN_z}\sum_{p=0}^{N_x-1}\sum_{k=0}^{N_z-1} g(x_{p},y,z_{k}) ~ \exp\left(- 2\upi \I \left(\tfrac{mp}{N_x}+\tfrac{kn}{N_z} \right) \right).
\end{aligned}
\end{equation}
The analysis of the spectral terms is following \citet{bolotnov2010} based on
the use of wavenumber bins.
A wavenumber bin is an annulus-shaped portion of the wavenumber space.
Each wavenumber vector $\vv{k}$ is assigned to a wavenumber bin according to
its norm $k$: if $k$ ranges between the lower bound $k_i$ and upper bound
$k_{i+1}$ of the bin \#$i$, it is assigned to the bin \#$i$.
Then, we associate the sum of the values of all wavenumbers
assigned to a bin to its wavenumber centre $k_{c,i}$.
This procedure removes the spectral directionality of the energy transfers
since only the wavenumber norm is taken into account.
The same wavenumber bins as in \citet{bolotnov2010} and \citet{dupuy2018turbulence}
are used:
\begin{equation}
k_i = k_{\text{min}}\left(\frac{k_{\text{max}}}{k_{\text{min}}}\right)^{\frac{i}{N_b}},
\end{equation}
with $N_b=24$ and where $k_{\text{min}}=1$ m$^{-1}$ and $k_{\text{max}}=73$ m$^{-1}$ are the minimum
and maximum bound across all bins.
The length of the bins is uniform in logarithmic scale.
Equal bin values therefore represent equal contributions to the integral over
the wavenumber space on a logarithmic scale.

\section{Results}\label{r003}

The energy exchanges associated with the half-trace of the velocity covariance
tensor are investigated using the semi-local scaling
\citep{huang1995compressible}, which is an appropriate scaling for the energy
exchanges \citep{patel2017scalar, pecnik2017scaling, dupuy2018turbulence}.
In contrast with the classical scaling, based only on wall values,
the semi-local scaling is based on a velocity scale $U_{\tau}^*$ and a length
scale $\m{\nu}/U_{\tau}^*$. This leads to a semi-local wall coordinate $y^*$, a
semi-local wavenumber $k^*$, a semi-local velocity $\vv{U}^*$ and a semi-local
friction Reynolds number $Re_{\tau}^*$,
\begin{align}
y^* ={}& \frac{y U_{\tau}^*}{\m{\nu}(y)}, \\
k^* ={}& \frac{k \m{\nu}(y)}{U_{\tau}^*}, \\
\vv{U}^* ={}& \frac{\vv{U}}{U_{\tau}^*}, \\
Re_{\tau}^* ={}& \frac{U_{\tau}^* h}{\m{\nu}(y)},
\end{align}
with $U_{\tau}^*$ the semi-local friction velocity, function of the wall shear stress
and the local value of the mean density,
\begin{equation}
U_{\tau}^* = \sqrt{\frac{\mu_{\omega}}{\m{\rho}(y)} \left(\der{\m{U}_x}{y}\right)_{\!\omega}},
\end{equation}
where the subscript $\omega$ denotes the value at the wall.
The terms of the evolution equation of the half-trace
of the velocity covariance tensor are scaled by
${U_{\tau}^*}^4/\m{\nu}(y)$.

The energy exchanges are investigated in the spatial and spectral domains.
In both cases, the analysis is carried out in two steps.
First, we study the effect of the mean friction Reynolds number on the
production, the conservative energy transfer
and the interaction with internal energy,
through the combined effect of mean-property
terms and thermal terms.
Then, we study its effect on the mean-property terms and the thermal
terms separately.

\subsection{Effect of the Reynolds number in the spatial domain}

\subsubsection{Profiles of the energy exchanges in the anisothermal configuration}

The temperature gradient creates at $Re_{\tau} = 180$ and $Re_{\tau} = 395$ an
asymmetry between the hot and cold sides.
The profiles of temperature and of the standard deviation of temperature
are provided in figure \ref{r008}.
The mean and fluctuation
profiles are modified by the coupling between turbulence and the temperature
gradient (figure \ref{r009}). This has been investigated in
previous studies \citep{serra2011, serra2012, toutant2013, sanchez2013}.
The temperature gradient also creates an
asymmetry between spatial profiles at the hot and cold sides of three
energy exchanges associated with the half-trace of the velocity
covariance tensor:
the production $\m{\check{\mathcal{P}}}$,
the conservative energy transfer $\m{\check{\varphi}}$ and
the interaction with internal energy $\m{\check{\varepsilon}}$ \citep{dupuy2018turbulence}. 
The energy exchanges have a greater magnitude at the cold side than at the hot
side, but occur over a narrower spatial range and closer to the wall (figure \ref{r010}).
These effects of the temperature gradient are Reynolds number dependent.

\begin{figure*}
\centerline{
\includegraphics[width=0.50\textwidth, trim={0 0 15 5}, clip, clip]{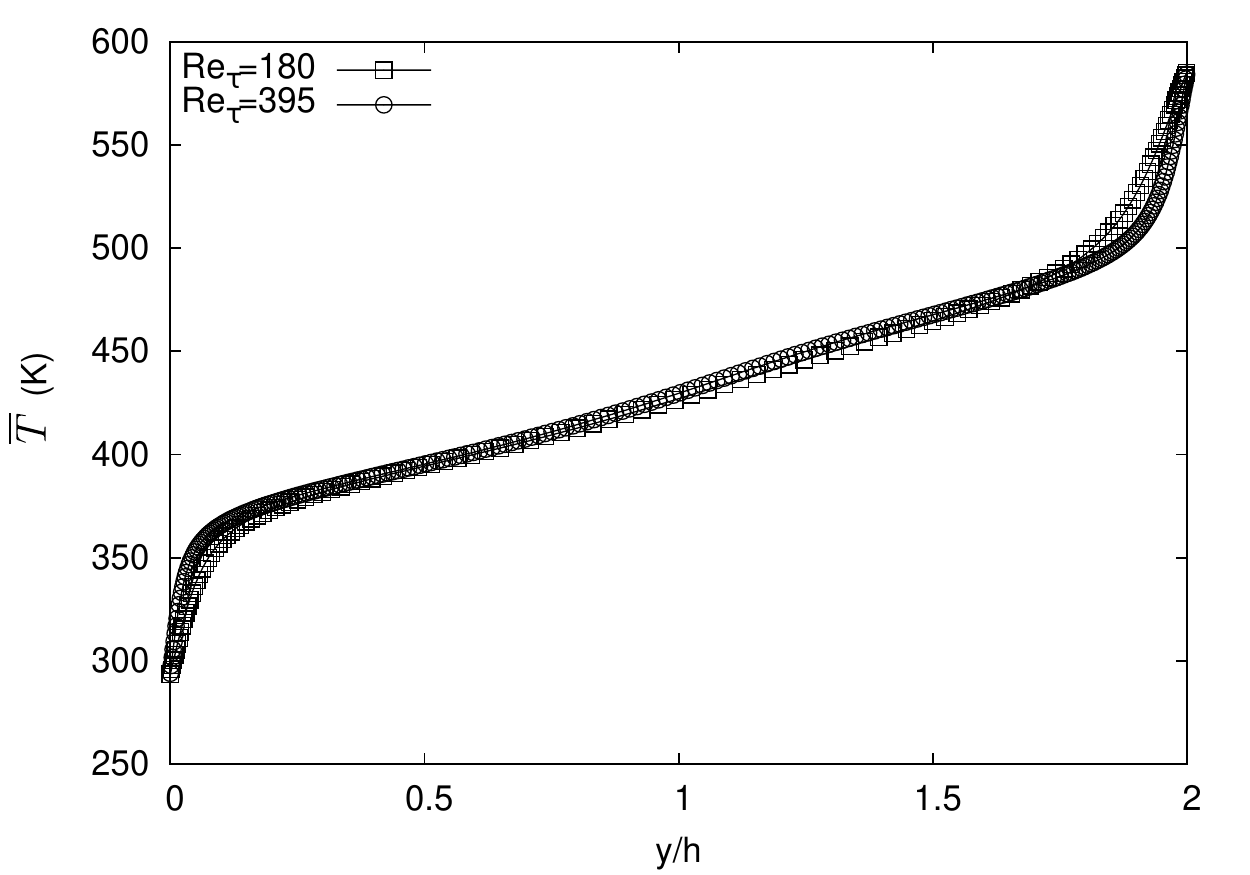}
\includegraphics[width=0.50\textwidth, trim={0 0 15 5}, clip, clip]{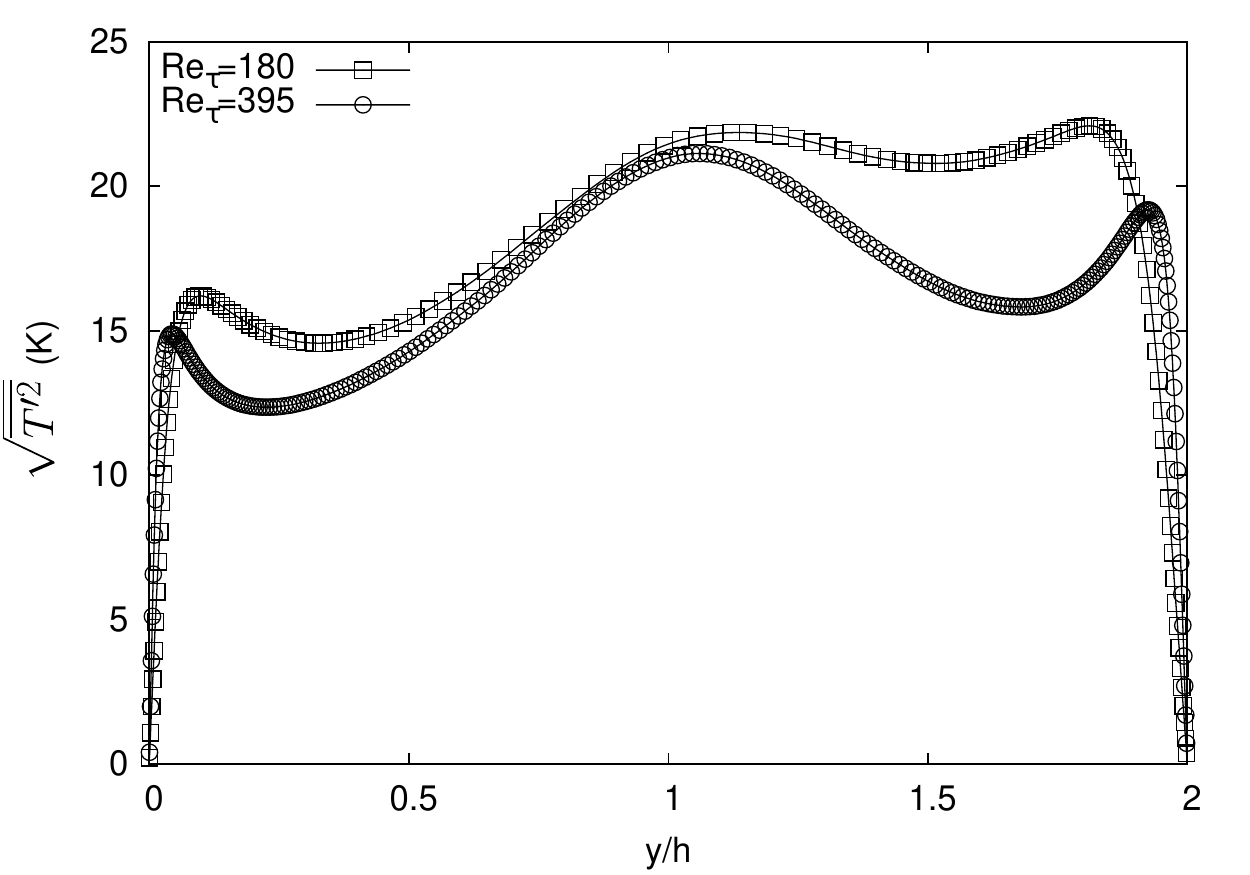}
}
\caption{
Profiles of the mean temperature (left) and the standard deviation of temperature (right) without scaling at $Re_{\tau}$ = 180 and $Re_{\tau}$ = 395.
\label{r008}}
\end{figure*}

\begin{figure*}
\centerline{
\includegraphics[width=0.50\textwidth, trim={0 0 15 5}, clip, clip]{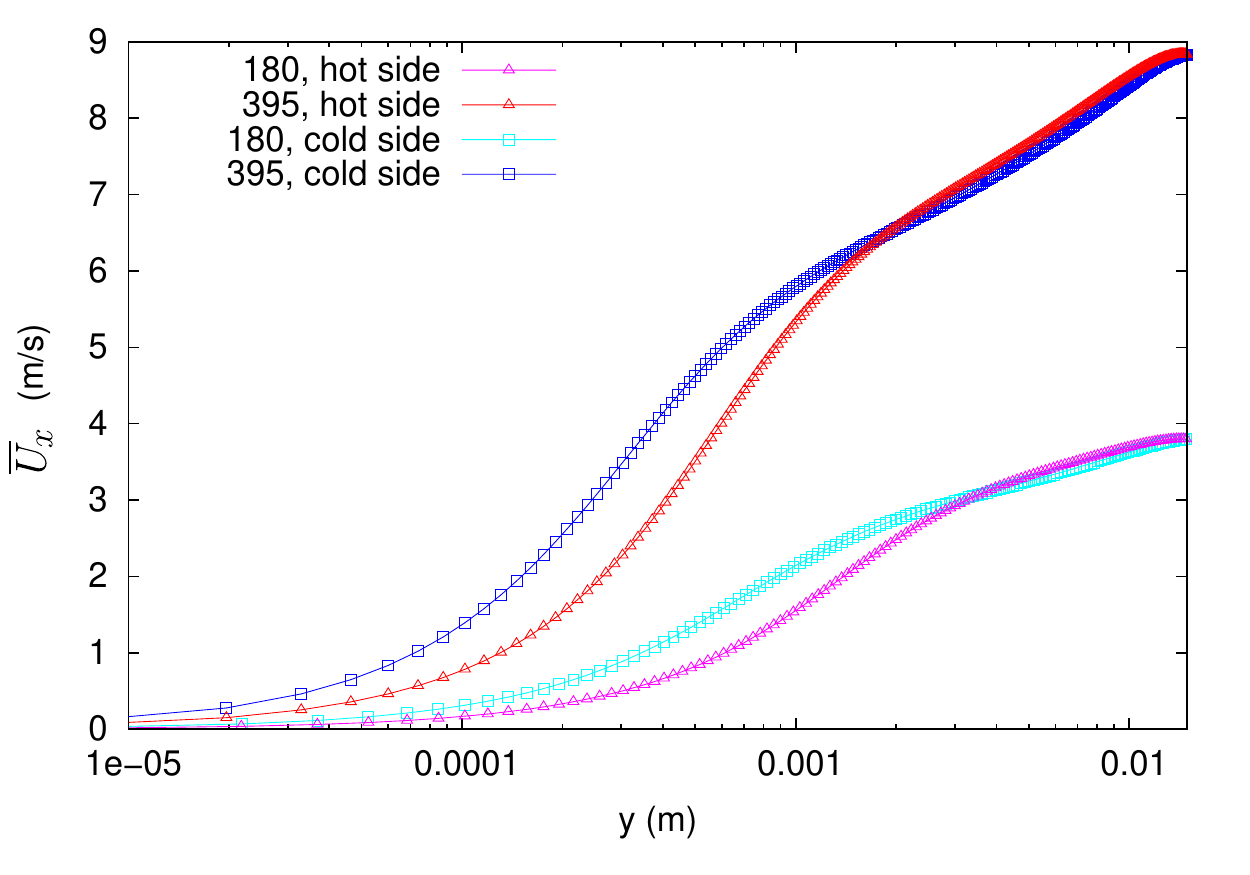}
\includegraphics[width=0.50\textwidth, trim={0 0 15 5}, clip, clip]{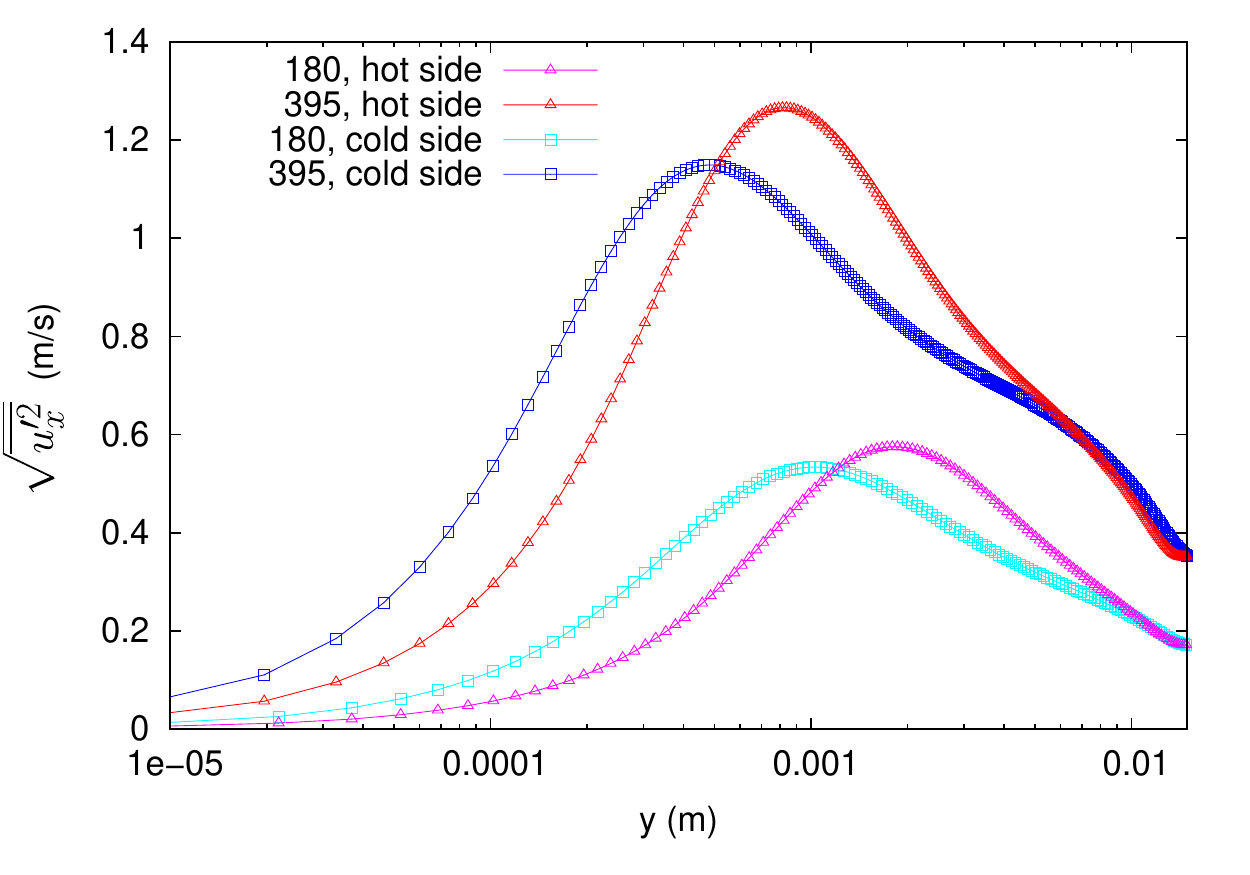}
}
\caption{
Profiles of the mean streamwise velocity (left) and the standard deviation of streamwise velocity (right) without scaling at $Re_{\tau}$ = 180 and $Re_{\tau}$ = 395.
\label{r009}}
\end{figure*}

\begin{figure*}
\centerline{
\includegraphics[width=0.50\textwidth, trim={0 0 15 5}, clip, clip]{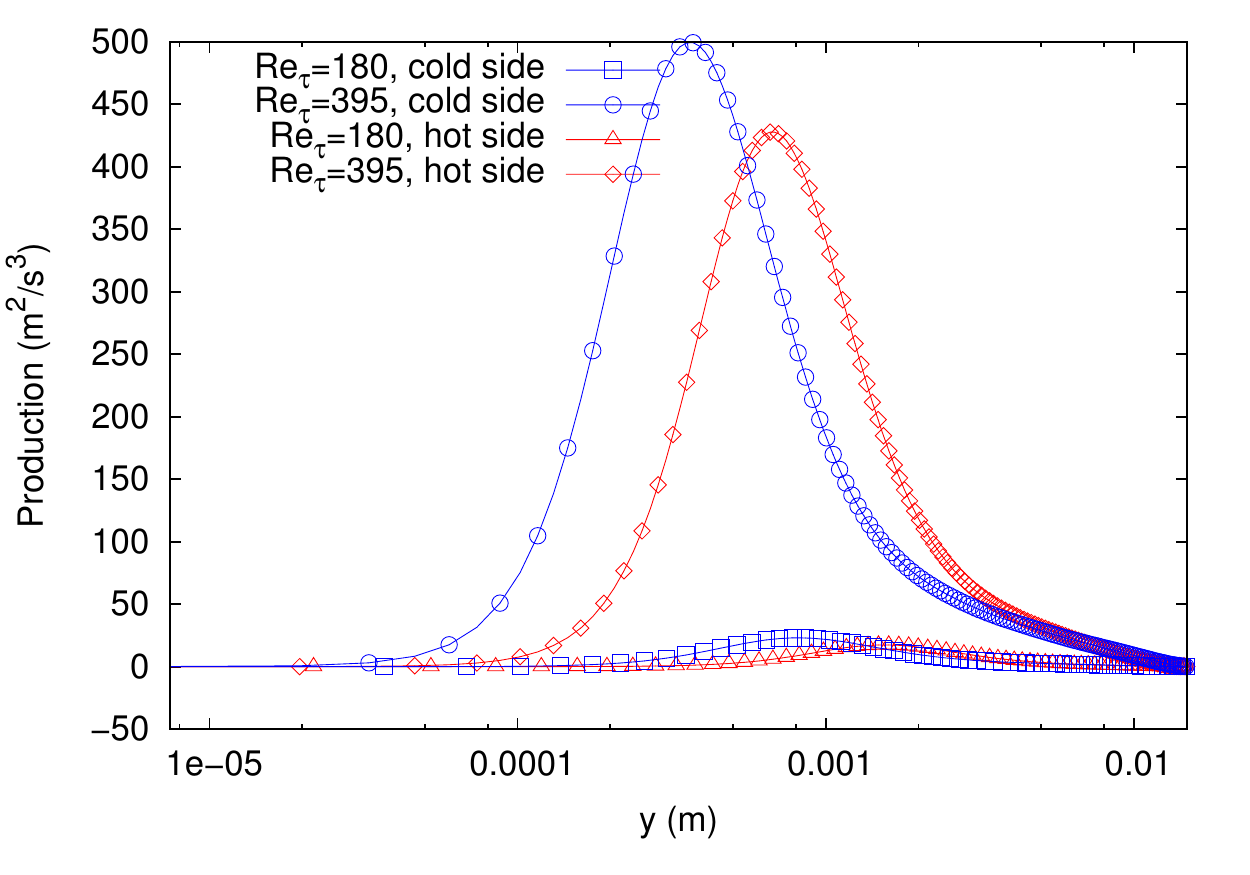}
}
\caption{
Profiles of the terms of the production $\m{\mathcal{P}}$ without scaling at $Re_{\tau}$ = 180 and $Re_{\tau}$ = 395.
\label{r010}}
\end{figure*}

The spatial profiles at $Re_{\tau} = 180$ and $Re_{\tau} = 395$
are provided in figure \ref{r011} with the
semi-local scaling.
The profiles of the two most significant parts of the conservative energy
transfer, the viscous transfer and the convection, are also given.
At higher mean friction Reynolds number, the asymmetry between the hot and cold
sides is reduced.
This is consistent with the assumption that the asymmetry between the hot and cold
sides is mainly due to two separate contributions \citep{dupuy2018turbulence}:
\begin{itemize}
\item the variations of the local velocity and length scales due to the variations
of the local fluid properties, to some extent taken into account by the
semi-local scaling;
\item a varying low Reynolds number effect between the hot and cold sides
depending on the semi-local friction Reynolds number (not taken into account by
the semi-local scaling).
\end{itemize}
Indeed, the profiles of the energy exchanges exhibit low Reynolds number
effects in the constant-property isothermal case within the Reynolds number range
of this study.
All energy exchanges investigated have a larger amplitude
at the higher Reynolds number with the semi-local scaling (figure \ref{r011}).
This Reynolds number dependence is more pronounced for the interaction with
internal energy and the viscous transfer than for the production and the
convection.
In addition, the peaks of production (figure \ref{r013}), of the
conservative energy transfer (figure \ref{r014}), of the viscous
transfer (figure \ref{r015}) and of convection (figure \ref{r016})
are slightly closer to the wall at $Re_{\tau} = 395$ than at $Re_{\tau} = 180$.
In the anisothermal channel, the semi-local friction Reynolds number varies
across the channel due to the temperature gradient.
At $Re_{\tau} = 180$, the semi-local friction Reynolds number ranges from 105 at the
hot wall to 260 at the cold wall.
At $Re_{\tau} = 395$, the semi-local friction Reynolds number ranges from 227 at the
hot wall to 556 at the cold wall.
The profiles are therefore subjected to a varying low Reynolds number effect
between the hot and cold sides, which cannot be expected to be taken into
account by any scaling based on the local fluid properties, given that the
classical wall scaling does not collapse the constant-property isothermal profiles
at various friction Reynolds number.

The semi-local scaling
properly accounts for the effect of the variations of the local fluid
properties on production at $Re_{\tau} = 180$ and
$Re_{\tau} = 395$ (figure \ref{r013}). Indeed, the maximum of production
with the semi-local scaling follows a low Reynolds number effect according to the
empirical relation of \citet{laadhari2002evolution}.
While the semi-local scaling is also relevant for the other energy exchanges,
the profiles of
the conservative energy transfer (figure \ref{r014}) and
the interaction with internal energy (figure \ref{r019})
show evidence that the approximation of the semi-local scaling does not
perfectly take into account the effect of the variations of the local fluid
properties. 
The value of the interaction with internal energy (figure \ref{r019})
at the wall is not equal to that of the constant-property isothermal profile at
the wall semi-local friction Reynolds number.
The hot and cold profiles of the viscous transfer
(figure \ref{r015}) deviates from all constant-property isothermal
profiles, even at the point near $y^*=3.5$ where they pass through the same
points.
The asymmetry between the hot and cold profiles of the convection (figure
\ref{r016}) is very small compared to a low Reynolds number effect,
especially at the positive extremum, closer from the wall.
These results confirms previous findings at $Re_{\tau} = 180$ \citep{dupuy2018turbulence}.

Hence, the asymmetry between the hot and cold sides is reduced
at $Re_{\tau} = 395$ compared to $Re_{\tau} = 180$ because
low Reynolds number effects are smaller the higher the Reynolds number.
This only affects one component of the effect of the temperature gradient however.
While a simulation at even higher Reynolds number would be free of low Reynolds
number effects, the results suggest that the hot and cold profiles would still
not collapse because the approximation of the semi-local scaling does not
perfectly account for the effect of the variations of the local fluid
properties.
This is investigated further by decomposing the energy exchanges into
mean-property terms and thermal terms.

\begin{figure*}
\centerline{
\subfigure[Production, total term $\m{\mathcal{P}}$ and mean-property term~$\m{\mathcal{P}_{I}}$ (m.-p. term). \label{r013}]{\includegraphics[width=0.50\textwidth, trim={0 10 15 5}, clip, clip]{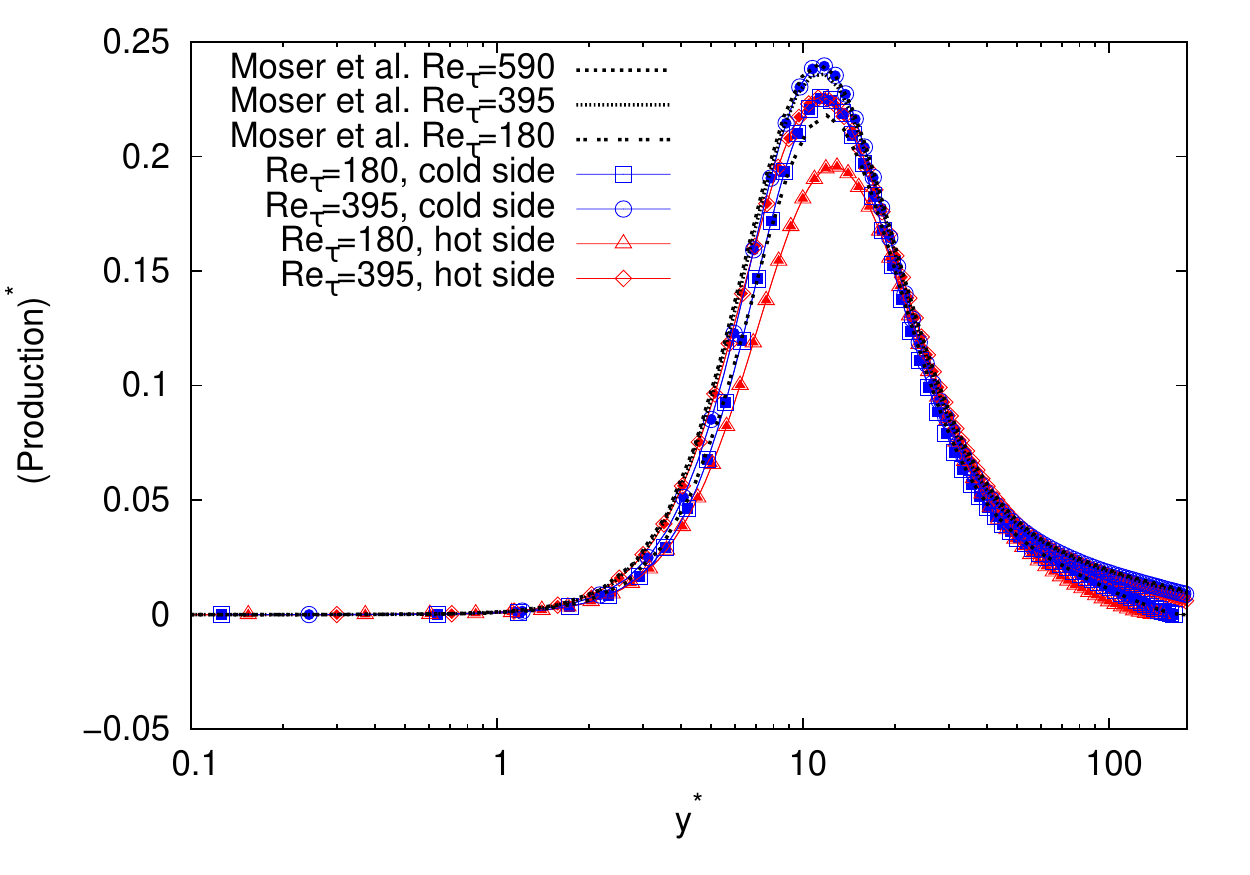}}
\subfigure[Conservative energy transfer, total term $\m{\varphi}$ and mean-property term~$\m{\varphi_{I}}$ (m.-p. term). \label{r014}]{\includegraphics[width=0.50\textwidth, trim={0 10 15 5}, clip, clip]{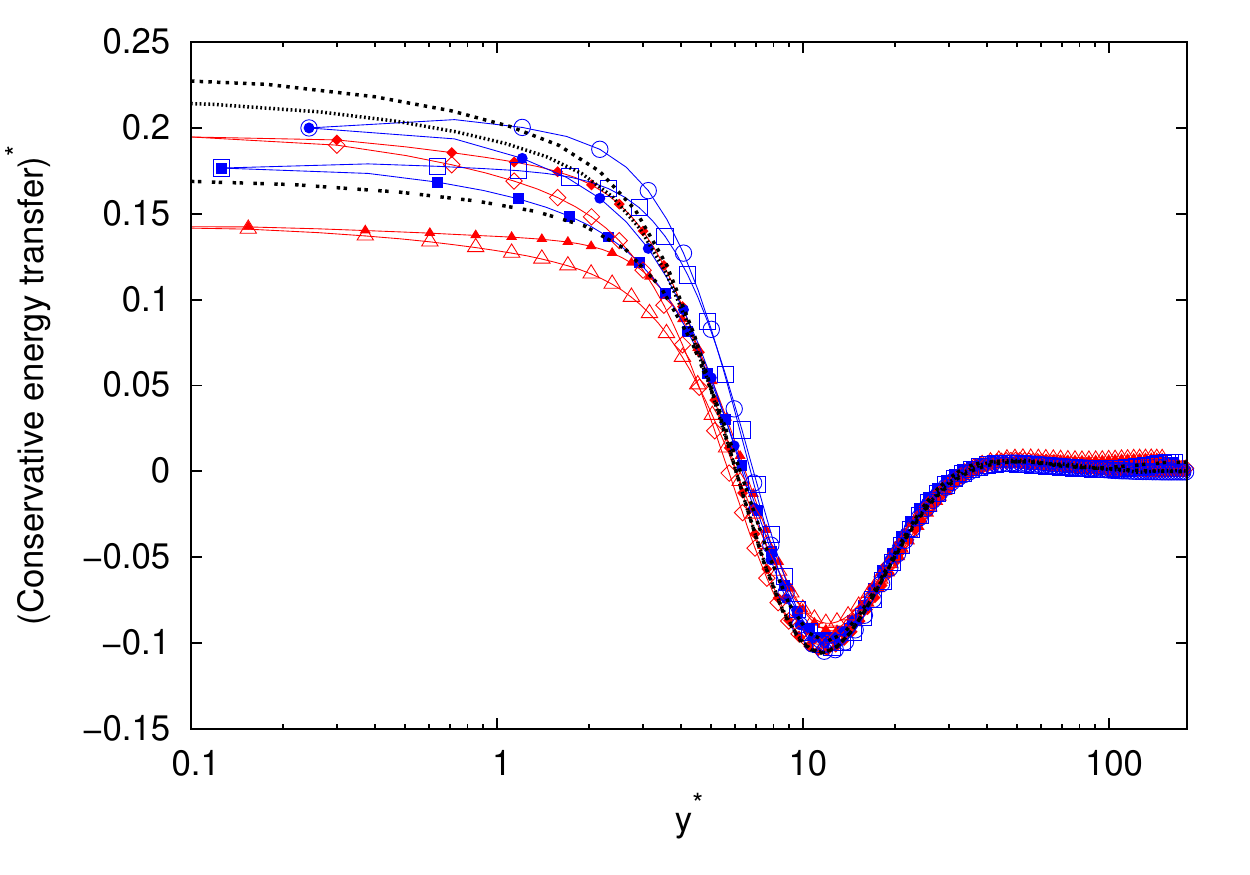}}
}
\centerline{
\subfigure[Interaction with internal energy, total term $\m{\varepsilon}$ and mean-property term~$\m{\varepsilon_{I}^{\nu}}$ (m.-p. term). \label{r019}]{\includegraphics[width=0.50\textwidth, trim={0 10 15 5}, clip, clip]{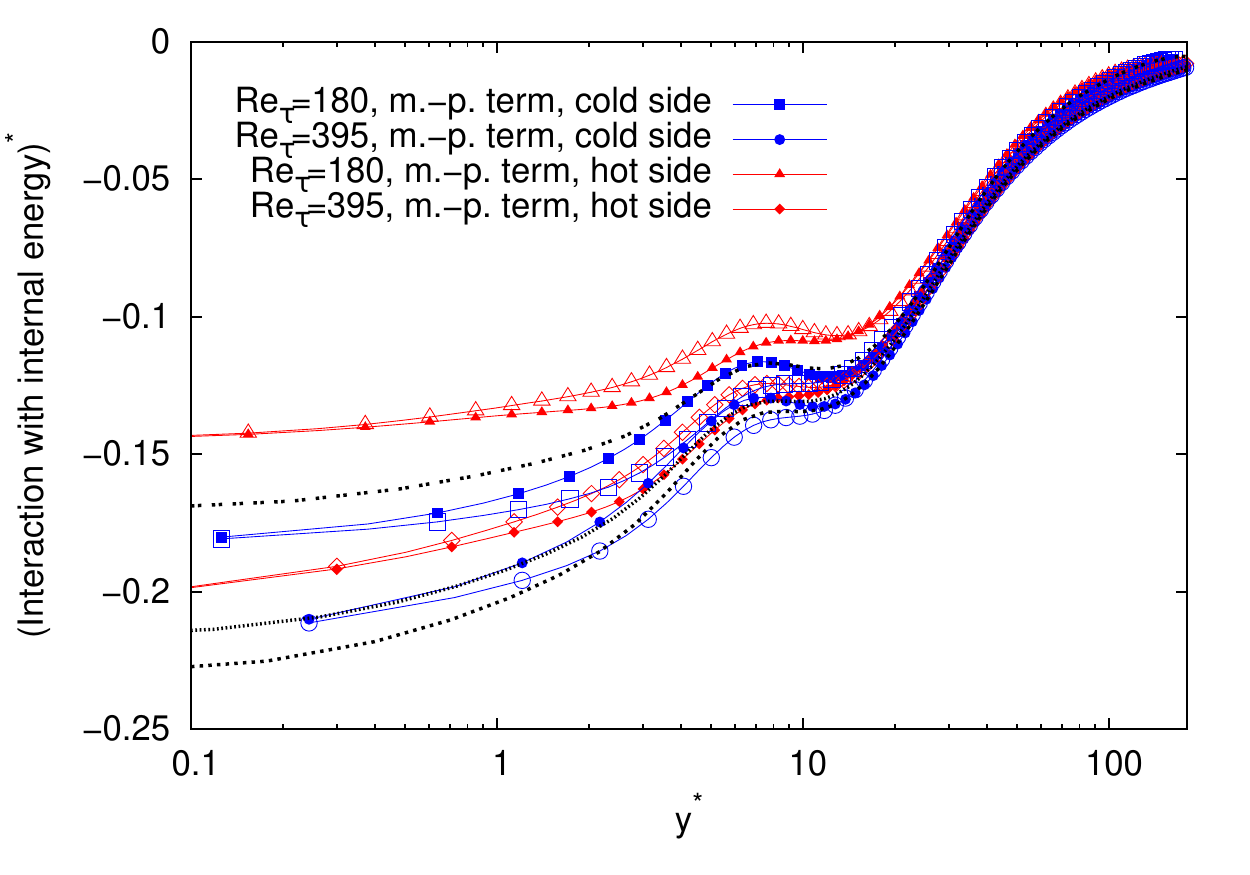}}
}
\centerline{
\subfigure[Viscous transfer, total term $\m{\varphi^{\nu}}$ and mean-property term~$\m{\varphi^{\nu}_{I}}$ (m.-p. term). \label{r015}]{\includegraphics[width=0.50\textwidth, trim={0 10 15 5}, clip, clip]{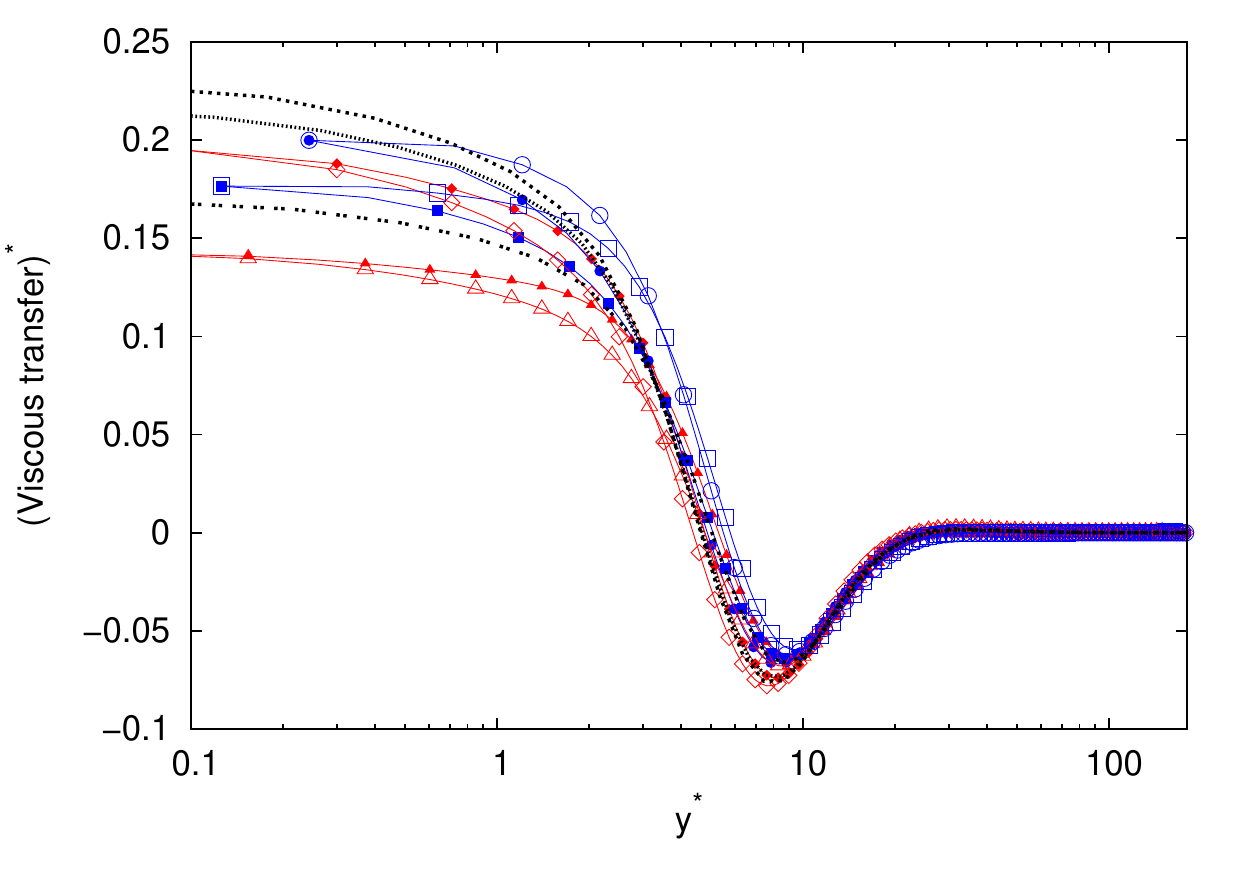}}
\subfigure[Convection, total term $\m{\varphi^{c}}$ and mean-property term~$\m{\varphi^{c}_{I}}$ (m.-p. term). \label{r016}]{\includegraphics[width=0.50\textwidth, trim={0 10 15 5}, clip, clip]{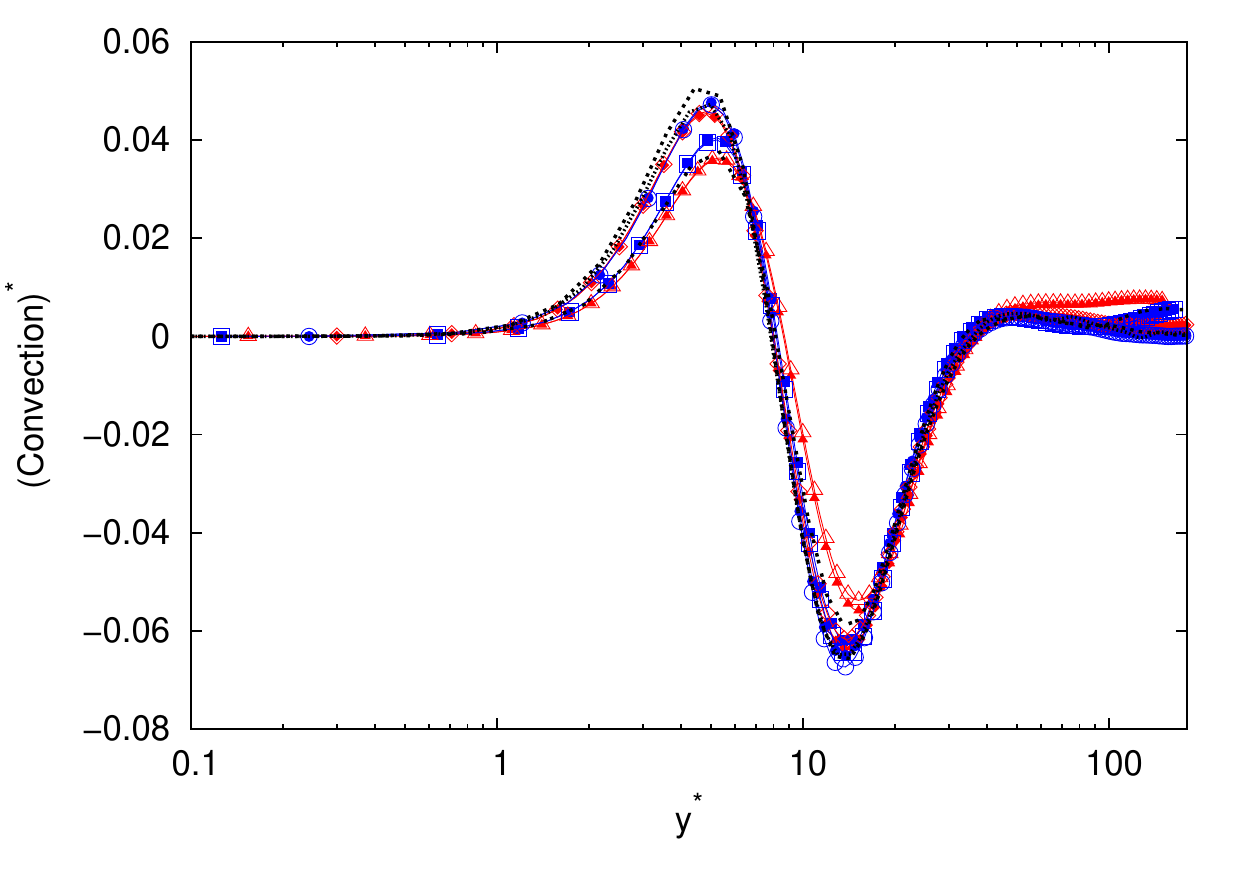}}
}
\caption{
Profiles of the terms of the budget of the half-trace of the velocity covariance tensor with the semi-local scaling at $Re_{\tau}$ = 180 and $Re_{\tau}$ = 395.
The profiles of the two most significant parts of the conservative energy
transfer are also given.
The results are compared to the reference data of \citet{moser1999a}
for a constant-property isothermal channel flow at the friction Reynolds numbers
180, 395 and 590.
The viscous transfer and the convection are the two most significant parts of the conservative energy transfer.
\label{r011}}
\end{figure*}

\subsubsection{Profiles of the mean-property terms in the anisothermal configuration}

The spatial profiles with the semi-local scaling of the mean-property terms
follows for the most part the same general behaviour as the total
energy exchanges and are subject to a similar effect of the
Reynolds number (figure \ref{r011}).
The mean-property production is identical to the total production because the
thermal production is negligible (figure \ref{r013}). 
At $Re_{\tau} = 180$, the profiles of the mean-property conservative energy
transfer (figure \ref{r014}) and the mean-property interaction with
internal energy (figure \ref{r019}) at the hot and cold sides are
closer to each other and to the constant-property isothermal profiles, especially
in an area around $y^*=4$ but not at the wall where the mean-property and total
terms are identical.
At $Re_{\tau} = 395$, the effect is similar but goes further, inverting the
ordering of the hot and cold profiles at some parts of the channel,
leading to a larger amplitude at the hot side.
This is due to the reduced asymmetry of the energy exchanges between the hot
and cold sides.

\subsubsection{Profiles of the thermal terms in the anisothermal configuration}

We provide in figure \ref{r027} the spatial profiles with the semi-local
scaling of the three most significant thermal terms at $Re_{\tau} = 180$ and $Re_{\tau} = 395$:
the thermal conservative energy transfer $\m{\varphi_{\varGamma}}$ (figure \ref{r028}),
the interaction with variable density kinetic energy $\m{\zeta}$ (figure \ref{r029}) and
the thermal interaction with internal energy $\m{\varepsilon_{\varGamma}}$ (figure \ref{r030}).
The interaction with variable density kinetic energy is a thermal energy
exchange while the thermal conservative energy transfer and the thermal
interaction with internal energy are parts of energy exchanges.
The thermal production is not discussed because it is negligible compared to
the mean-property production.
The three investigated thermal terms show some similarity as their profiles are
composed of two peaks which are of opposite sign at the hot and cold sides, and
have a larger magnitude at the cold side than at the hot side.

\begin{figure*}
\centerline{
\subfigure[Thermal conservative energy transfer $\m{\varphi_{\varGamma}}$. \label{r028}]{\includegraphics[width=0.50\textwidth, trim={0 10 15 5}, clip, clip]{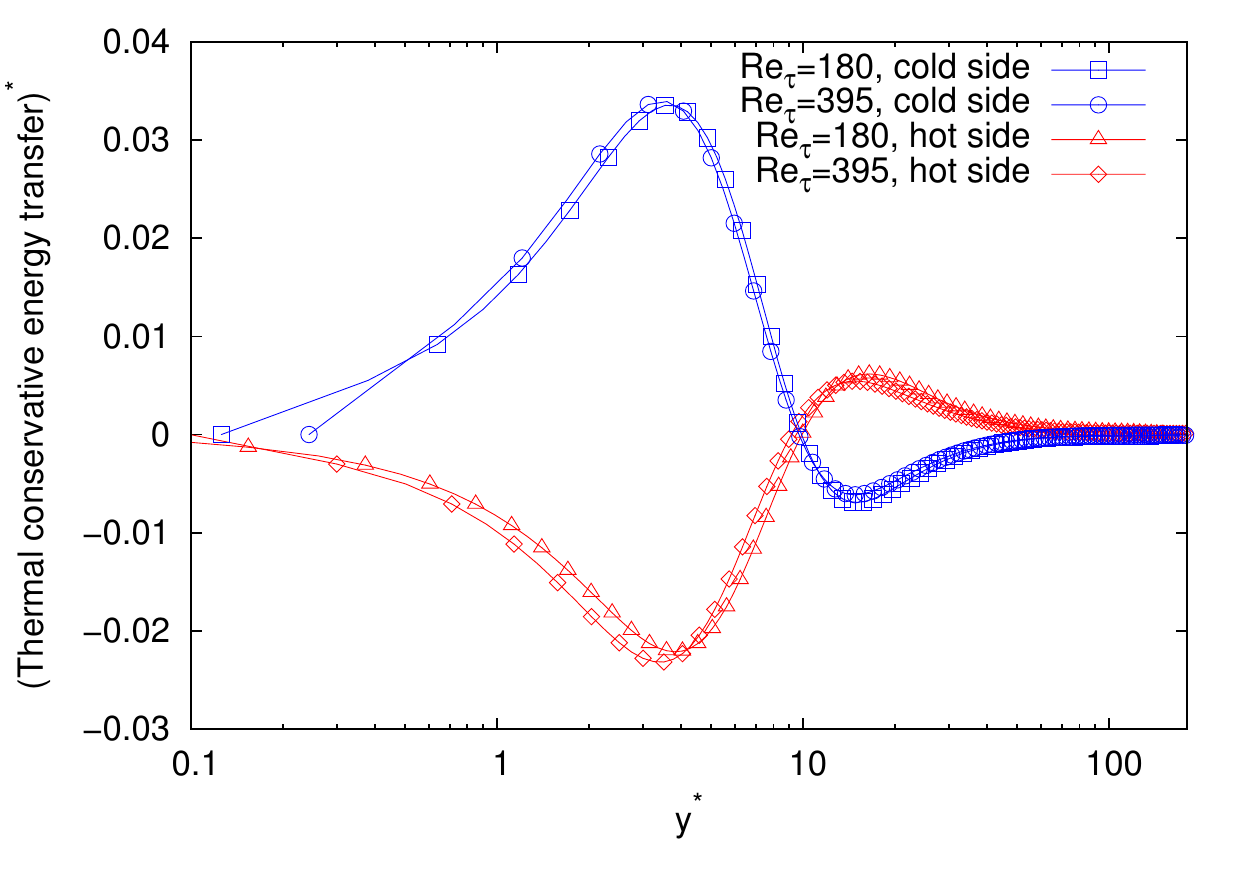}}
\subfigure[Interaction with variable density kinetic energy $\m{\zeta}$. \label{r029}]{\includegraphics[width=0.50\textwidth, trim={0 10 15 5}, clip, clip]{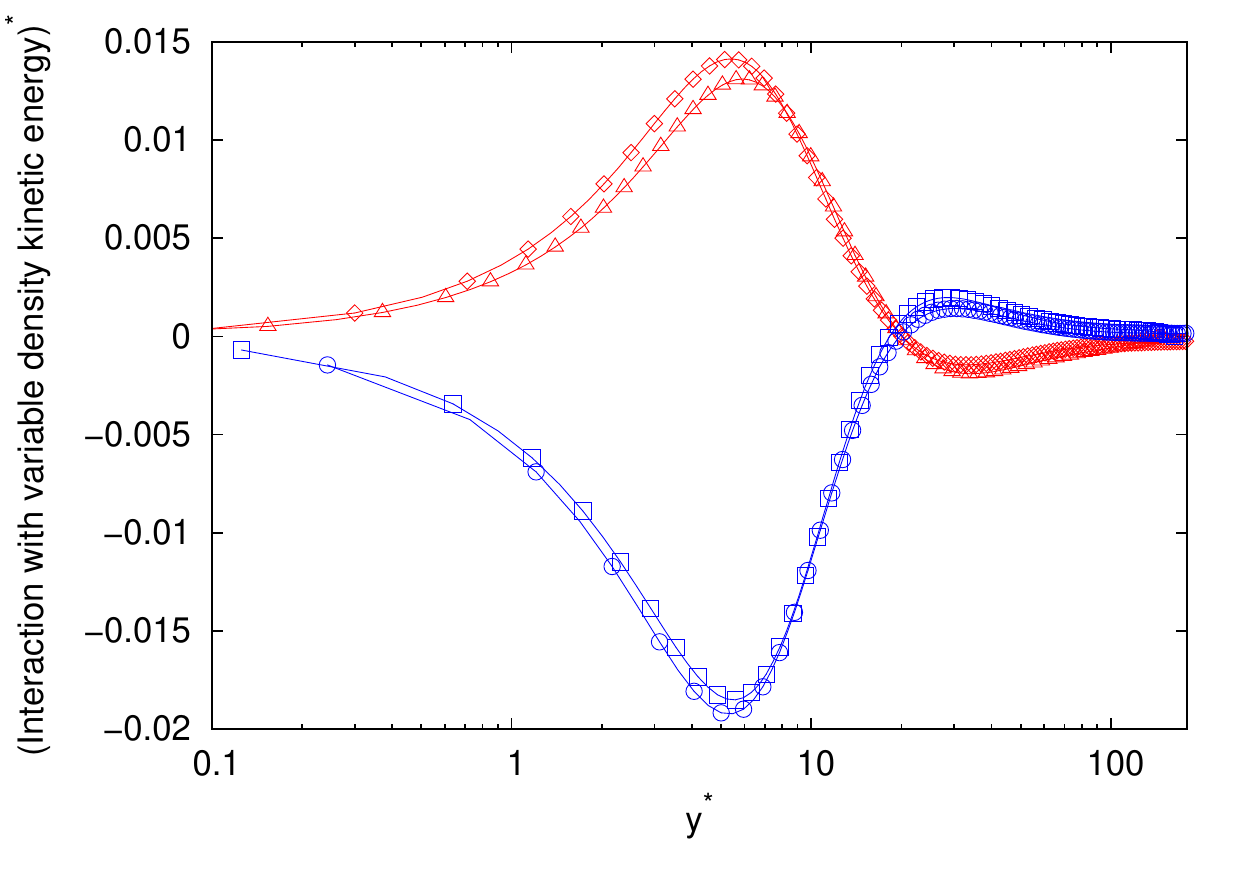}}
}
\centerline{
\subfigure[Thermal interaction with internal energy $\m{\varepsilon_{\varGamma}}$. \label{r030}]{\includegraphics[width=0.50\textwidth, trim={0 10 15 5}, clip, clip]{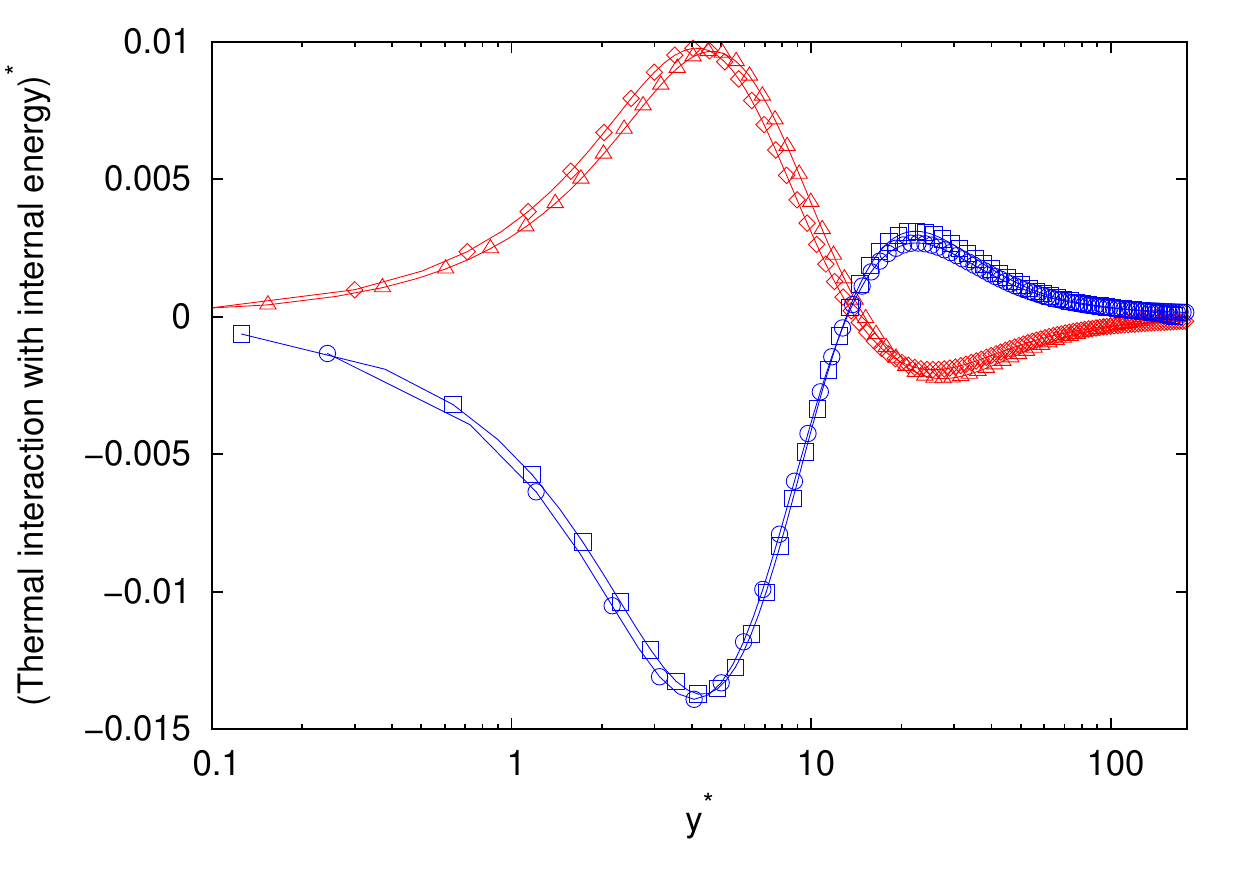}}
}
\caption{
Profiles of the thermal terms of the budget of the half-trace of the velocity covariance tensor with the semi-local scaling at $Re_{\tau}$ = 180 and $Re_{\tau}$ = 395.
\label{r027}}
\end{figure*}

The profiles of the thermal terms at the two mean friction Reynolds number are
very similar. The effects of the Reynolds number on the thermal terms, very
small, are given hereafter.
At $Re_{\tau} = 395$, the position of the extrema is shifted towards the wall
compared to $Re_{\tau} = 180$, reducing the position asymmetry between the hot
and cold sides for the extremum closer to the centre of the channel.
The amplitude of the extremum closer to the wall is increased slightly whereas
the amplitude of the extremum closer to the centre of the channel is decreased.
These effects are not significant and do not affect notably the amplitude
asymmetry between the hot and cold sides.
All in all, the Reynolds number dependence of the thermal terms is very slight,
suggesting that the profiles are almost universal within the Reynolds number
range of this study.
Therefore, we may infer the approximate conclusion that low Reynolds number
effects only affect the mean-property terms but are negligible on thermal
terms. In other words, the thermal terms are functions of the variations of the mean local fluid
properties only and do not substantially depend on the Reynolds number.

Since the magnitude of the mean-property terms increases with the mean
friction Reynolds number while thermal terms are largely unaffected, the
relative contributions of the thermal terms to the energy exchanges decline at
high Reynolds number.  On the other hand, the relative importance of the
thermal terms on the asymmetry between the hot and cold sides become larger,
given that the asymmetry between the hot and cold sides is reduced for the
mean-property terms.
This follows from the assumption that a varying low Reynolds number effect is a
significant part of the effect of the temperature gradient on the
mean-property terms whereas the thermal terms are only tied to the variations
of the mean local fluid properties. Increasing the mean friction Reynolds
number reduces the low Reynolds number effects, which gives less weight to the
mean-property terms on the asymmetry between the hot and cold sides.
However, as noted above, the mean-property terms are also in a large part
related to the variations of the local fluid properties since the decomposition
does not completely separate the two effects.

\subsection{Effect of the Reynolds number in the spectral domain}

\subsubsection{Spectra of the energy exchanges in the anisothermal configuration}

The effect of the mean friction Reynolds number on the spectral energy
exchanges is investigated.
The spectral behaviour of the energy exchanges has been described in
\citet{dupuy2018turbulence} in the constant-property isothermal case and in the
anisothermal case.
The half-trace of the velocity covariance tensor is in the
spectral domain predominantly governed by four spectral energy
exchanges: the production $\m{\check{\mathcal{P}}}$, the interscale transport $\m{\check{\Xi}}$, the conservative energy
transfer $\m{\check{\varphi}}$ and the interaction with internal energy $\m{\check{\varepsilon}}$.
The peak production of turbulence kinetic energy occurs around $y^*=12$ and
$k^*=0.07$. The interscale transport redistributes the produced energy among
scales, mainly towards small scales but also towards large scales. The energy
is, in addition, redistributed spatially by the conservative energy transfer.
A pseudo-energy transfer can be identified from the peak production area,
where the conservative energy transfer is positive, to the wall and the center
of the channel, where the conservative energy transfer is negative.
A cut of the spectra at the location of the spatial peak of production, $y^* \approx 12$
(figure \ref{r031}) shows that no energy exchange predominates.
The produced energy is split almost evenly in three energy processes: its
transfer away and towards the wall, its redistribution towards large and small scales
and its conversion into internal energy.
Near the wall ($y^* < 2$),
the production and the purely spectral transfer are negligible.
The interaction with internal energy and the conservative energy transfer are
thus in balance and occurs at the same wavenumber (figure \ref{r032}).
At the centre of the channel, the production is negligible.
Energy coming from the production area is transported towards small scales where
it is dissipated. The conservative energy
transfer and the interaction with internal energy thus do not occur at the same
wavenumber (figure \ref{r033}).
While the Reynolds number differences are addressed by the semi-local scaling at
the spatial peak of production or near the wall, the semi-local scaling is not
appropriate at the centre of the channel.
Since the relevant length scale of turbulence near the centre of the channel
is the half-height of the channel $h$ \citep{hoyas2008reynolds},
we suggest the use a semi-local outer-flow scaling ($^h$) based on the length scale $h$
and a velocity scale $U_{\tau y}^*$ to scale the amplitude of the energy exchanges, with
\begin{equation}
U_{\tau y}^* = \sqrt{\frac{1}{\m{\rho}(y)} \left[ \frac{2h-y}{2h} \mu_{1} \left(\der{\m{U}_x}{y}\right)_{\!1} - \frac{y}{2h} \mu_{2} \left(\der{\m{U}_x}{y}\right)_{\!2}\right]}.
\end{equation}
The linear combination of the wall shear stresses at the hot and cold sides ensures
that hot and cold profiles are equal at the centre of the channel but takes into account
the variations of the mean friction Reynolds number.
With this scaling, the peak of the conservative energy transfer has a
smaller semi-local wavenumbers $k^*$ at $Re_{\tau} = 395$ than at $Re_{\tau} = 180$,
whereas dissipation occurs at similar semi-local wavenumbers $k^*$ (figure \ref{r033}).
Note however that the unscaled wavenumbers of the conservative energy transfer are similar
at $Re_{\tau} = 395$ and $Re_{\tau} = 180$, whereas the scales of dissipation
are smaller at high Reynolds number without scaling.

\begin{figure*}
\centerline{
\subfigure[Production $\m{\check{\mathcal{P}}}$. \label{figscoupeprod}]{\includegraphics[width=0.50\textwidth, trim={0 10 15 5}, clip, clip]{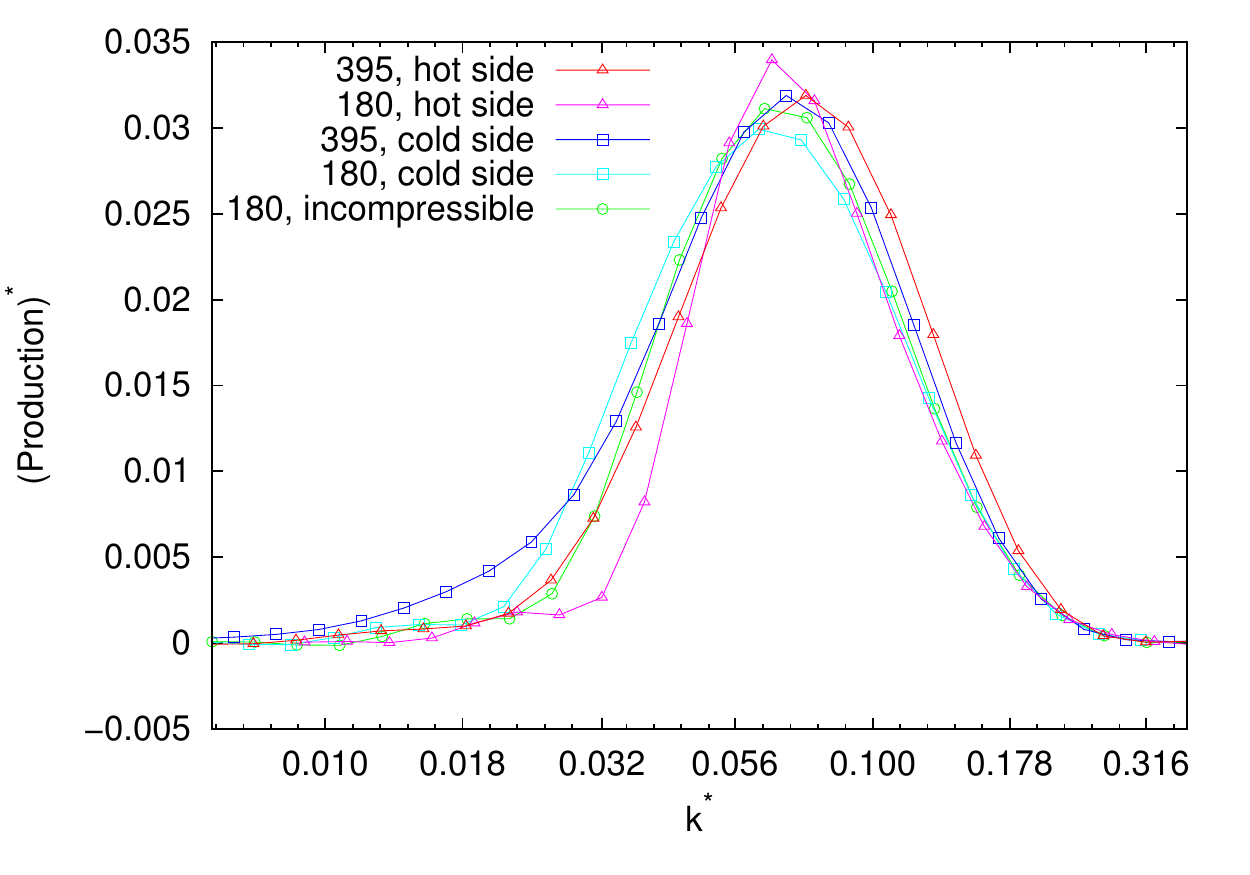}}
\subfigure[Conservative energy transfer $\m{\check{\varphi}}$. \label{figscoupecons}]{\includegraphics[width=0.50\textwidth, trim={0 10 15 5}, clip, clip]{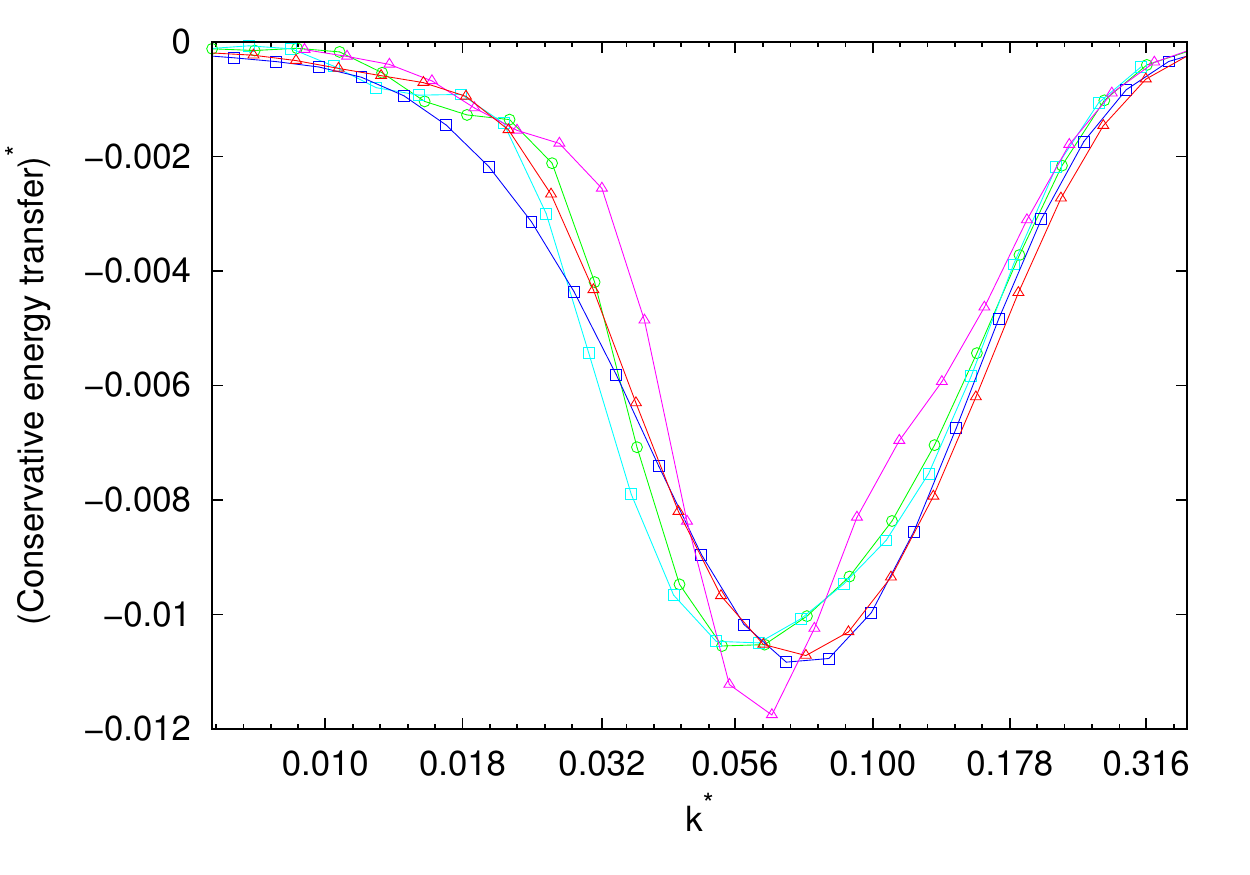}}
}
\centerline{
\subfigure[Purely spectral transfer $\m{\check{\Xi}}$. \label{figscoupepure}]{\includegraphics[width=0.50\textwidth, trim={0 10 15 5}, clip, clip]{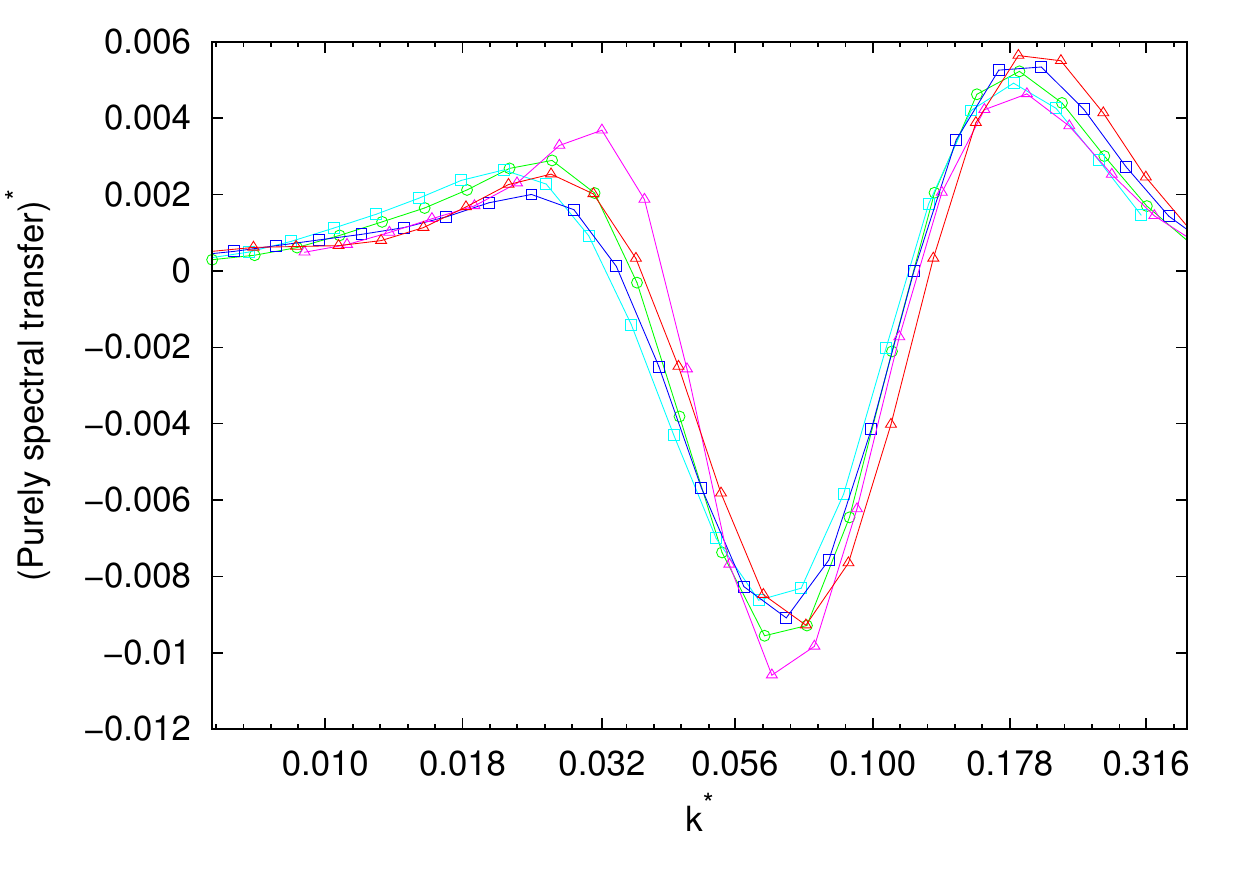}}
\subfigure[Interaction with internal energy $\m{\check{\varepsilon}}$. \label{figscoupeinteractionavecinternal}]{\includegraphics[width=0.50\textwidth, trim={0 10 15 5}, clip, clip]{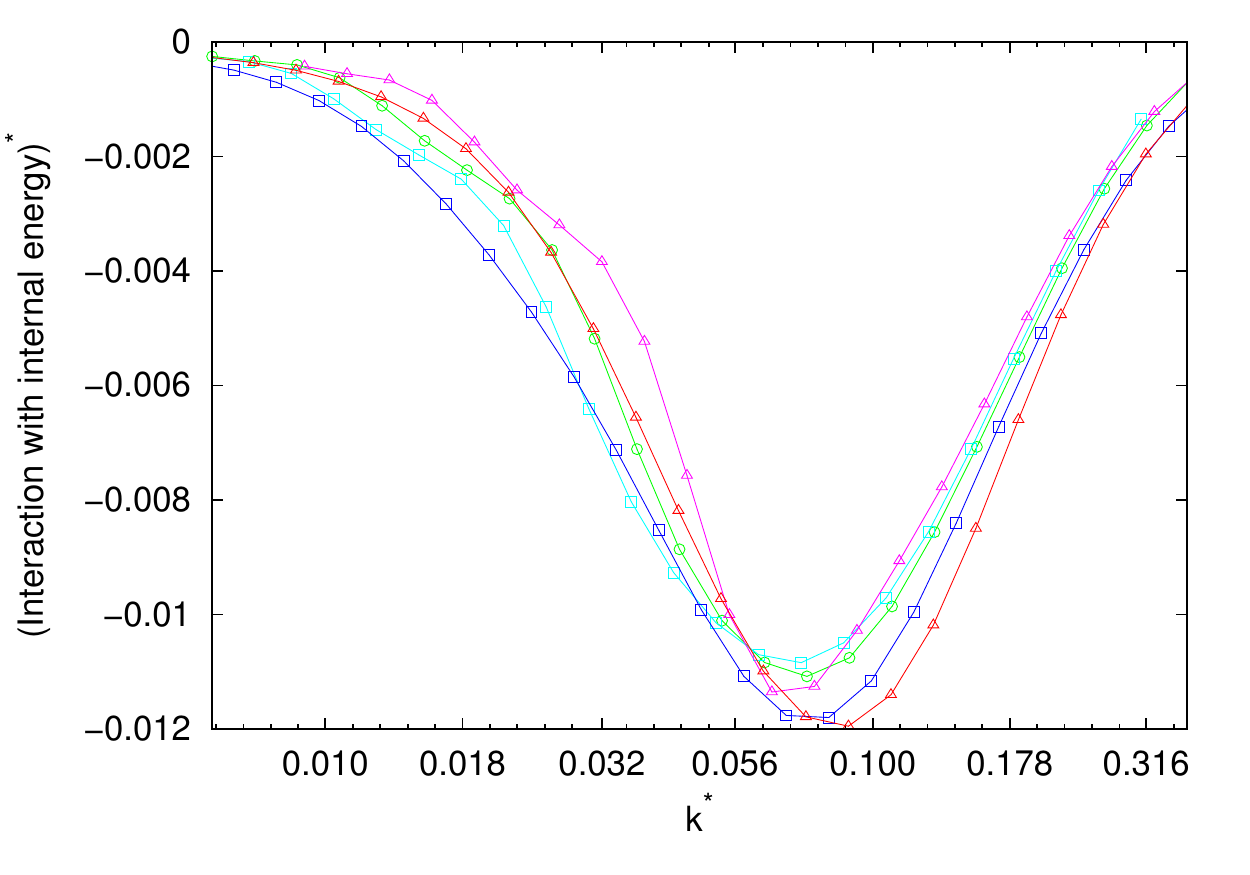}}
}
\caption{
Profiles of the terms of the budget of the half-trace of the velocity
covariance tensor with the semi-local scaling at $Re_{\tau}$ = 180
and $Re_{\tau}$ = 395 as a function of wavenumber at the location of the spatial peak of production~($y^* \approx 12$).
\label{r031}}
\end{figure*}

\begin{figure*}
\centerline{
\subfigure[Conservative energy transfer $\m{\check{\varphi}}$. \label{figscoupeconspp}]{\includegraphics[width=0.50\textwidth, trim={0 10 15 5}, clip, clip]{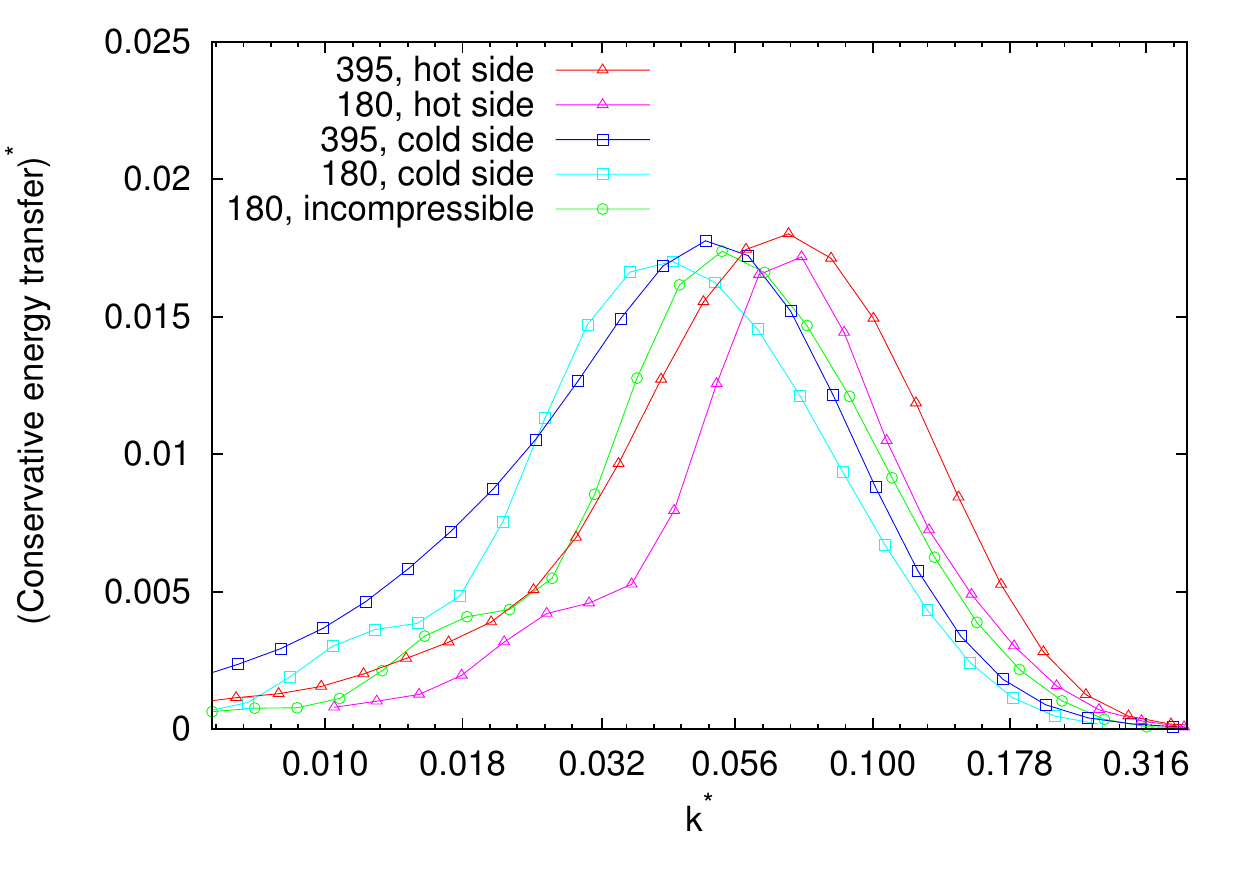}}
\subfigure[Interaction with internal energy $\m{\check{\varepsilon}}$. \label{figscoupeinteractionavecinternalpp}]{\includegraphics[width=0.50\textwidth, trim={0 10 15 5}, clip, clip]{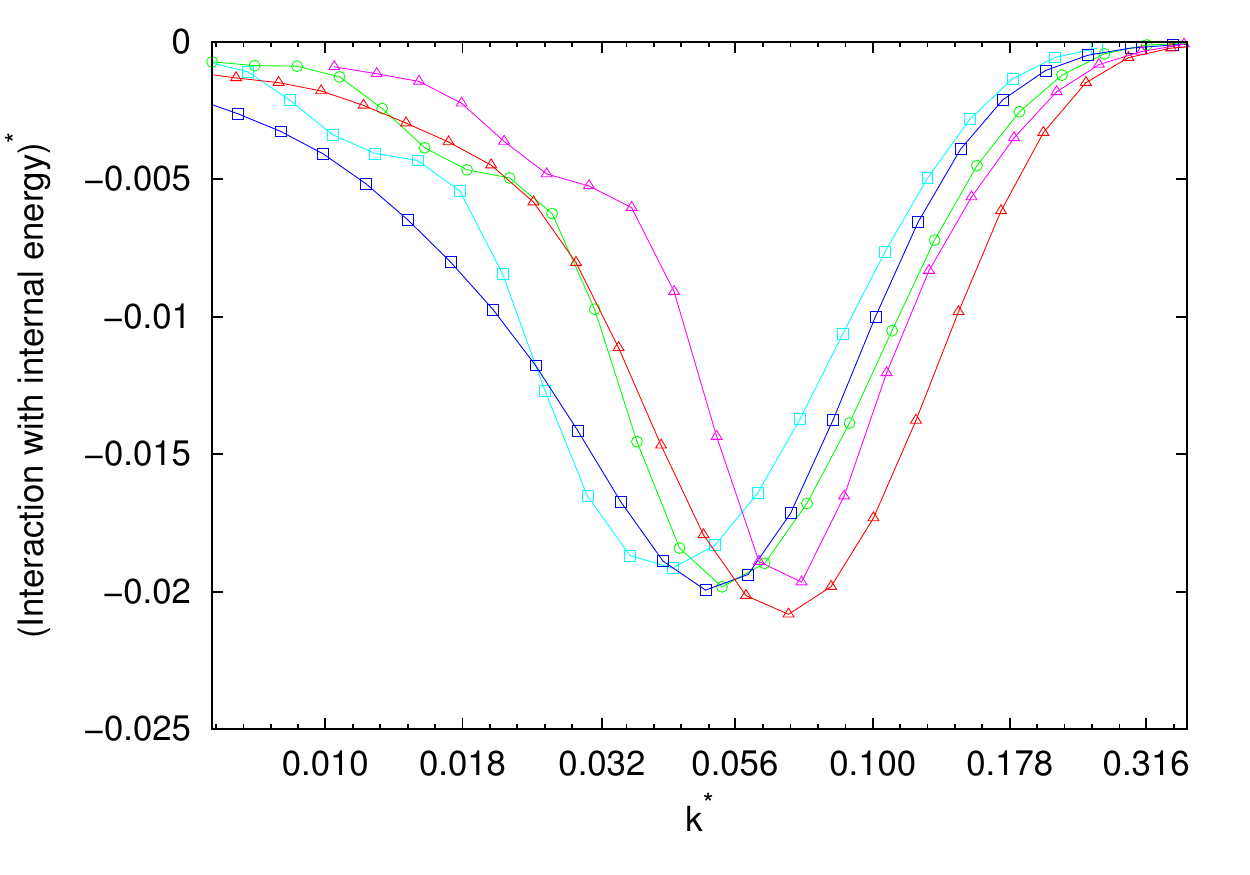}}
}
\caption{
Profiles of the terms of the budget of the half-trace of the velocity
covariance tensor with the semi-local scaling at $Re_{\tau}$ = 180
and $Re_{\tau}$ = 395 as a function of wavenumber very near to the wall
(at $Re_{\tau}$ = 180, $y^* = 0.25$ for the constant-property profile, $y^* = 0.38$ at the cold side and $y^* = 0.15$ at the hot side;
at $Re_{\tau}$ = 395, $y^* = 0.73$ at the cold side and $y^* = 0.30$ at the hot side).
The production and the purely spectral transfer are negligible.
\label{r032}}
\end{figure*}

\begin{figure*}
\centerline{
\subfigure[Conservative energy transfer $\m{\check{\varphi}}$. \label{figscoupeconsac}]{\includegraphics[width=0.50\textwidth, trim={0 10 15 5}, clip, clip]{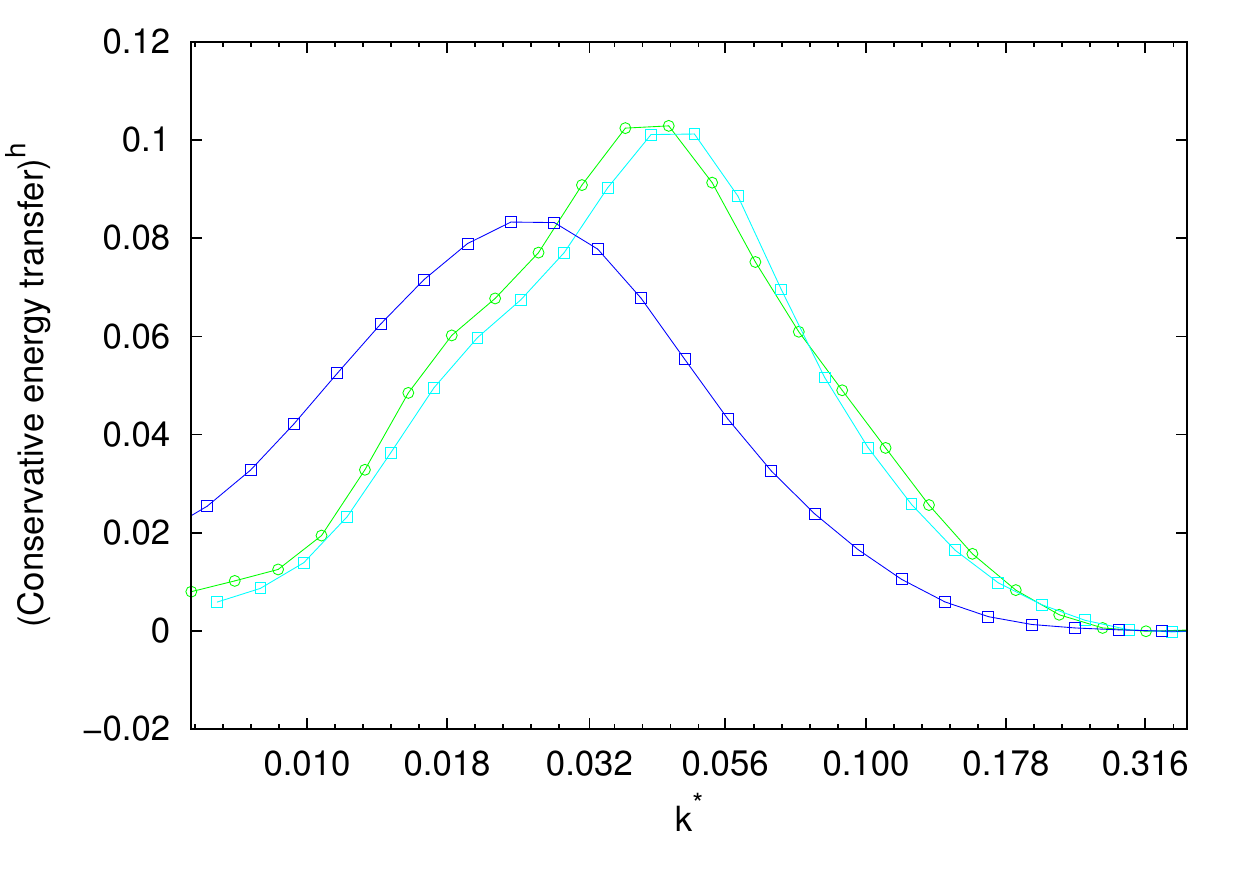}}
}
\centerline{
\subfigure[Purely spectral transfer $\m{\check{\Xi}}$. \label{figscoupepureac}]{\includegraphics[width=0.50\textwidth, trim={0 10 15 5}, clip, clip]{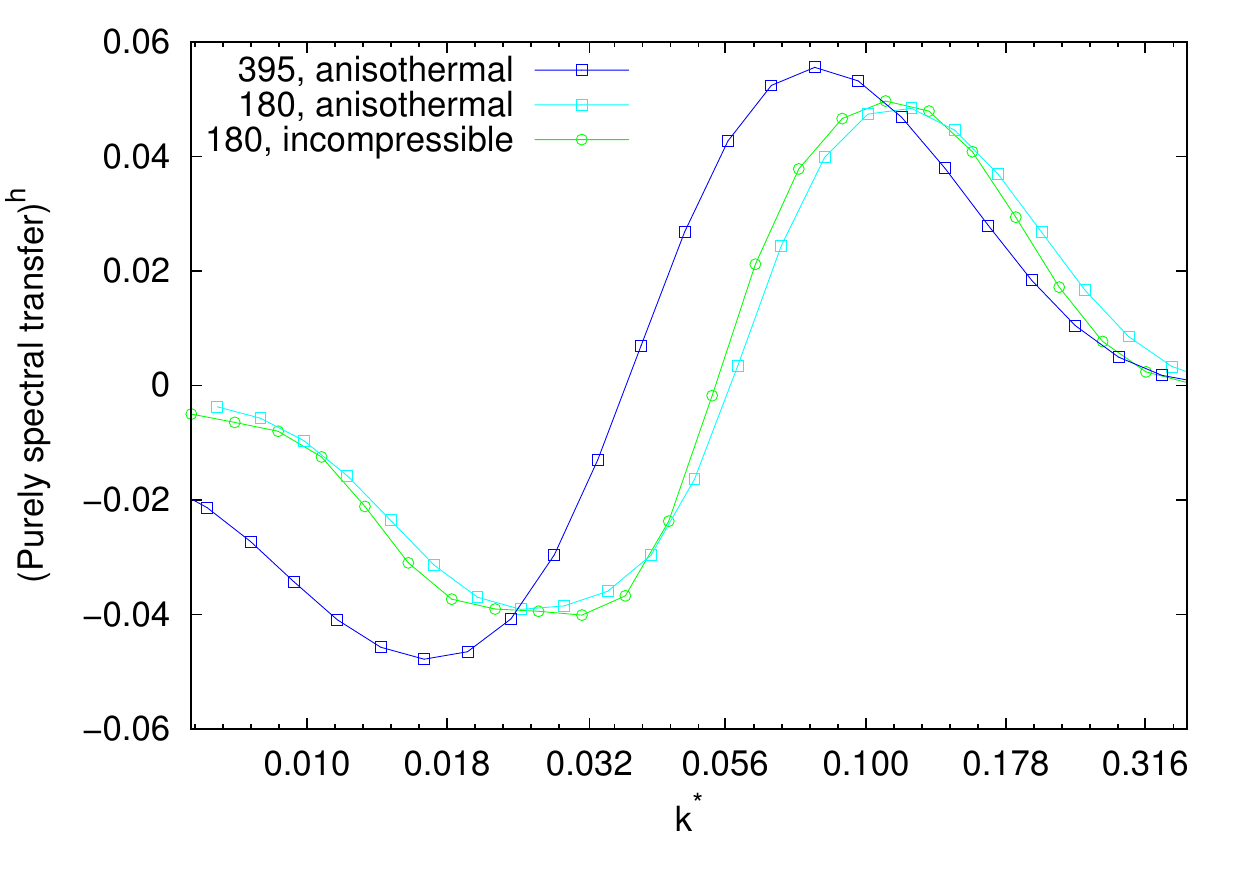}}
\subfigure[Interaction with internal energy $\m{\check{\varepsilon}}$. \label{figscoupeinteractionavecinternalac}]{\includegraphics[width=0.50\textwidth, trim={0 10 15 5}, clip, clip]{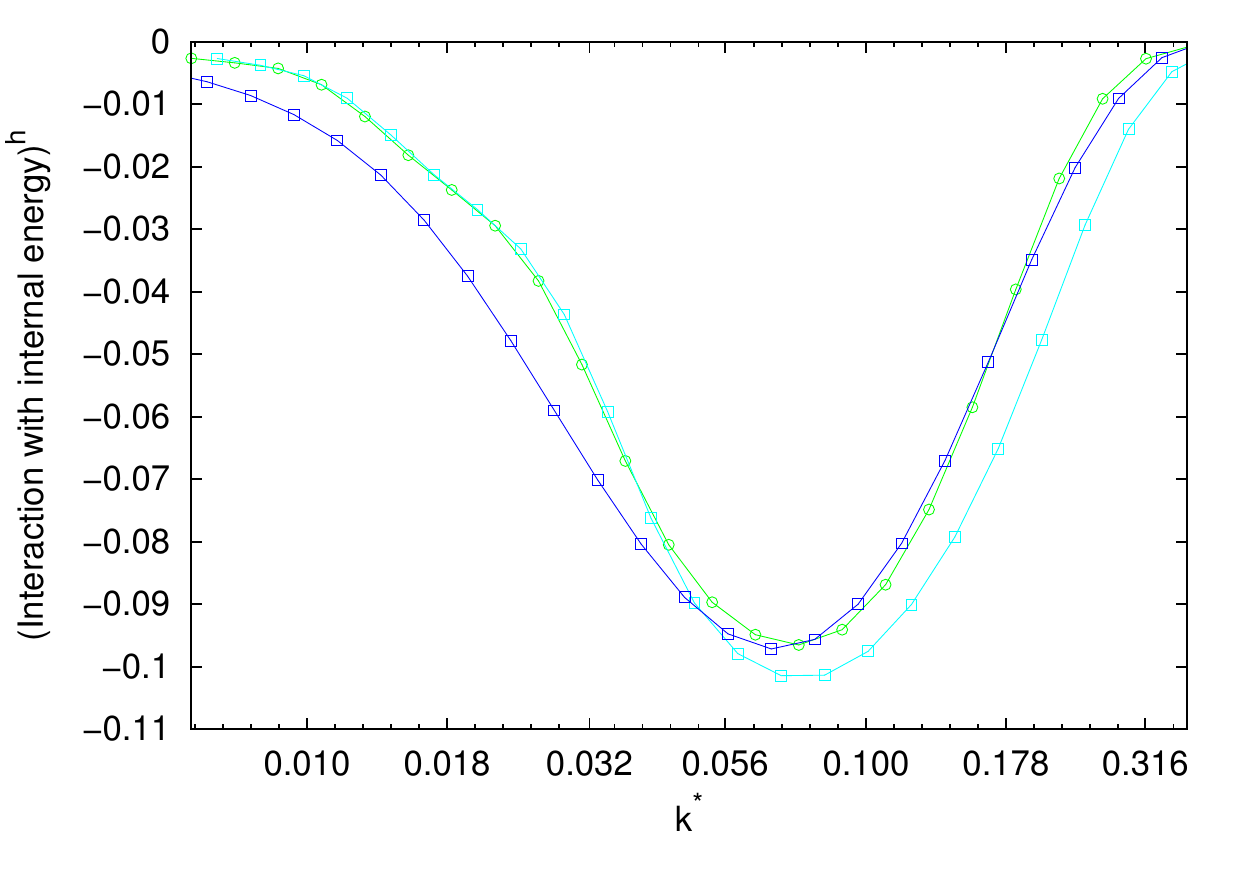}}
}
\caption{
Profiles of the terms of the budget of the half-trace of the velocity
covariance tensor scaled by ${U_{\tau y}^*}^3/h$ at $Re_{\tau}$ = 180
and $Re_{\tau}$ = 395 as a function of wavenumber at the centre of the channel.
The production is negligible. In the anisothermal case, the hot and cold
profiles are equal.
\label{r033}}
\end{figure*}

The temperature gradient alters the energy exchanges. With the
semi-local scaling, the energy exchanges have a lower
spectral magnitude at the cold side than at the hot side, but handle a larger
range of scales. Besides, the peaks of each energy exchanges are located at
smaller wavenumbers and closer to the wall at the cold side than at the hot
side.
This is shown in table \ref{r035}, which reports the spatial and
spectral position and the amplitude of the peaks of each energy exchange.
In addition, the spectra are provided in the set
of figures \ref{r036} to \ref{r044}.
The spectra of the viscous transfer and the convection, the two
most significant parts of the conservative energy transfer, are given in
figures \ref{r038} and \ref{r039}.
With the semi-local scaling, the general behaviour of the spectral energy
exchanges are similar at $Re_{\tau} = 180$ and $Re_{\tau} = 395$.
As in the spatial domain, we consider that the spectral energy exchanges in
the anisothermal configuration are in a large part due to the combination of
the variations of the local fluid properties and a varying low Reynolds number
effect. These two effects also influence the scales handled by the energy exchanges.
A large part of the wavenumber asymmetry related to the variations of the
local fluid properties is taken into account by the semi-local scaling.
In accordance with the spatial results, the spectral extrema are closer to the
wall at higher mean friction Reynolds number.
This also applies to the interscale transport (figure \ref{r043}), which has no contribution in the spatial domain.
The spectra also suggest that the range of scales handled by the energy
exchanges increases with the mean friction Reynolds number, mainly towards large but also towards
small scales.

\begin{figure*}
\centerline{
\subfigure[Hot side, semi-local scaling.  \label{figs395015h}]{\includegraphics[width=0.49\textwidth, trim={200 150 150 400}, clip, clip]{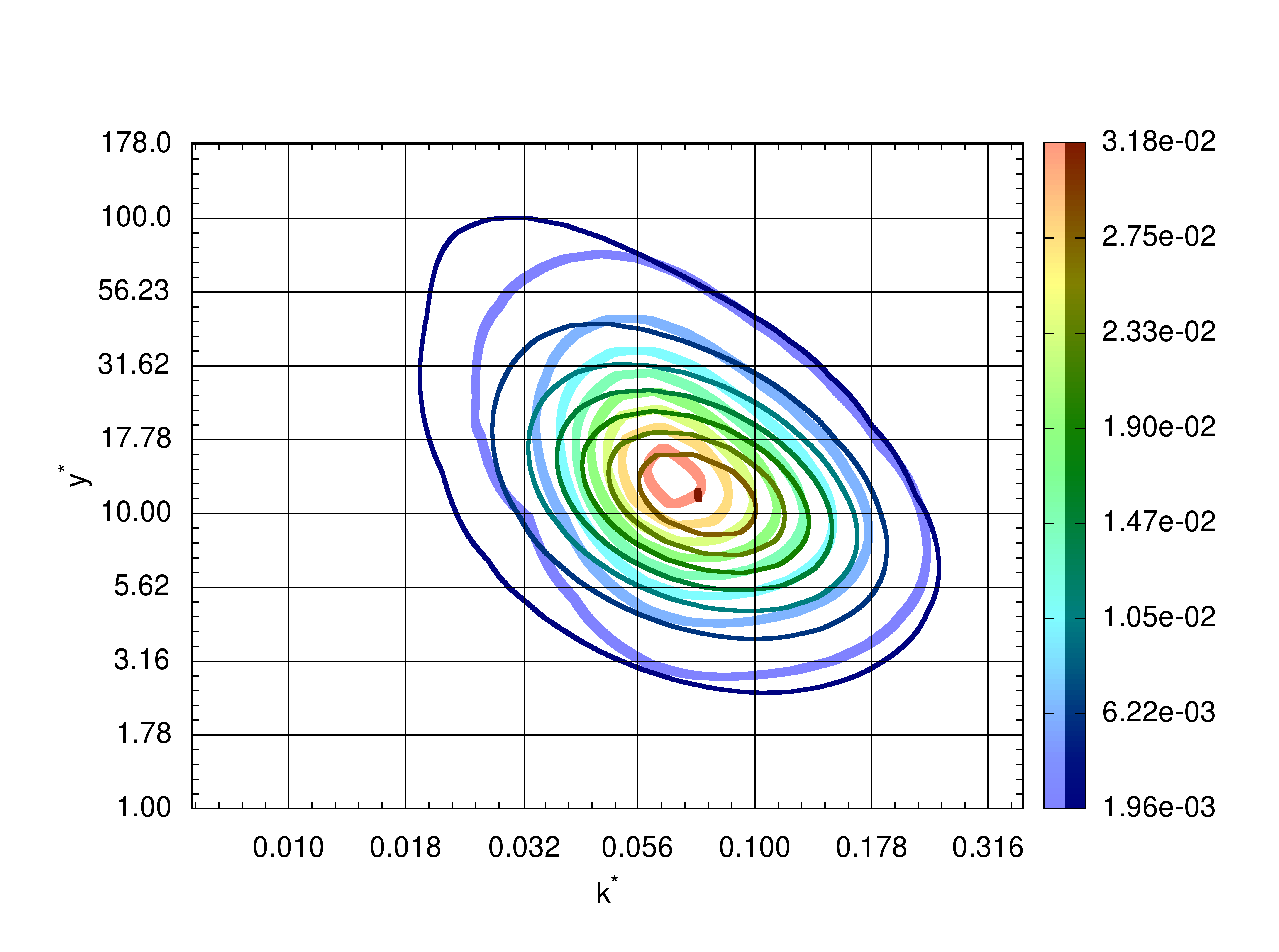}}
\subfigure[Cold side, semi-local scaling.  \label{figs395015c}]{\includegraphics[width=0.49\textwidth, trim={200 150 150 400}, clip, clip]{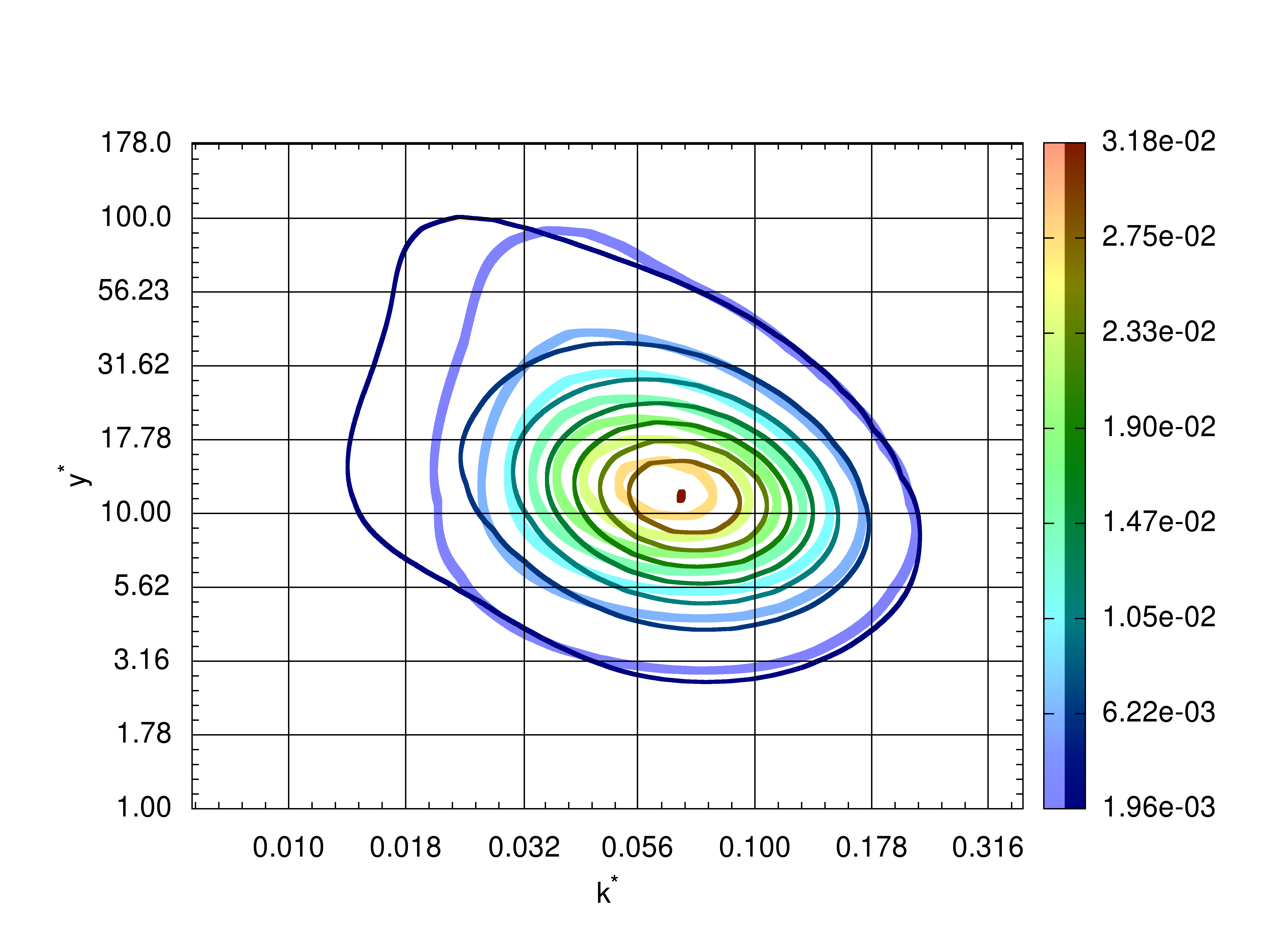}}
}
\caption{
Spectra of the production $\m{\check{\mathcal{P}}}$ with semi-local scaling at $Re_{\tau}$ = 180 (thick lines, light colour) and $Re_{\tau}$ = 395 (thin line, dark colour).
\label{r036}}
\end{figure*}

\begin{figure*}
\centerline{
\subfigure[Hot side, semi-local scaling.  \label{figs395052h}]{\includegraphics[width=0.49\textwidth, trim={200 150 150 400}, clip, clip]{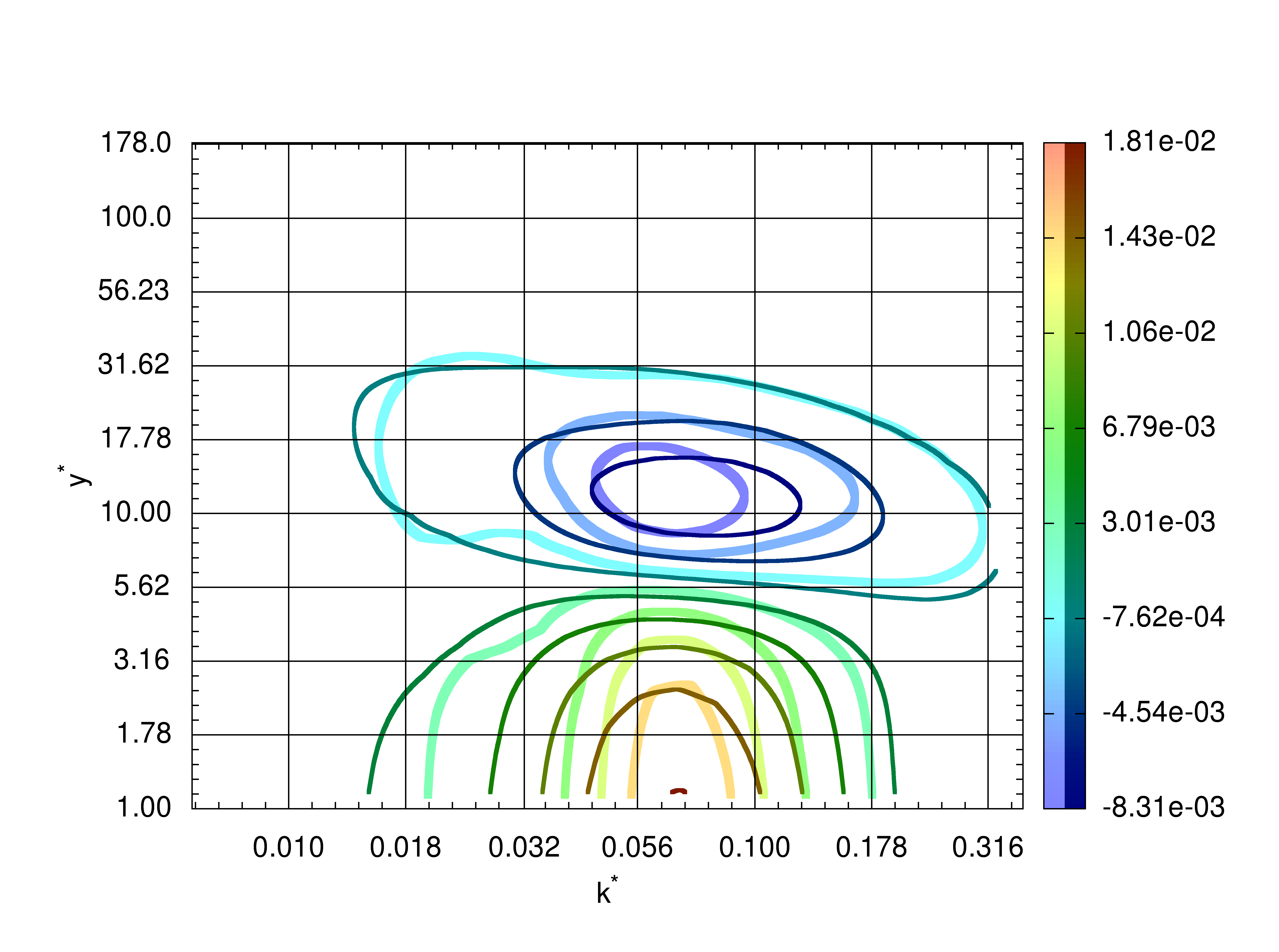}}
\subfigure[Cold side, semi-local scaling.  \label{figs395052c}]{\includegraphics[width=0.49\textwidth, trim={200 150 150 400}, clip, clip]{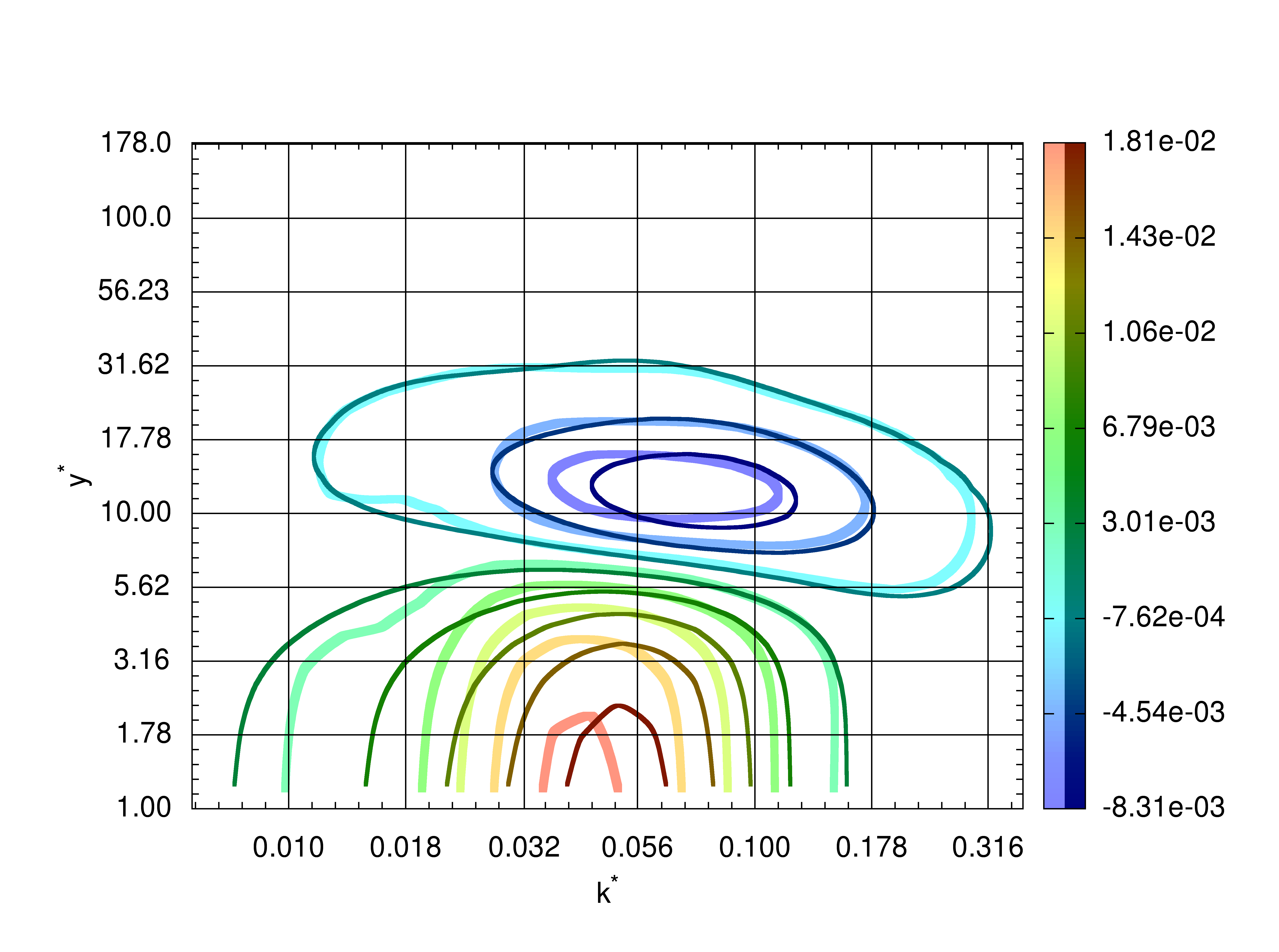}}
}
\caption{
Spectra of the conservative energy transfer $\m{\check{\varphi}}$ with semi-local scaling at $Re_{\tau}$ = 180 (thick lines, light colour) and $Re_{\tau}$ = 395 (thin line, dark colour).
\label{r042}}
\end{figure*}

\begin{figure*}
\centerline{
\subfigure[Hot side, semi-local scaling.  \label{figs395065h}]{\includegraphics[width=0.49\textwidth, trim={200 150 150 400}, clip, clip]{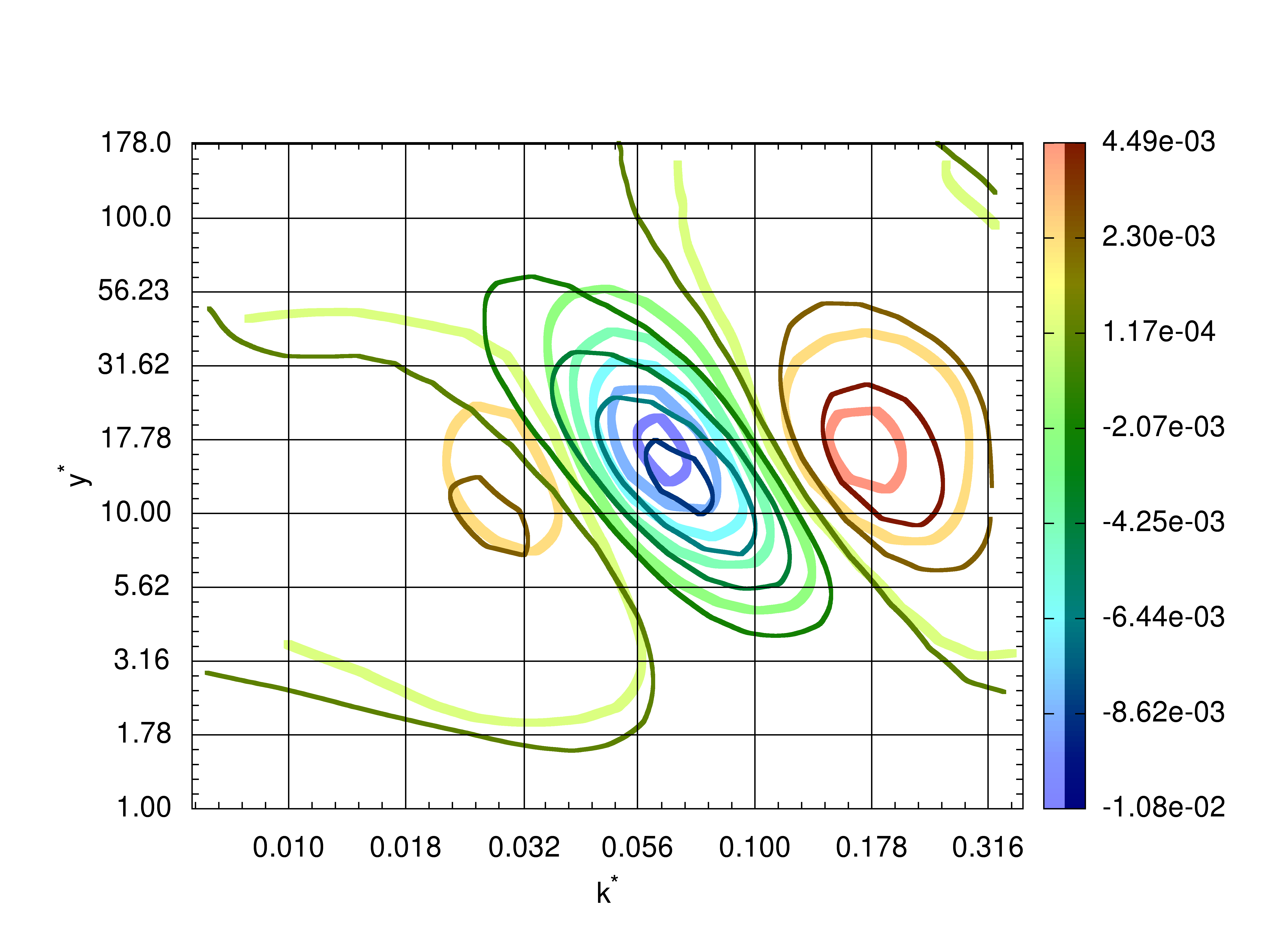}}
\subfigure[Cold side, semi-local scaling. \label{figs395065c}]{\includegraphics[width=0.49\textwidth, trim={200 150 150 400}, clip, clip]{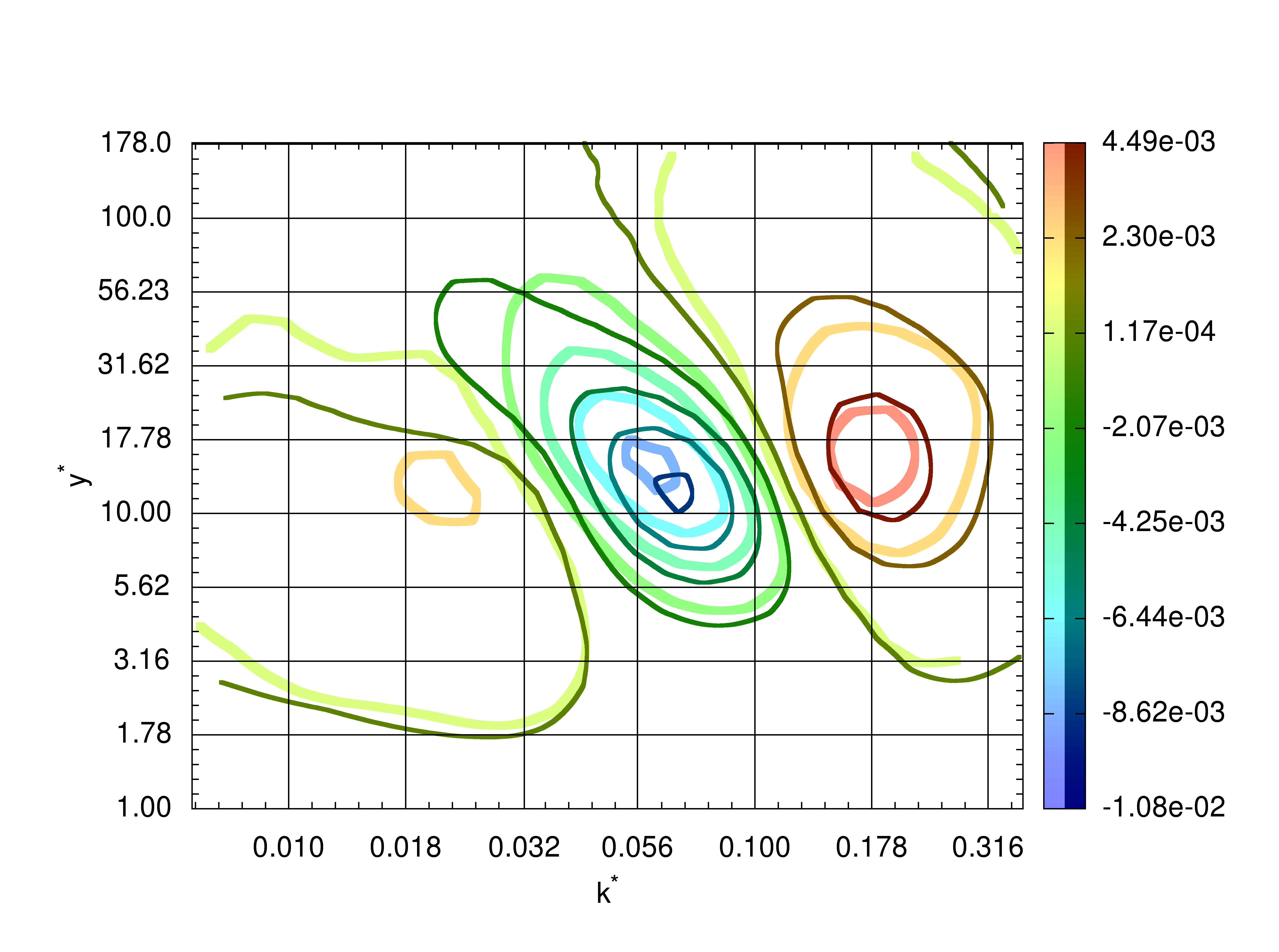}}
}
\caption{
Spectra of the purely spectral transfer $\m{\check{\Xi}}$ with semi-local scaling at $Re_{\tau}$ = 180 (thick lines, light colour) and $Re_{\tau}$ = 395 (thin line, dark colour).
\label{r043}}
\end{figure*}

\begin{figure*}
\centerline{
\subfigure[Hot side, semi-local scaling.  \label{figs395050h}]{\includegraphics[width=0.49\textwidth, trim={200 150 150 400}, clip, clip]{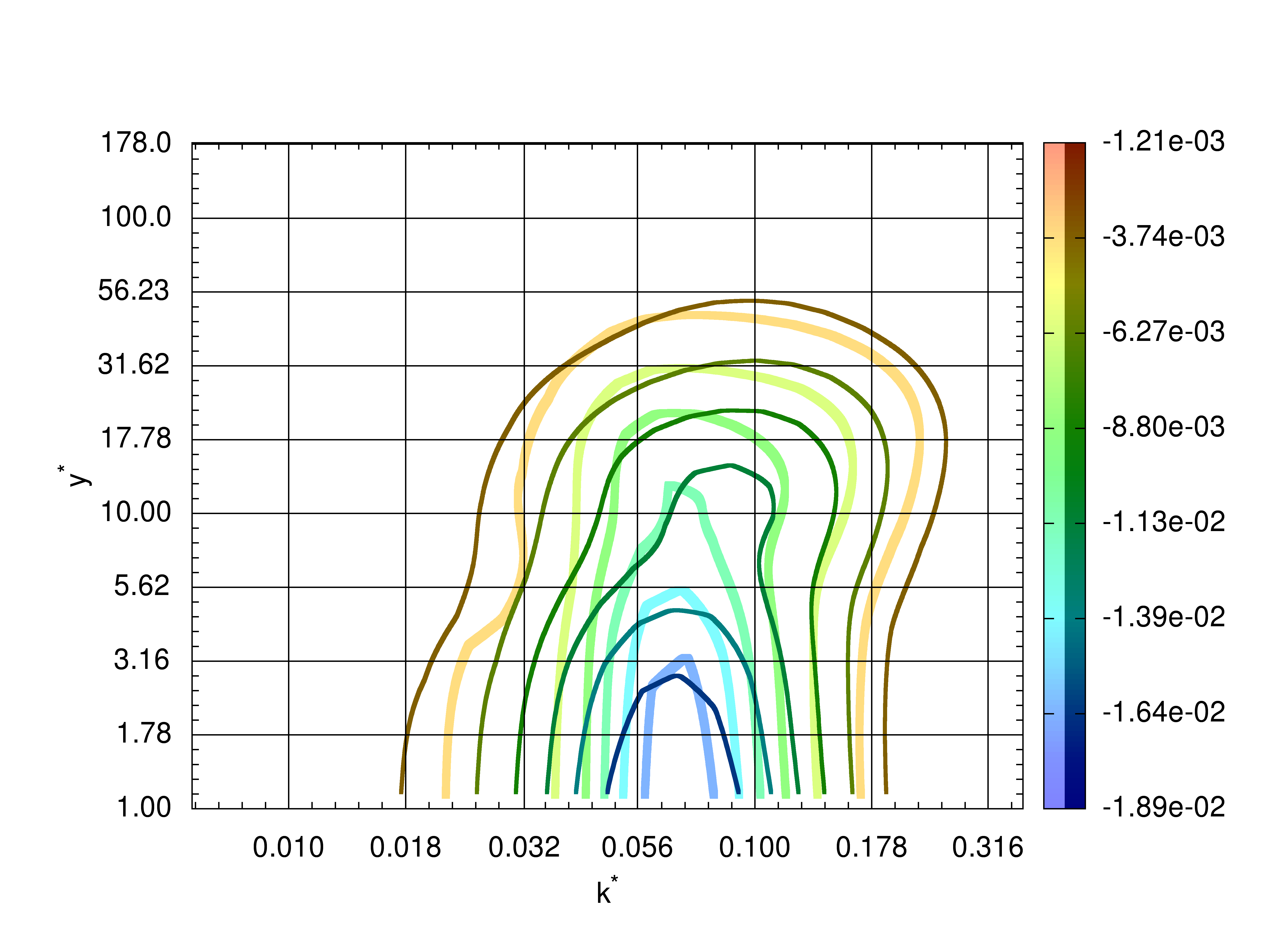}}
\subfigure[Cold side, semi-local scaling.  \label{figs395050c}]{\includegraphics[width=0.49\textwidth, trim={200 150 150 400}, clip, clip]{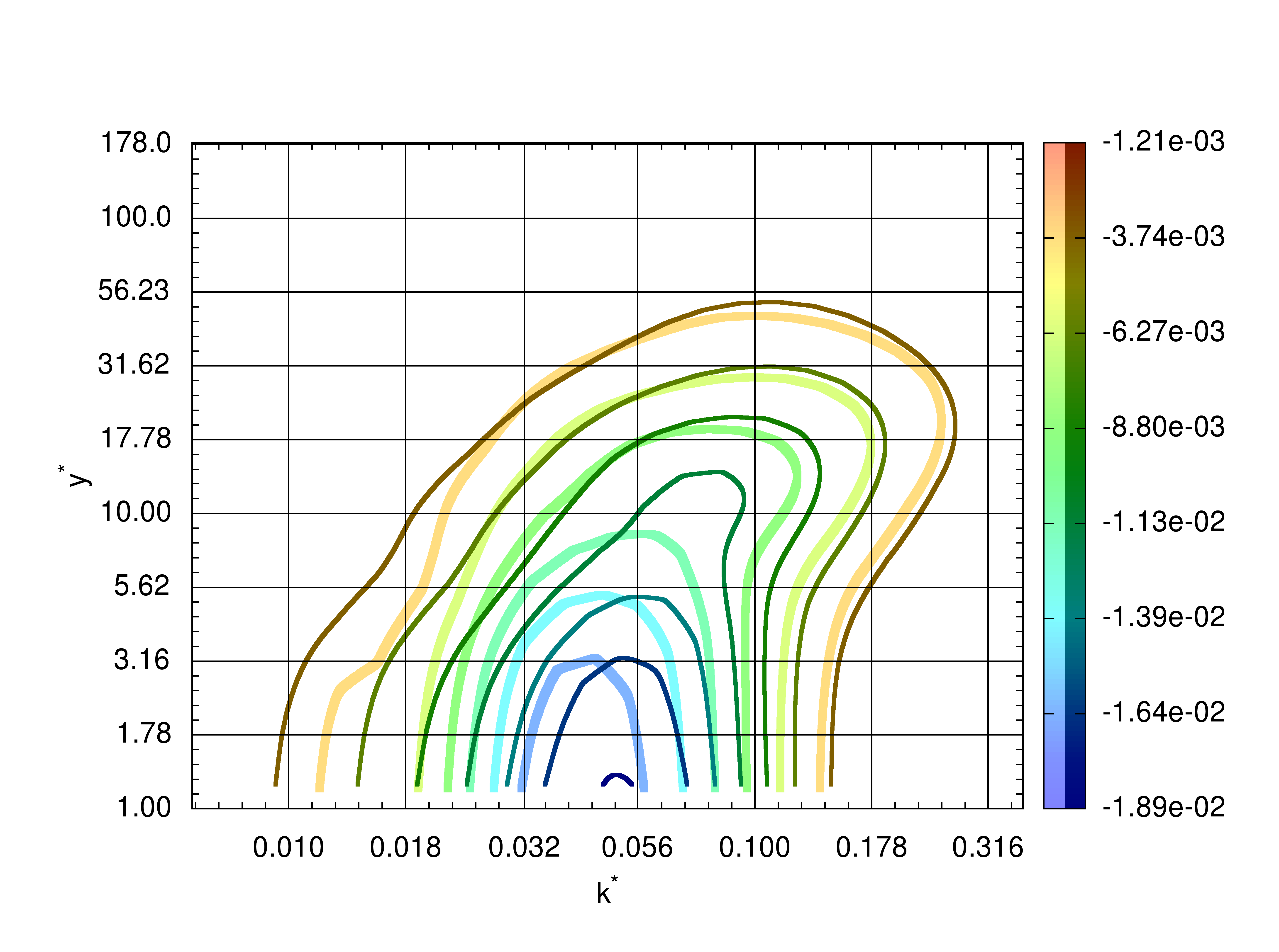}}
}
\caption{
Spectra of the interaction with internal energy $\m{\check{\varepsilon}}$ with semi-local scaling at $Re_{\tau}$ = 180 (thick lines, light colour) and $Re_{\tau}$ = 395 (thin line, dark colour).
\label{r044}}
\end{figure*}

\begin{figure*}
\centerline{
\subfigure[Hot side, semi-local scaling.  \label{figs395023h}]{\includegraphics[width=0.49\textwidth, trim={200 150 150 400}, clip, clip]{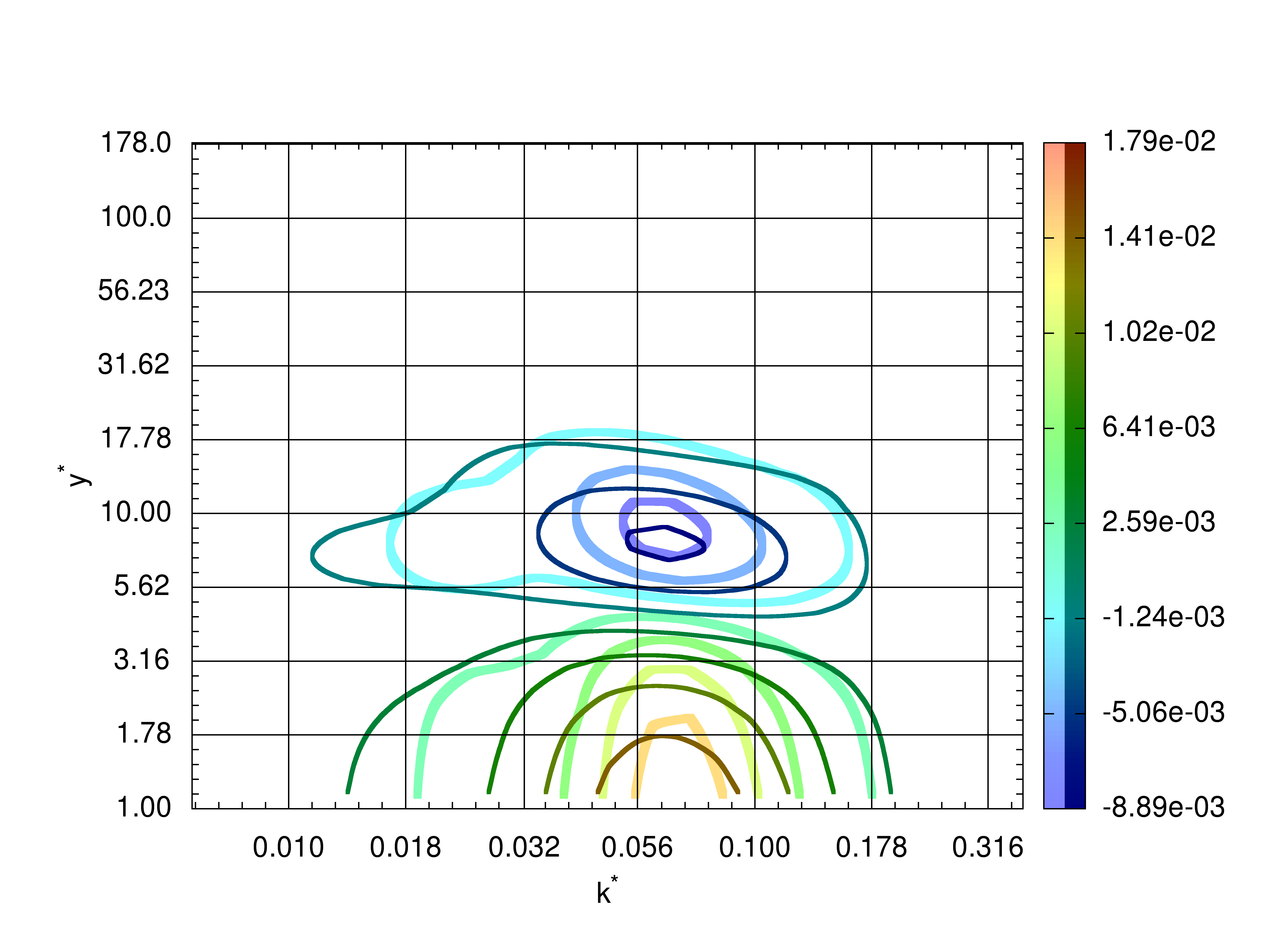}}
\subfigure[Cold side, semi-local scaling. \label{figs395023c}]{\includegraphics[width=0.49\textwidth, trim={200 150 150 400}, clip, clip]{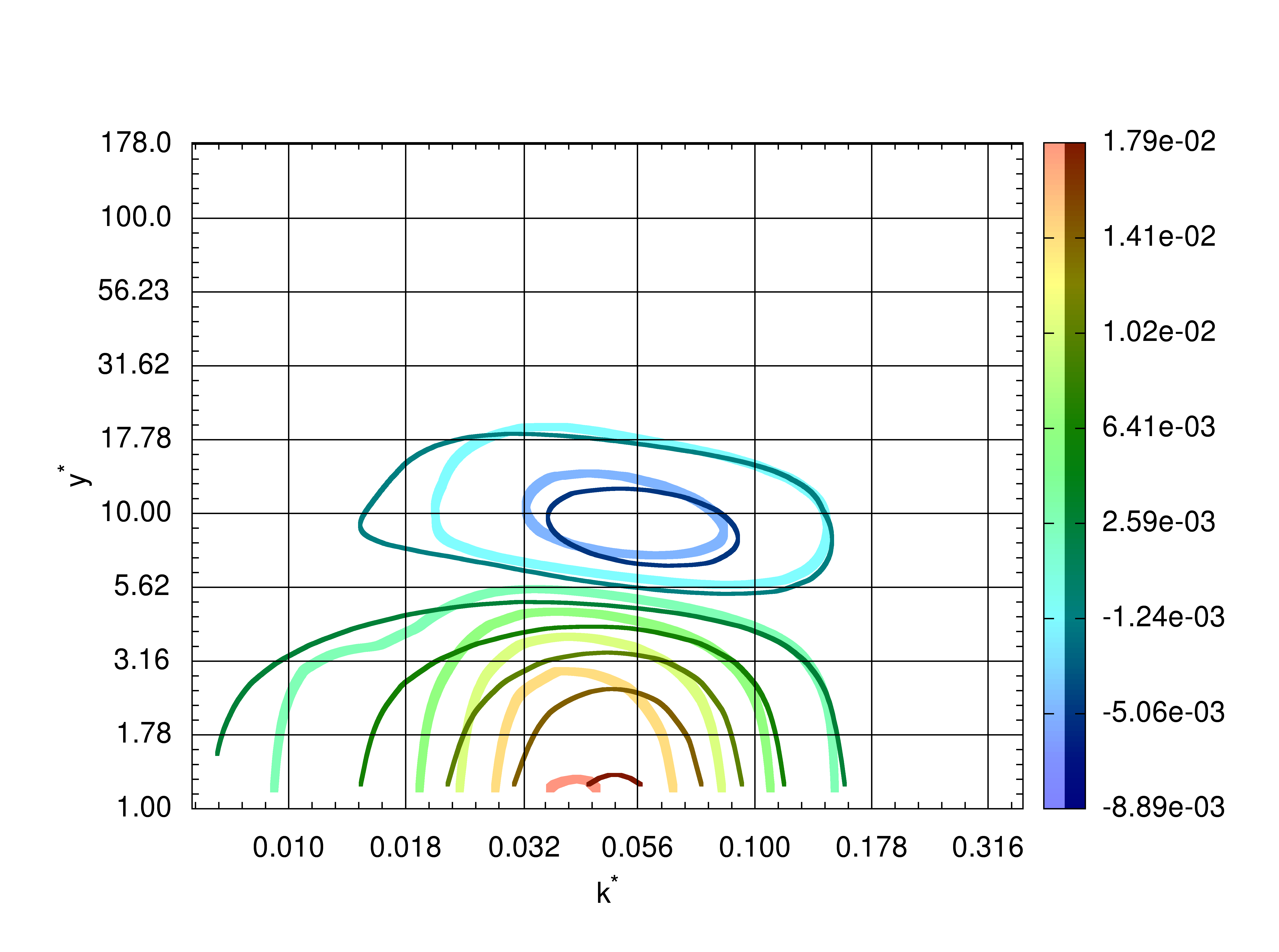}}
}
\caption{
Spectra of the viscous transfer (part of the conservative energy transfer) $\m{\check{\varphi}^{\nu}}$ with semi-local scaling at $Re_{\tau}$ = 180 (thick lines, light colour) and $Re_{\tau}$ = 395 (thin line, dark colour).
\label{r038}}
\end{figure*}

\begin{figure*}
\centerline{
\subfigure[Hot side, semi-local scaling.  \label{figs395042h}]{\includegraphics[width=0.49\textwidth, trim={200 150 150 400}, clip, clip]{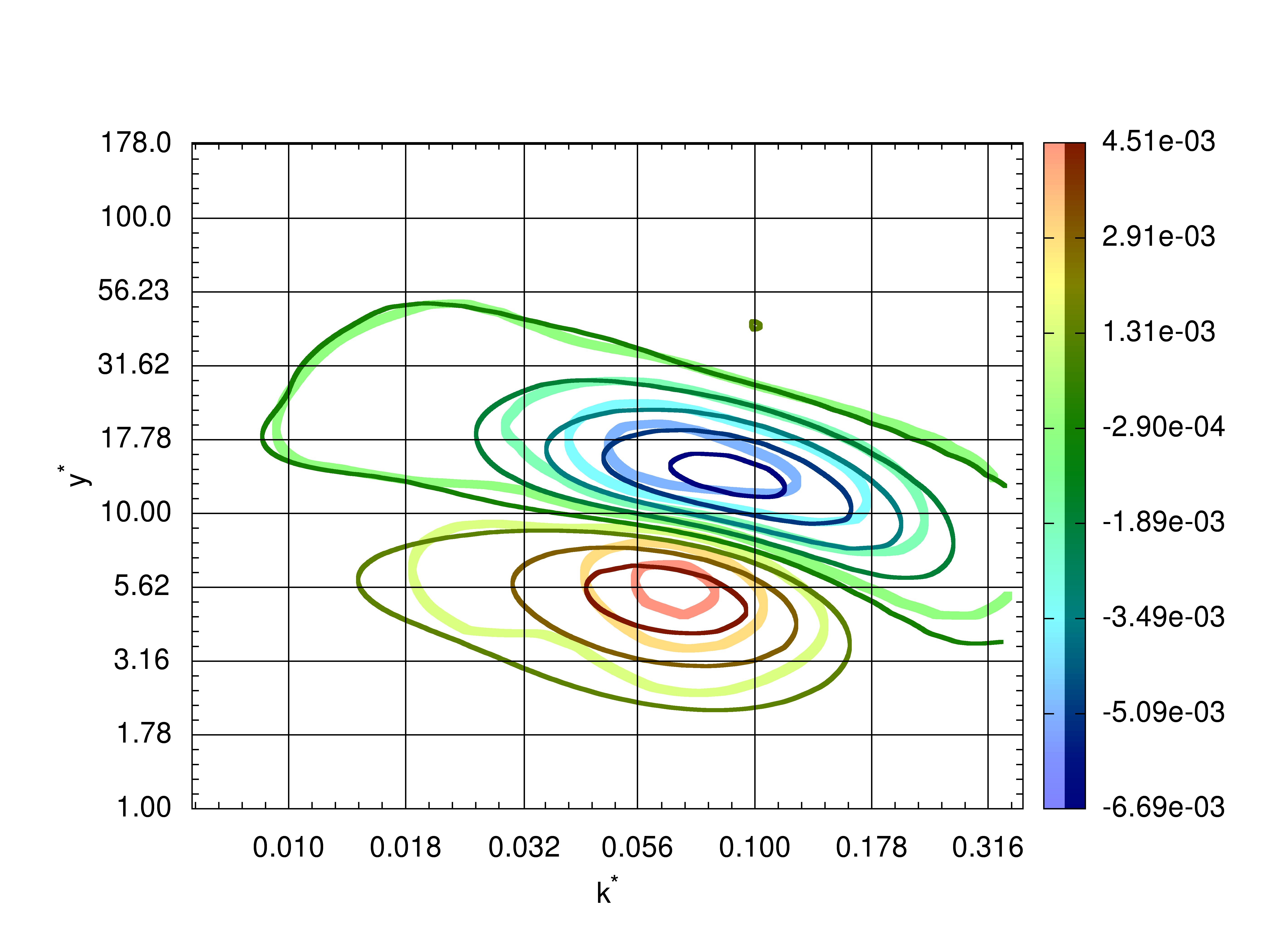}}
\subfigure[Cold side, semi-local scaling. \label{figs395042c}]{\includegraphics[width=0.49\textwidth, trim={200 150 150 400}, clip, clip]{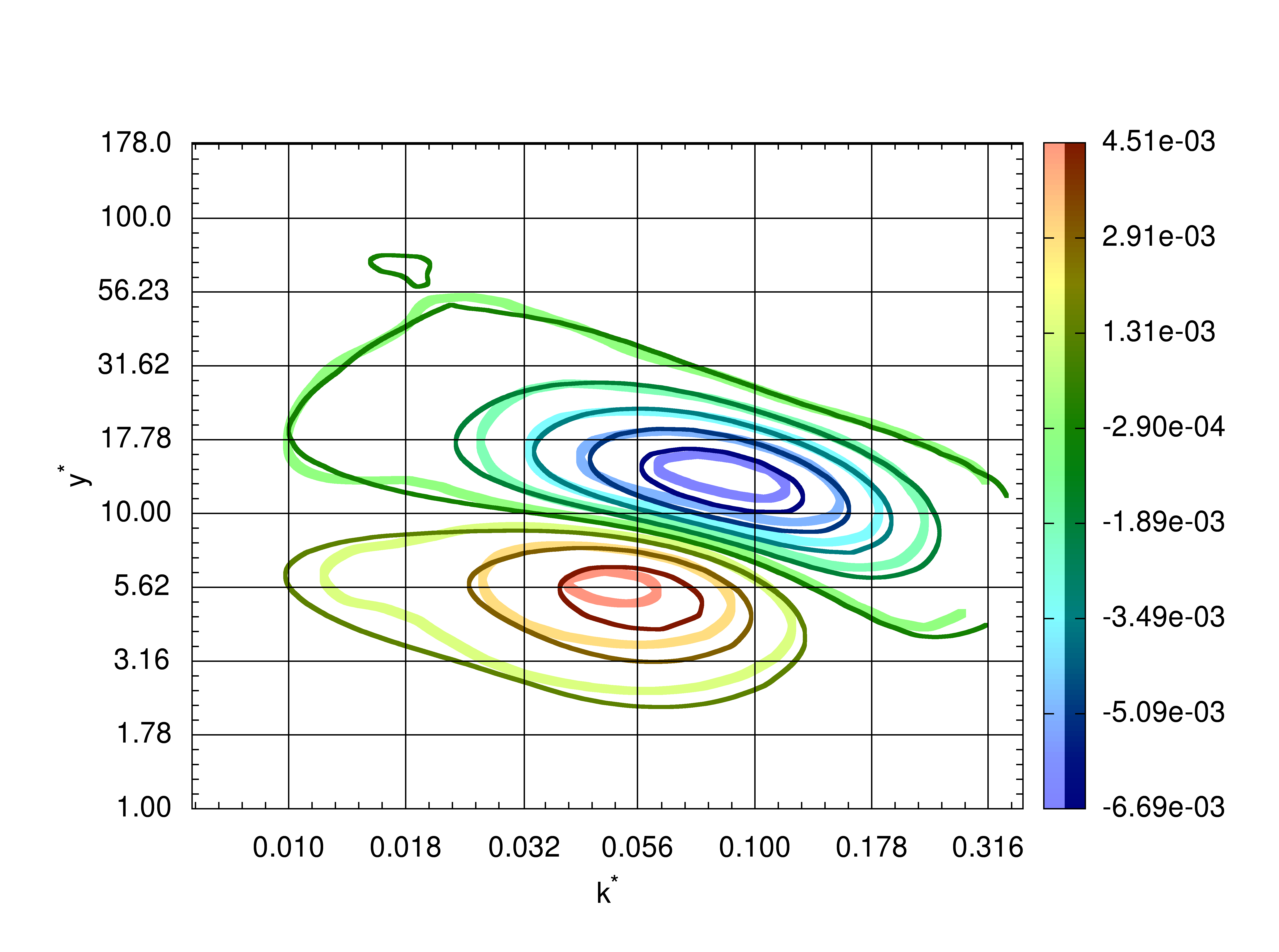}}
}
\caption{
Spectra of the convection (part of the conservative energy transfer) $\m{\check{\varphi}^{c}}$ with semi-local scaling at $Re_{\tau}$ = 180 (thick lines, light colour) and $Re_{\tau}$ = 395 (thin line, dark colour).
\label{r039}}
\end{figure*}

Besides these general effects relevant to all energy exchanges,
the production peak (figure \ref{r036})
is shifted towards small scales
at $Re_{\tau} = 395$ compared to $Re_{\tau} = 180$
at the hot and cold sides. Its amplitude is increased at the cold side and decreased at the hot side, reducing the
asymmetry between the hot and cold sides in terms of spectral magnitude.
The negative peak of the conservative energy transfer (figure \ref{r042})
follows the same pattern, thereby staying centred on the peak of production.
The positive peak behaves like the extremum of the interaction with internal
energy, described later.
The negative extremum of the interscale transport (figure \ref{r043})
is shifted towards
small scales at the higher mean friction Reynolds number while the maximum at large scales is shifted towards large
scales at the hot side and towards small scales at the cold side, reducing the
asymmetry between the hot and cold sides.
The negative area and the positive area at large scales have a smaller
amplitude at $Re_{\tau} = 395$ than at $Re_{\tau} = 180$ whereas the positive area
at small scales has a larger amplitude. The direct interscale transport towards
small scales is therefore strengthened compared to the transport towards large
scales.
The extremum of the interaction with internal energy (figure \ref{r044})
has a larger amplitude at the higher mean friction Reynolds number.
It is also shifted towards large scales at the hot side and towards small scales at
the cold side, reducing the wavenumber asymmetry between the hot and cold
side. This suggests that the wavenumber asymmetry between the hot and cold
sides at $Re_{\tau} = 180$ results in part from a low Reynolds number
effect, that is from the semi-local friction Reynolds number asymmetry between the
hot and cold side.
The same remark can be made regarding the amplitude asymmetry of the
production peak and the wavenumber asymmetry of the positive area at large
scales of the interscale transport, although carefulness is required.
A more precise knowledge of the different effects of the friction Reynolds
number on the spectra in the constant-property isothermal case would be required
for a more accurate separation of the different effects of the
temperature gradient.

\begin{table*}
\centerline{\begin{tabular}{lccclccl}%
\multicolumn{2}{l}{                                       }                           & \multicolumn{3}{c}{$\overbrace{\hspace*{3.0cm}}^{\displaystyle Re_{\tau}=180}$}     & \multicolumn{3}{c}{$\overbrace{\hspace*{3.0cm}}^{\displaystyle Re_{\tau}=395}$}  \\
                                                                    & Side            & $k^*$  & $y^*$ & Amplitude              & $k^*$  & $y^*$ & Amplitude\\[.5em]
\hline\\[-.5em]
\multirow{2}{6cm}{Production}                                         & C               & $0.062$  & $12.4$  & $\phantom{-}\numprint{3.00e-2}$    & $0.069$  & $11.2$  & $\phantom{-}\numprint{3.19e-2}$   \\
                                                                    & H               & $0.065$  & $13.5$  & $\phantom{-}\numprint{3.42e-2}$    & $0.076$  & $11.6$  & $\phantom{-}\numprint{3.19e-2}$   \\
\hline\\[-1em]
\multirow{4}{6cm}{Conservative energy transfer}                       & C               & $0.043$  & $0.64$  & $\phantom{-}\numprint{1.92e-2}$    & $0.050$  & $1.21$  & $\phantom{-}\numprint{1.97e-2}$   \\
                                                                    & H               & $0.074$  & $0.26$  & $\phantom{-}\numprint{1.93e-2}$    & $0.070$  & $0.50$  & $\phantom{-}\numprint{2.00e-2}$\\[.5em]
                                                                    & C               & $0.052$  & $12.4$  & $\numprint{-1.05e-2}$              & $0.070$  & $11.7$  & $\numprint{-1.08e-2}$             \\
                                                                    & H               & $0.066$  & $11.9$  & $\numprint{-1.18e-2}$              & $0.076$  & $11.2$  & $\numprint{-1.07e-2}$             \\
\hline\\[-1em]
\multirow{6}{6cm}{Purely spectral transfer}                           & C               & $0.021$  & $12.4$  & $\phantom{-}\numprint{2.65e-3}$    & $0.023$  & $9.76$  & $\phantom{-}\numprint{2.25e-3}$   \\
                                                                    & H               & $0.032$  & $13.5$  & $\phantom{-}\numprint{3.73e-3}$    & $0.026$  & $10.4$  & $\phantom{-}\numprint{2.59e-3}$\\[.5em]
                                                                    & C               & $0.063$  & $14.0$  & $\numprint{-9.13e-3}$              & $0.070$  & $11.7$  & $\numprint{-9.08e-3}$             \\
                                                                    & H               & $0.064$  & $16.5$  & $\numprint{-1.21e-2}$              & $0.075$  & $12.4$  & $\numprint{-9.33e-3}$\\[.5em]
                                                                    & C               & $0.186$  & $14.6$  & $\phantom{-}\numprint{5.29e-3}$    & $0.174$  & $14.8$  & $\phantom{-}\numprint{5.77e-3}$   \\
                                                                    & H               & $0.187$  & $15.8$  & $\phantom{-}\numprint{5.07e-3}$    & $0.181$  & $15.3$  & $\phantom{-}\numprint{6.24e-3}$   \\
\hline\\[-1em]
\multirow{2}{6cm}{Interaction with internal energy}                   & C               & $0.043$  & $0.38$  & $\numprint{-1.91e-2}$              & $0.049$  & $0.73$  & $\numprint{-1.99e-2}$             \\
                                                                    & H               & $0.074$  & $0.15$  & $\numprint{-1.97e-2}$              & $0.070$  & $0.30$  & $\numprint{-2.08e-2}$             \\
\hline\\[-1em]
\multirow{4}{6cm}{Viscous transfer}                                   & C               & $0.043$  & $0.64$  & $\phantom{-}\numprint{1.88e-2}$    & $0.050$  & $1.21$  & $\phantom{-}\numprint{1.84e-2}$   \\
                                                                    & H               & $0.074$  & $0.26$  & $\phantom{-}\numprint{1.90e-2}$    & $0.070$  & $0.50$  & $\phantom{-}\numprint{1.91e-2}$\\[.5em]
                                                                    & C               & $0.050$  & $9.56$  & $\numprint{-7.81e-3}$              & $0.056$  & $8.80$  & $\numprint{-7.50e-3}$             \\
                                                                    & H               & $0.067$  & $8.71$  & $\numprint{-1.06e-2}$              & $0.065$  & $7.95$  & $\numprint{-9.45e-3}$             \\
\hline\\[-1em]
\multirow{4}{6cm}{Convection}                                         & C               & $0.047$  & $5.58$  & $\phantom{-}\numprint{4.75e-3}$    & $0.054$  & $5.48$  & $\phantom{-}\numprint{5.51e-3}$   \\
                                                                    & H               & $0.070$  & $5.62$  & $\phantom{-}\numprint{5.20e-3}$    & $0.067$  & $5.15$  & $\phantom{-}\numprint{5.62e-3}$\\[.5em]
                                                                    & C               & $0.090$  & $13.5$  & $\numprint{-7.29e-3}$              & $0.084$  & $13.2$  & $\numprint{-8.09e-3}$  \\
                                                                    & H               & $0.064$  & $15.8$  & $\numprint{-6.52e-3}$              & $0.089$  & $13.6$  & $\numprint{-7.30e-3}$  \\
\hline\\[-1em]
\multirow{4}{6cm}{Thermal conservative energy transfer}               & C               & $0.038$  & $3.87$  & $\phantom{-}\numprint{3.77e-3}$    & $0.052$  & $3.59$  & $\phantom{-}\numprint{3.33e-3}$   \\
                                                                    & H               & $0.071$  & $4.03$  & $\numprint{-3.08e-3}$              & $0.057$  & $3.51$  & $\numprint{-2.49e-3}$\\[.5em]
                                                                    & C               & $0.045$  & $15.1$  & $\numprint{-9.19e-4}$              & $0.059$  & $13.8$  & $\numprint{-6.94e-4}$             \\
                                                                    & H               & $0.053$  & $16.5$  & $\phantom{-}\numprint{9.90e-4}$    & $0.052$  & $14.5$  & $\phantom{-}\numprint{6.51e-4}$   \\
\hline\\[-1em]
\multirow{4}{6cm}{Interaction with variable density kinetic energy}   & C               & $0.040$  & $5.94$  & $\numprint{-2.04e-3}$              & $0.054$  & $5.48$  & $\numprint{-1.88e-3}$             \\
                                                                    & H               & $0.069$  & $5.92$  & $\phantom{-}\numprint{1.80e-3}$    & $0.066$  & $5.44$  & $\phantom{-}\numprint{1.51e-3}$\\[.5em]
                                                                    & C               & $0.057$  & $27.5$  & $\phantom{-}\numprint{2.09e-4}$    & $0.075$  & $27.2$  & $\phantom{-}\numprint{1.73e-4}$   \\
                                                                    & H               & $0.050$  & $31.6$  & $\numprint{-2.56e-4}$              & $0.059$  & $30.3$  & $\numprint{-1.66e-4}$             \\
\hline\\[-1em]
\multirow{4}{6cm}{Thermal interaction with internal energy}           & C               & $0.039$  & $4.53$  & $\numprint{-1.55e-3}$              & $0.052$  & $4.07$  & $\numprint{-1.37e-3}$             \\
                                                                    & H               & $0.070$  & $4.79$  & $\phantom{-}\numprint{1.34e-3}$    & $0.056$  & $4.30$  & $\phantom{-}\numprint{1.05e-3}$\\[.5em]
                                                                    & C               & $0.160$  & $19.6$  & $\phantom{-}\numprint{3.53e-4}$    & $0.150$  & $19.8$  & $\phantom{-}\numprint{3.15e-4}$   \\
                                                                    & H               & $0.150$  & $24.8$  & $\numprint{-2.39e-4}$              & $0.147$  & $22.7$  & $\numprint{-2.22e-4}$             \\
\end{tabular}}
\caption{Wavenumber, wall-normal coordinate and amplitude of the local spectral
extrema of each term of the equation of the half-trace of the velocity
covariance tensor with the semi-local scaling at the hot (H)  and
cold (C) sides at $Re_{\tau} = 180$ and $Re_{\tau} = 395$ in the anisothermal
configuration. \label{r035}}
\end{table*}

\subsubsection{Spectra of the thermal terms in the anisothermal configuration}

\begin{figure*}
\centerline{
\subfigure[Hot side, semi-local scaling.  \label{figs395074h}]{\includegraphics[width=0.49\textwidth, trim={200 150 150 400}, clip, clip]{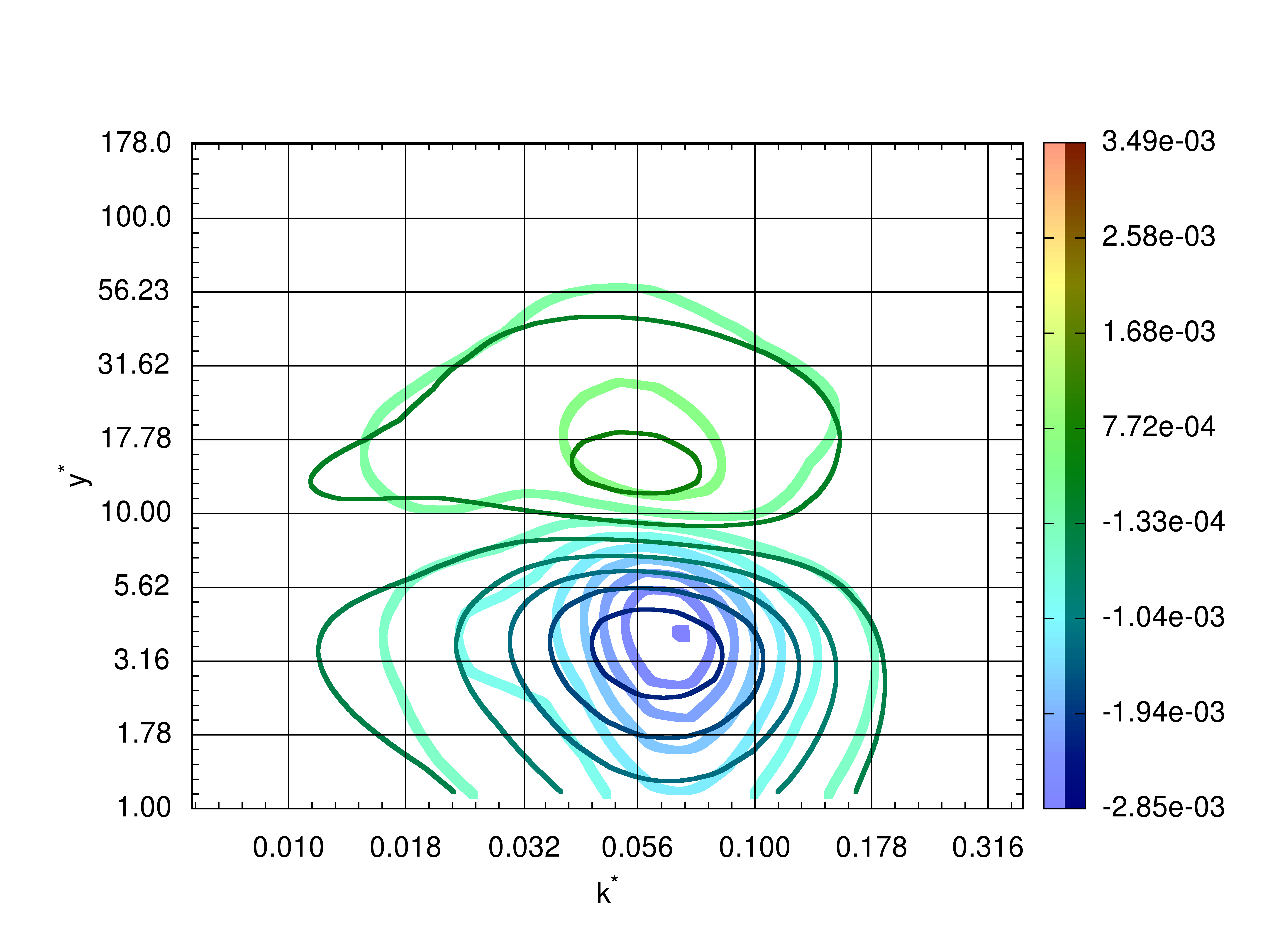}}
\subfigure[Cold side, semi-local scaling. \label{figs395074c}]{\includegraphics[width=0.49\textwidth, trim={200 150 150 400}, clip, clip]{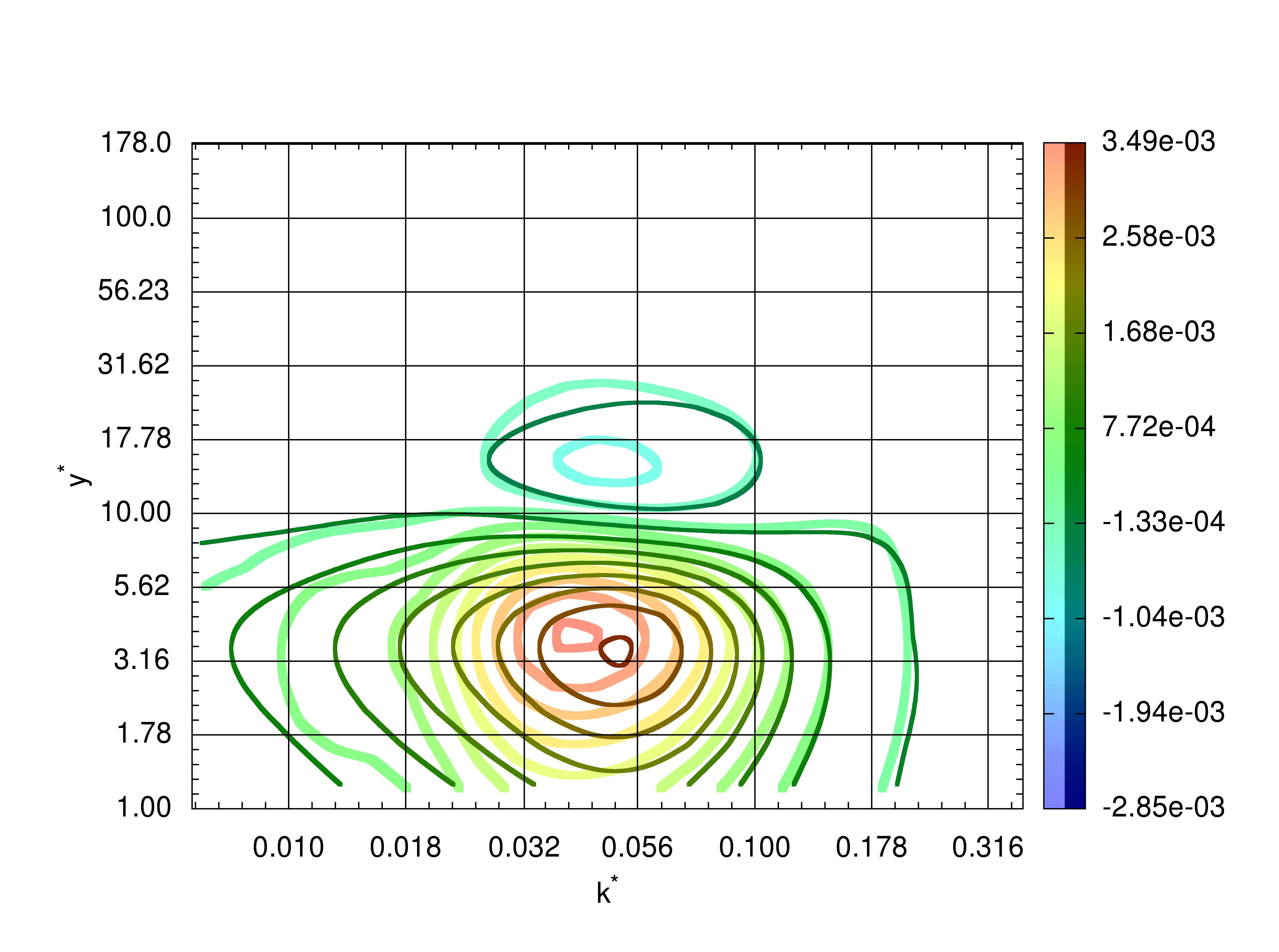}}
}
\caption{
Spectra of the thermal conservative energy transfer $\m{\check{\varphi}_{\varGamma}}$ with semi-local scaling at $Re_{\tau}$ = 180 (thick lines, light colour) and $Re_{\tau}$ = 395 (thin line, dark colour).
\label{r045}}
\end{figure*}

\begin{figure*}
\centerline{
\subfigure[Hot side, semi-local scaling.  \label{figs395051h}]{\includegraphics[width=0.49\textwidth, trim={200 150 150 400}, clip, clip]{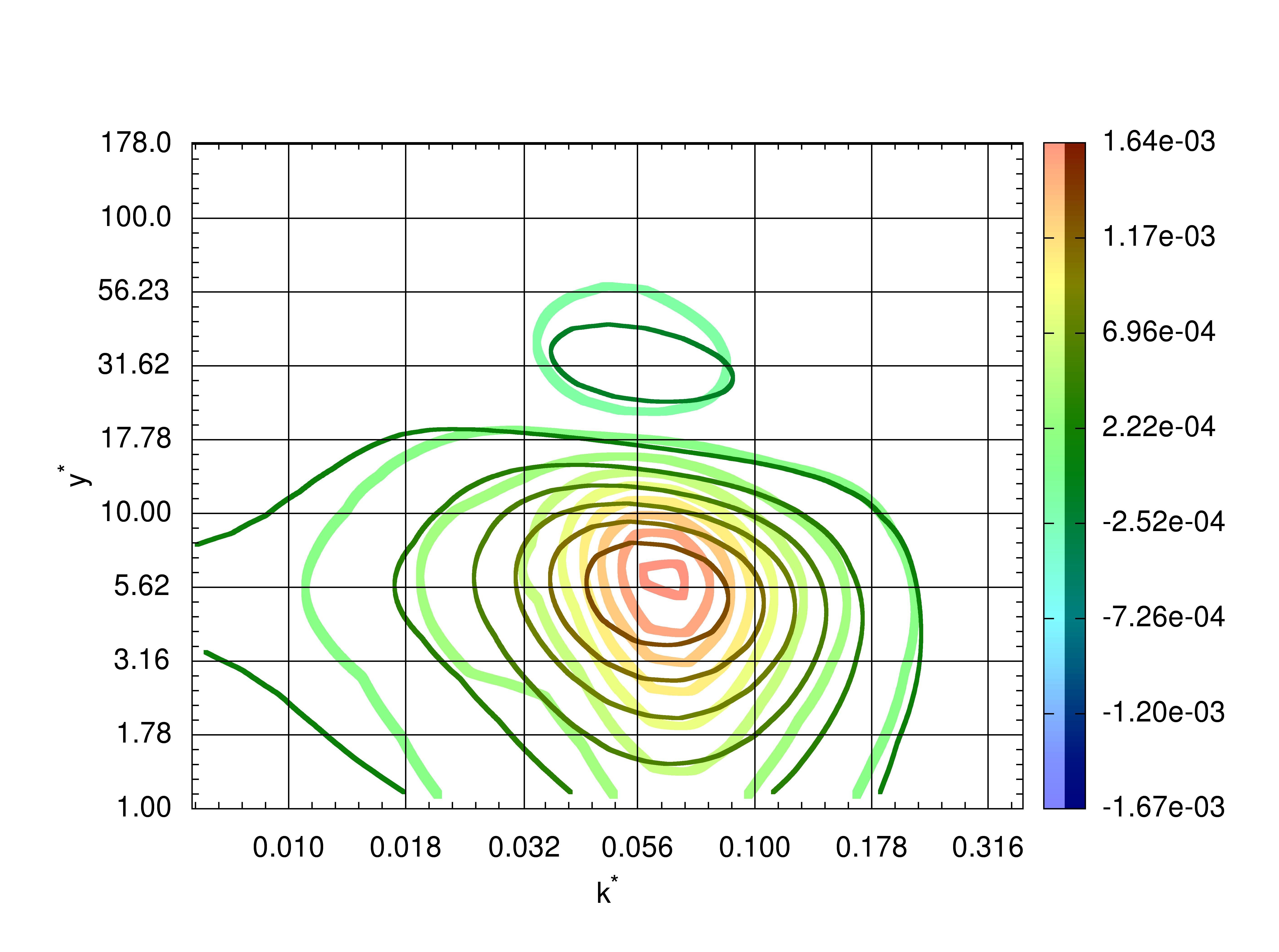}}
\subfigure[Cold side, semi-local scaling.  \label{figs395051c}]{\includegraphics[width=0.49\textwidth, trim={200 150 150 400}, clip, clip]{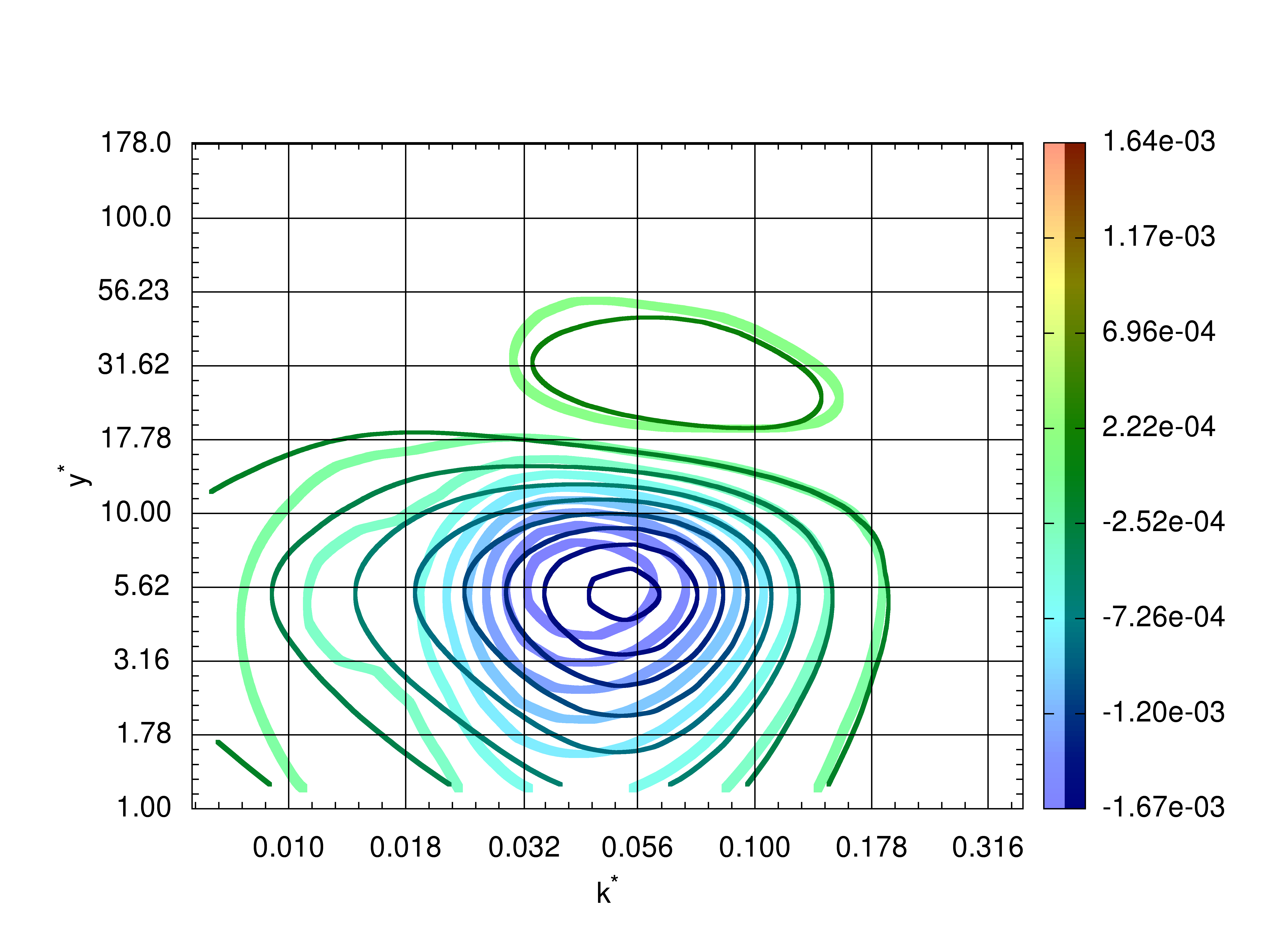}}
}
\caption{
Spectra of the interaction with variable density kinetic energy $\m{\check{\zeta}}$ with semi-local scaling at $Re_{\tau}$ = 180 (thick lines, light colour) and $Re_{\tau}$ = 395 (thin line, dark colour).
\label{figs395051}}
\end{figure*}
\begin{figure*}
\centerline{
\subfigure[Hot side, semi-local scaling.  \label{figs395075h}]{\includegraphics[width=0.49\textwidth, trim={200 150 150 400}, clip, clip]{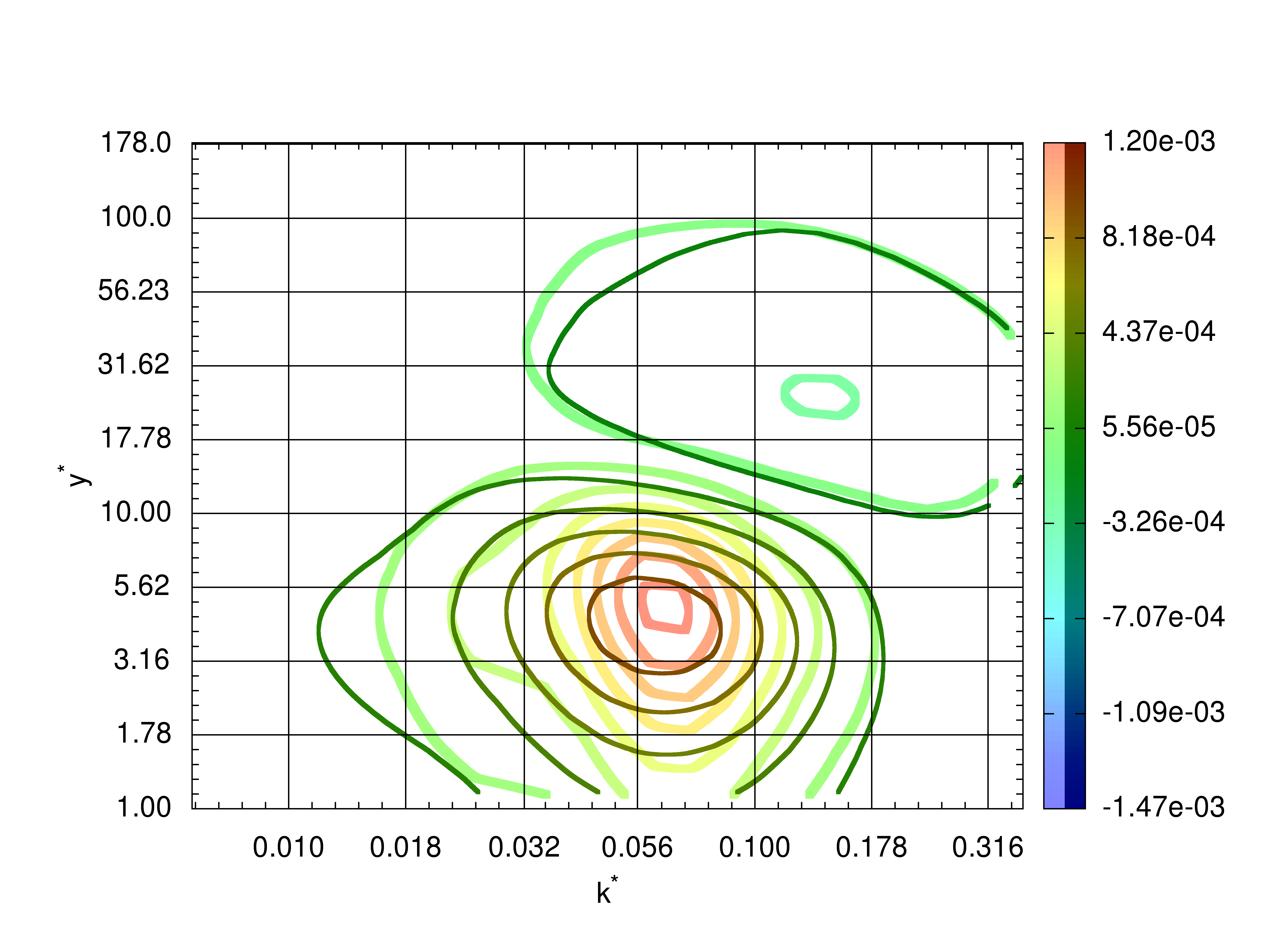}}
\subfigure[Cold side, semi-local scaling. \label{figs395075c}]{\includegraphics[width=0.49\textwidth, trim={200 150 150 400}, clip, clip]{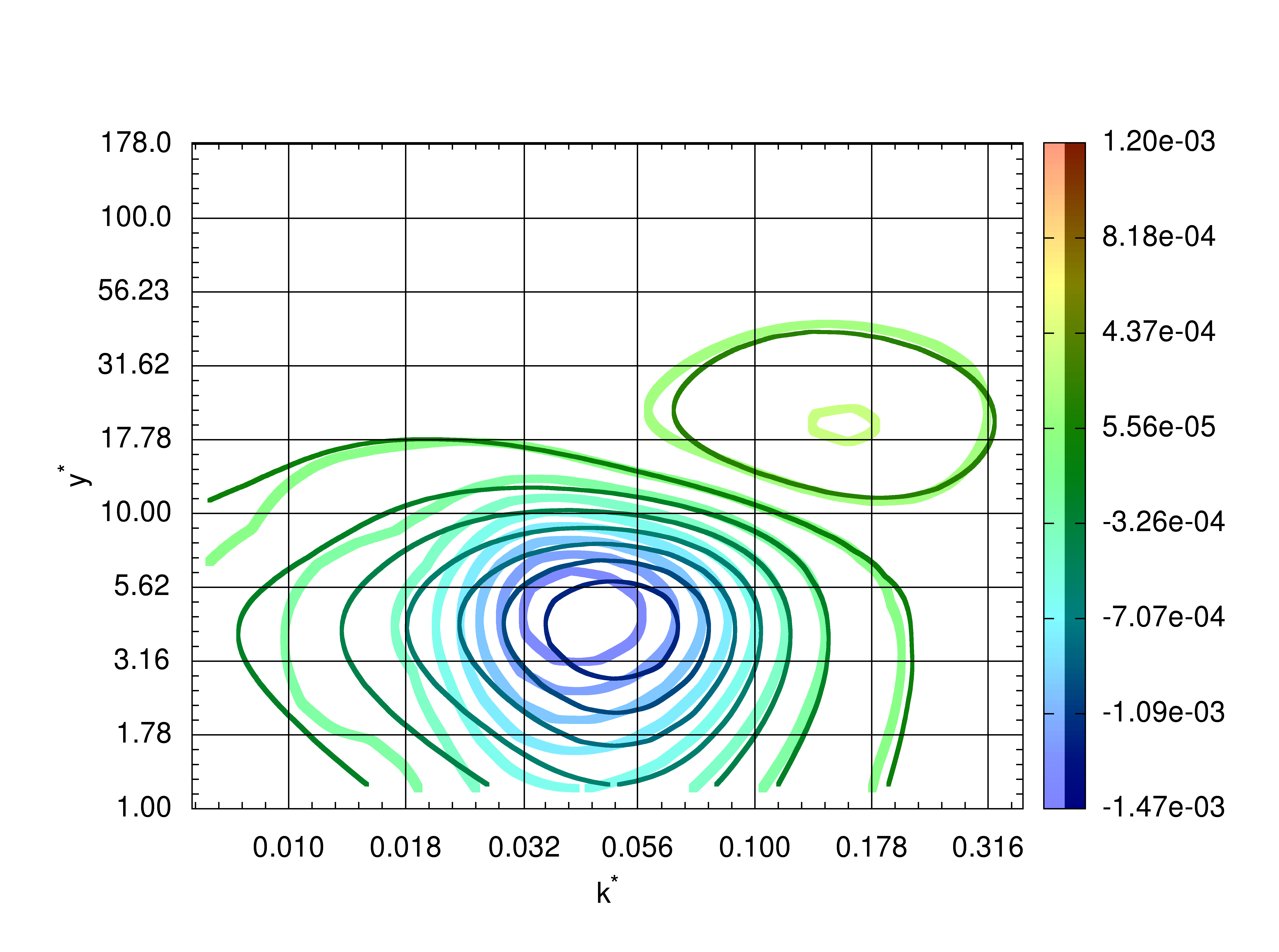}}
}
\caption{
Spectra of the thermal interaction with internal energy $\m{\check{\varepsilon}_{\varGamma}}$ with semi-local scaling at $Re_{\tau}$ = 180 (thick lines, light colour) and $Re_{\tau}$ = 395 (thin line, dark colour).
\label{r048}}
\end{figure*}

The spectra of the three most significant thermal terms include the spectra of
the interaction with variable density kinetic energy $\m{\check{\zeta}}$, a thermal energy
exchange, and the thermal conservative energy transfer $\m{\check{\varphi}_{\varGamma}}$ and the thermal
interaction with internal energy $\m{\check{\varepsilon}_{\varGamma}}$, the thermal parts of energy
exchanges.
They are given with the semi-local scaling  at $Re_{\tau} = 180$ and
$Re_{\tau} = 395$ in the set of figures \ref{r045} to \ref{r048}.
The effect of the mean friction Reynolds number on the spatial profiles of the
thermal terms was found to be small. In the spectral domain, more differences
are identified.
For the three thermal terms (figures \ref{r045} to \ref{r048}),
the amplitude of the extremum close to the wall is decreased at the hot and cold sides at $Re_{\tau} = 395$ compared to $Re_{\tau} = 180$,
by around 20\% at the hot side and by around 10\% at the cold side. This is rather unexpected since the amplitude of the spatial profiles is largely unaffected by the mean friction Reynolds number. The lower spectral magnitude
are counterbalanced by the larger range of scales handled.
In addition to a spatial shift towards the wall affecting all terms, the
extremum close to the wall is shifted towards large scales at the hot side and towards small
scales at the cold side. This reduces the wavenumber asymmetry between the hot
and cold sides and suggests that this asymmetry is tied to a low Reynolds
number effect.
The mean friction Reynolds number seems to modify the spectral behaviour of the
thermal terms but does not alter significantly their spatial profiles within
the Reynolds number range of this study.

\subsubsection{Kinetic energy spectrum}

\begin{figure*}
\centerline{
\includegraphics[width=0.50\textwidth, trim={0 10 15 5}, clip, clip]{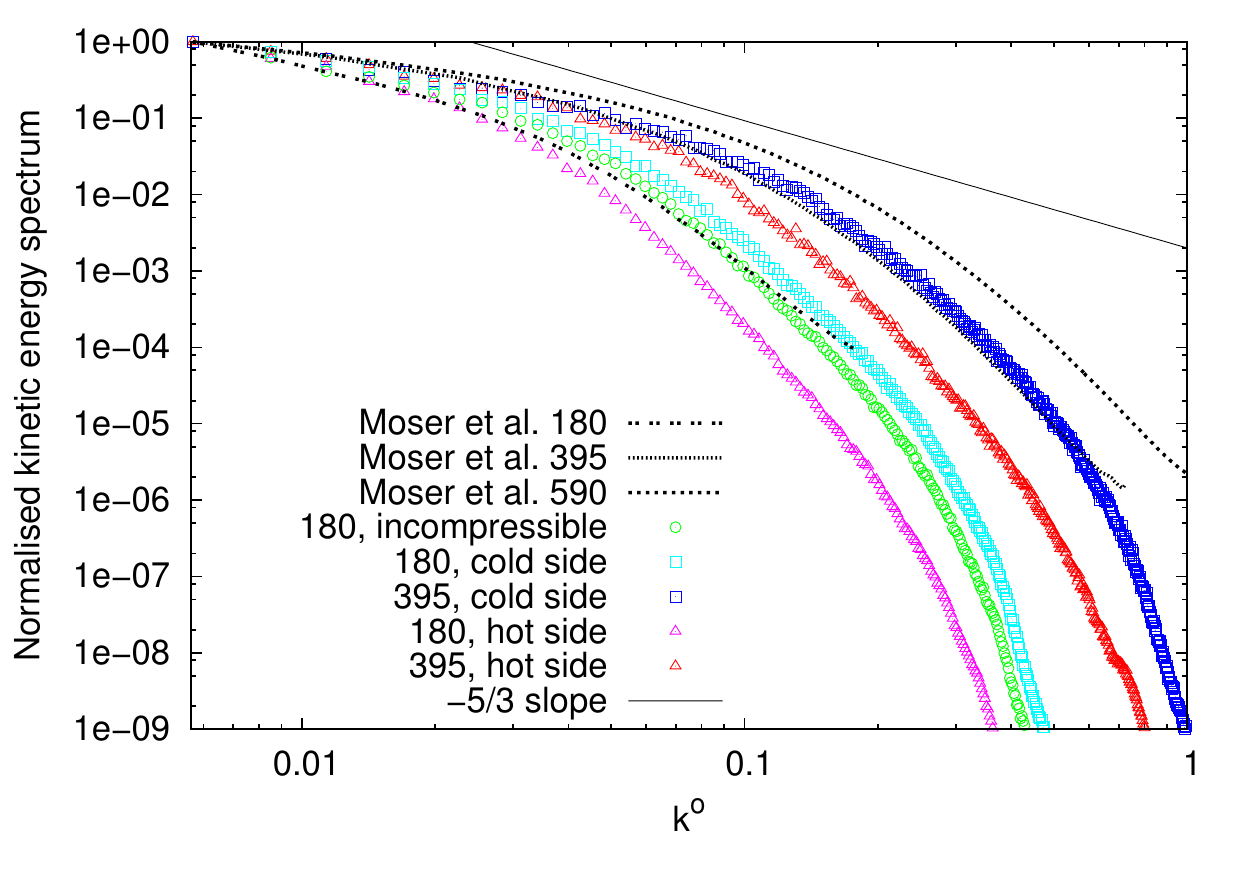}
}
\caption{
Normalised streamwise spectrum of the half-trace of the velocity covariance tensor at $Re_{\tau}$ = 180
and $Re_{\tau}$ = 395 at the location of the spatial peak of production~($y^* \approx 12$).
\label{r049}}
\end{figure*}

The energy exchanges drive the spectrum of the half-trace of the velocity
covariance tensor, which is also asymmetrised between the hot and
cold sides.
Figure \ref{r049} gives the streamwise spectra in the anisothermal
channels at $Re_{\tau}$ = 180 and $Re_{\tau}$ = 395 at the location of the
spatial peak of production. To compare the slope of the different spectra at
the hot and cold sides, the spectra are normalised to one at the smallest
wavenumber common to all simulations.
The slope of the spectrum is increased at the hot side and decreased at the cold
side. The effect is consistent with an effect of the semi-local friction
Reynolds number as shown by the reference data of \citet{moser1999a}.
We are not able to distinguish an effect of the variations of the fluid properties
on the slope of the spectrum.

\section{Conclusion}

The effect of the Reynolds number on the energy exchanges associated with the
half-trace of the velocity covariance tensor is investigated in
the spatial and spectral domains in a strongly anisothermal low Mach fully
developed turbulent channel flow.
The study is based on the direct numerical simulations of the channel at the
mean friction Reynolds number of $Re_{\tau} = 180$ and $Re_{\tau} = 395$.
The temperature gradient creates an asymmetry between the energy exchanges at
the hot and cold side.
The asymmetry can be attributed to the combined effect
of the variations of the local fluid properties and a varying low Reynolds
number effect, based on the semi-local friction Reynolds number which varies
across the channel.
The effect of the variations of the local fluid properties is to some extent
taken into account by the semi-local scaling, defined using the wall
shear-stress and the mean local fluid properties. The low Reynolds number
effects, not taken into account by the semi-local scaling, are investigated
using reference data in the constant-property isothermal channel.
The effects of the semi-local friction Reynolds number variations are smaller
at higher mean friction Reynolds number, reducing the asymmetry between
the hot and cold sides.
With the semi-local scaling, the energy exchanges occur closer to the wall at
$Re_{\tau} = 395$ than at $Re_{\tau} = 180$, handle a larger range of scales
and have a larger spatial amplitude.
The energy exchanges are decomposed to isolate the terms specific to flows with
variable fluid properties, called thermal terms. The mean friction Reynolds
number modifies the spectral behaviour of the thermal terms but
does not affect significantly their spatial profile.

\section*{Acknowledgments}
The authors gratefully acknowledge the CEA for the development of the TRUST
platform and TrioCFD. This work was granted access to the HPC resources of CINES under the
allocations 2017-A0022A05099 and 2018-A0042A05099 made by GENCI.

\let\stdthebibliography\thebibliography
\renewcommand{\thebibliography}{%
\let\chapter\section
\stdthebibliography}
  {
  \small
  \bibliography{./biblio}
  }
}

\end{document}